 \documentclass[aps,showpacs,dvips,showkeys]{revtex4}

\usepackage{graphicx}
\usepackage{color}
\usepackage{amsmath}

\def \be  {\begin{equation}}
\def \ee  {\end{equation}}
\def \ee  {\end{equation}}
\def \bea {\begin{eqnarray}}
\def \eea {\end{eqnarray}}
\def \le   {\left}
\def \ri   {\right}

\newcommand{\nn}{\nonumber}

\newcommand{\dslash}{\ensuremath{\partial\hspace{-1.2ex} /}}
\newcommand{\bsig}{\ensuremath{\bar{\sigma}}}

\newcommand{\lsm}{LSM}

\begin{document}

\preprint{ECTP-2014-03\hspace*{0.5cm}and\hspace*{0.5cm}WLCAPP-2014-03}

\title{Polyakov SU(3) extended linear $\sigma$-model: Sixteen mesonic states in chiral phase-structure}

\author{Abdel Nasser Tawfik\footnote{http://atawfik.net/}}
\affiliation{Egyptian Center for Theoretical Physics (ECTP), Modern University for Technology and Information (MTI), 11571 Cairo, Egypt}
\affiliation{World Laboratory for Cosmology And Particle Physics (WLCAPP), Cairo, Egypt}

\author{Abdel Magied Diab} 
\affiliation{World Laboratory for Cosmology And Particle Physics (WLCAPP), Cairo, Egypt}

\begin{abstract}

In the mean field approximation, derivative of the grand potential, non-strange and strange condensates and deconfinement phase-transition in thermal and dense hadronic medium are verified in the  SU(3) Polyakov linear $\sigma$-model (P\lsm). The chiral condensates $\sigma_x$ and $\sigma_y$ are analysed towards determining the chiral phase-transition. The temperature- and density-dependence of the chiral mesonic phase-structures is taken as free parameters and fitted experimentally. They are classified according to the scalar meson nonets; (pseudo)-scalar and (axial)-vector.  For the deconfinement phase-transition, the effective Polyakov loop-potentials $\phi$ and $\phi^*$ are implemented.
The in-medium effects on the masses of sixteen mesonic states are investigated. The results are presented for two different forms for the effective Polyakov-loop potential and compared with other models, which include and exclude the anomalous terms. It is found that the Polyakov-loop potential has considerable effects on the chiral phase-transition so that the restoration of the chiral symmetry-breaking becomes sharper and faster. Assuming that the Matsubara frequencies contribute to the meson masses, we have normalized all mesonic states with respect to the lowest frequency. By doing this, we characterize temperatures and chemical potentials, at which the different meson states dissolve to {\it free} quarks. Different dissolving temperatures and chemical potentials are estimated. The different meson states survive the {\it typically-averaged} QCD phase boundary, which is defined by the QCD critical temperatures at varying chemical potentials. The thermal behavior of all meson masses has been investigated in large-$N_c$ limit. It is found that at high $T$, the scalar meson masses are $T$-independent (except $\pi$ and $\sigma$). For the pseudoscalar meson masses, the large $N_c$ limit unifies the $T$-dependences of the various states into a universal bundle. The same is also observed for axial and axialvector meson masses.

\end{abstract}


\maketitle

\section{Introduction}
\label{intro}

The systematic study of strongly interacting matter at finite density allows analysing special theories that probably agree with the heavy-ions experiments aiming to tackle the quantum chromodynamic (QCD) phase-transition between combined nuclear matter and the quark-gluon plasma QGP and improving our understanding of the evolution of the early Universe. All these can be probed in experiments like STAR at the Relativistic Heavy-Ion Collider RHIC (BNL), ALICE at the Large Hadron Collider LHC (CERN), Compressed  Baryon Matter (CBM) at the Facility for Antiproton and Ion Reaserch (GSI) and Baryonic Matter at the Nuclotron (BM@N) at the Nuclotron-based  Ion Collider fAcility (JINR). In-medium effects on thermodynamics quantities is presented in the numerical solutions of difference effective models, especially the QCD-like ones. There are two main first-principle models, the Polyakov  Nambu-Jona-Lasinio (PNJL) and Polyakov linear $\sigma$ model (P\lsm) or the Polyakov quark meson (PQM) model.

As the finite quark masses break the chiral symmetry of QCD, explicitly, one has to resort numerical calculations in order to determine the chiral phase-transition, such as SU(3)$_r \times$ SU(3)$_\ell $ linear $\sigma$-model \cite{Gell Mann:1960}. Thus, SU(3)$_l \times$ SU(3)$_r \times$ U(1)$_A \rightarrow$ SU(3)$_V \times$ SU(3)$_A$. Long time ago, the quark constituents of scalar mesons have been debated \cite{reffff2a,reffff2b}. Accordingly, the determination of all meson states is possible  $\langle\bar{q}q\rangle=\langle\bar{q}_r q_\ell +\bar{q}_\ell q_r\rangle\ne 0$~\cite{C. Vafa:2007}. The chiral structure of the four categories of the meson states is classified through quantum numbers, orbital angular momentum $J$, parity $P$ and charge conjugate $C$, which can be constructed from $u-$ and $d-$ and $s-$quarks, into scalars ($J^{PC}=0^{++}$) and pseudoscalars ($J^{PC}=0^{-+}$), vectors ($J^{PC}=1^{-}$) and axial-vectors ($J^{PC}=1^{++}$). As the chiral symmetry is explicitly broken, the deconfinement phase-transition likely affects the mass spectrum and shows under which conditions certain state degenerates with another one and when the thermal and dense evolution goes through phase transition.  

In the present work, the in-medium effects on the masses of different meson states are analysed, systematically. We study the effects of finite temperature on sixteen meson states at vanishing and finite baryon-chemical potentials and also their density-dependence at finite temperatures. To this end, extending LSM to PLSM, in which information about the confining gluonic sector is also embedded in form of the Polyakov-loop potential is very crucial. The Polyakov- loop potential is extracted from pure Yang-Mills lattice simulations \cite{Polyakov:1978vu, Susskind:1979up, Svetitsky:1982gs,Svetitsky:1985ye}. In investigating the chiral phase-transition, \lsm$\,$ at finite temperature has been implemented \cite{Lenaghan:1999si, Petropoulos:1998gt}. Furthermore, U(N$_f$)$_r  \times$ U($N_f$)$_l$ \lsm$\,$ with $N_f=2$, $3$ or even $4$ quark flavors has been analysed \cite{l, Hu:1974qb, Schechter:1975ju, Geddes:1979nd}. 

\lsm$\,$  thermodynamic properties like pressure, equation of state, speed of sound, specific heat and trace anomaly can be evaluated at finite and vanishing baryon-chemical potential \cite{Schaefer:2008ax,Mao:2010,Schaefer B:2009,Tawfik2014lsm,Tawfik:2014bna} and under effects of an external magmatic field \cite{Tawfik:2014hwa}. Furthermore, the normalized and non-normalized higher-order moments of the particle multiplicity are investigated \cite{Schaefer:2009ab,Schaefer a:2011,Tawfik2014lsm}. With the inclusion of Polyakov-loop correction, the chiral phase-structure of the scalar and pseudoscalar meson states at finite and vanishing temperatures have been evaluated \cite{Schaefer:2009} with and without axial anomaly  \cite{V. Tiwari:2009,V. Tiwari:2013}. At finite isospin chemical potential, a three-flavor NJL  model for scalar and pseudoscalar mesonic states was presented in Ref. \cite{NJL:2013}. In the three-flavor PNJL model \cite{P. Costa:PNJL}, it is found that the inclusion of Polyakov-loop potential in the NJL model considerably affects the meson masses. Results from $2+1$  lattice QCD for pseudoscalar and vector meson states \cite{FodorMssA,FodorMssB,HotQCD,PACS-CS} and QCD thermodynamics including meson masses at vanishing temperature have been reported \cite{FodorT0}. The results deduced from Hot QCD \cite{HotQCD} and PACS-CS \cite{PACS-CS} are compared with the Particle Data Group (PDG) \cite{PDG:2012}. An excellent agreement was presented in Ref. \cite{NJL:2013,Schaefer:2009,V. Tiwari:2009,V. Tiwari:2013,HotQCD,PACS-CS,PDG:2012}.  

In general, P\lsm$\,$ has a wide range of implications. Not only the thermodynamics \cite{Mao:2010,Schaefer thermo:2009,Tawfik2014lsm,Schaffner:2013 thermo} but it can also describe the higher-order moments of the particle multiplicity \cite{Tawfik2014lsm,Schaefer:2009ab}, the hadron vacuum phenomenology \cite{Dirk Hparameters:2010,Rischke:2010 vacuum,Rischke:2012,Wolf:2011 vacuum,Rischke:2011 vacuum,Rischke:2010 decay} and the effects of the chiral and deconfinement phase-transitions \cite{Schaefer:2007 phase,Schaffner:2013 chiral,Tawfik2014phase} besides the chiral phase-structure of hadrons (the spectrum of hadrons in both thermal- and hadronic dense-medium) \cite{Rischke:2007,Lenaghan:2000ey,Schaefer:2009,V. Tiwari:2009,V. Tiwari:2013}, the decay width and the scattering length of hadronic states \cite{Rischke:2008 decay,Parganlija:2008,Rischke:2010 decay,Dirk Hparameters:2010,Rischke:2012,Rischke:2011 vacuum}.

In the present work, we introduce a systematic study using the chiral symmetric linear $\sigma$-model. We included in it scalar, pseudoscalar, vector, and axial-vector fields and estimate the representation of all these four categories in dependence on the temperature $T$ and the baryon-chemical potential $\mu$. This allows to define the characteristics of the chiral phase-structure for all these meson states in thermal and dense medium and determine the critical temperature and density, at which each meson state breaks into its free quarks.

The present paper is organized as follows. Section \ref{Model} gives details about the SU(3) Polyakov linear $\sigma$-model P\lsm$\,$, where the Lagrangian of the scalar and pseudoscalar fields are extended to include vector and axial-vector fields as well and interaction between mesonic sector in the presence of U(1)$_A$ symmetry breaking. The Ployakov-loop correction to the Lagrangian of P\lsm$\,$ is introduced in section \ref{PLOYAKOV}. The mean field approximation is outlined in section \ref{subsec:mean field}. The phase transition including quark condensates and order parameters shall be estimated in section \ref{ChiralphaseT}. Topics like deconfinement (crossover) phase-transition and order parameter due to chiral symmetry breaking shall be studied as well. In section  \ref{Meson}, we introduce the Ployakov-loop potential to \lsm$\,$ and investigate sixteen mesonic states in thermal (section \ref{T_dependence}) and hadronic dense medium (section \ref{Mu_dependence}). The critical temperature and the baryon chemical potential, at which each bound hadron state should dissolve into free quarks (QGP) shall be introduced in section \ref{QGP}.  Section \ref{sec:disc} is devoted to the conclusions.

\section{SU(3) Polyakov linear $\sigma$-model}
\label{Model}

The Lagrangian of \lsm$\,$  with $N_f =3$  quark flavors and $N_c =3$ color degrees of freedom, where the quarks couple to the Polyakov-loop dynamics $\Phi$-field represents a complex $(3\times 3)$-matrix for the SU(3)$_L \times$ SU(3)$_R$ symmetric \lsm$\,$ Lagrangian ${\cal L}_{chiral}= {\cal L}_q + {\cal L}_m$, where the fermionic part reads 
\bea
\label{eq:quarkL}
\mathcal{L}_{q} = \bar{q}\left[i \dslash - g\; T_a\,\left(\sigma_a + i\,  \gamma_5\, \pi_a + \gamma _{\mu} V_a ^{\mu}+\gamma _{\mu}\gamma _{5} A_a ^{\mu} \right)\,\right] q,
\eea
with $\mu$ is an additional Lorentz index \cite{Koch1997}, $g$ is the flavor-blind Yukawa coupling of the quarks to the mesonic contribution ${\cal L}_{m}={\cal L}_{SP}+ {\cal L}_{VA}+ {\cal L}_{Int}+  \mathcal{L}_{U(1)_A}$ represented to ${\cal L}_{SP}$ scalars ($J^{PC}=0^{++}$) and pseudoscalars ($J^{PC}=0^{-+}$), ${\cal L}_{VA}$ to vectors ($J^{PC}=1^{-}$) and axial-vectors ($J^{PC}=1^{++}$) mesons and $ {\cal L}_{Int}$ being the interaction between them. Finally the Lagrangian of  the anomaly term is given by $\mathcal{L}_{U(1)_A}$  
\cite{Gell Mann:1960,Gasiorowicz:1969,Rudaz:1994,Parganlija:2008,Pisarski:1995,Wolf:anomaly}.
\bea
  \mathcal{L}_{SP}&=&\mathrm{Tr}(\partial_{\mu}\Phi^{\dag}\partial^{\mu}\Phi-m^2
\Phi^{\dag} \Phi)-\lambda_1 [\mathrm{Tr}(\Phi^{\dag} \Phi)]^2
-\lambda_2 \mathrm{Tr}(\Phi^{\dag}
\Phi)^2 + \mathrm{Tr}[H(\Phi+\Phi^{\dag})], \label{scalar nonets} \\ 
  \mathcal{L}_{AV}&=&-\frac{1}{4}\mathop{\mathrm{Tr}}(L_{\mu\nu}^{2}+R_{\mu\nu}^{2}
)+\mathop{\mathrm{Tr}}\left[  \left( \frac{m_{1}^{2}}{2}+\Delta\right)  (L_{\mu}^{2}+R_{\mu}^{2})\right] \nn \\
&+&i \frac{g_{2}}{2} (\mathop{\mathrm{Tr}}\{L_{\mu\nu}[L^{\mu},L^{\nu}]\}+\mathop{\mathrm{Tr}}\{R_{\mu\nu}[R^{\mu},R^{\nu}]\}){\nonumber}\\
&+& g_{3}[\mathop{\mathrm{Tr}}(L_{\mu}L_{\nu}L^{\mu}L^{\nu}
)+\mathop{\mathrm{Tr}}(R_{\mu}R_{\nu}R^{\mu}R^{\nu})]+g_{4}
[\mathop{\mathrm{Tr}}\left(  L_{\mu}L^{\mu}L_{\nu}L^{\nu}\right)
+\mathop{\mathrm{Tr}}\left(  R_{\mu}R^{\mu}R_{\nu}R^{\nu}\right)
]{\nonumber}\\
&+&g_{5}\mathop{\mathrm{Tr}}\left(  L_{\mu}L^{\mu}\right)
\,\mathop{\mathrm{Tr}}\left(  R_{\nu}R^{\nu}\right)  +g_{6}
[\mathop{\mathrm{Tr}}(L_{\mu}L^{\mu})\,\mathop{\mathrm{Tr}}(L_{\nu}L^{\nu
})+\mathop{\mathrm{Tr}}(R_{\mu}R^{\mu})\,\mathop{\mathrm{Tr}}(R_{\nu}R^{\nu
})],\label{vector nonets}
\\
  \mathcal{L}_{Int}&=&\frac{h_{1}}{2}\mathop{\mathrm{Tr}}(\Phi^{\dagger}\Phi
)\mathop{\mathrm{Tr}}(L_{\mu}^{2}+R_{\mu}^{2})+h_{2}%
\mathop{\mathrm{Tr}}[\vert L_{\mu}\Phi \vert ^{2}+\vert \Phi R_{\mu} \vert ^{2}]+2h_{3}%
\mathop{\mathrm{Tr}}(L_{\mu}\Phi R^{\mu}\Phi^{\dagger}),\label{INT}
\\ 
\mathcal{L}_{U(1)_A}&=&c[\mathrm{Det}(\Phi)+\mathrm{Det}(\Phi^{\dag})]+c_0 [\mathrm{Det}(\Phi)-\mathrm{Det}(\Phi^{\dag})]^2 +c_1 [\mathrm{Det}(\Phi)+\mathrm{Det}(\Phi^{\dag})]\,\mathrm{Tr} [\Phi \Phi^{\dag}]. \hspace*{10mm}
 \label{eq:Lagrangian}
\eea
The first Lagrangian, Eq. (\ref{scalar nonets}), represents to kinetic and potential terms for the scalar meson nonets. The third term stands for the explicit symmetry breaking defined in Eq. (\ref{symmtery}). This Lagrangian creates scalar and pseudoscalar mesonic states defined in $\Phi$ nonets, Eq. (\ref{fieldmatrix}). While the second Lagrangian, Eq. (\ref{vector nonets}), represents the vector meson nonets involving explicit symmetry breaking in the second term defined Eq. (\ref{symmtery}). The  $3 \times 3$ matrix of the vector meson nonets involves vector and axial-vector fields, Eq. (\ref{fieldmatrix}).  This creates the vector and axial-vector mesonic states and the interactions between the (pseudo)-scalar and (axial)-vector introduced in Eq. (\ref{INT}). As the symmetry is broken, explicitly and spontaneously, the anomaly term $\mathcal{L}_{\text{U(1)}_A}$  in SU(3)$_r \times$ SU(3)$_\ell$ should be introduced into the effective Lagrangian and $c, c_0, c_1$ are the parameters to be determined, experimentally \cite{Rischke:2012}. The first two terms approximate the original axial anomaly term \cite{Schechter:1980,Schechter:2008}, while the third term  is a mixed one. It is proportionally to the first term. The concept of choosing the first anomaly term is essential, in which the other terms are used to compare with other effects of the different anomaly terms on the hadronic structure \cite{Wolf:anomaly}. 

To describe experimental data, large order terms  with local chiral symmetry should be included \cite{Rischke:2012}. It is worthwhile to highlight that  $\mathcal{L}_{\text{U(1)}_A}$ symmetry in the QCD Lagrangian is anomalous \cite{S. Weinberg}, known as QCD vacuum anomaly \cite{S. Weinberg,Schaefer:2009}, i.e. broken by quantum effects. Without anomaly a ninth pseudoscalar Goldstone boson corresponding to the spontaneous breaking of the chiral U(3)$_{\ell} \times$ U(3)$_r$ symmetry should unfold \cite{S. Weinberg,Schaefer:2009}. It is apparent that the hadron theory is not fundamental. Thus, it is assumed  to be valid at mass scale of $1-2~$GeV \cite{Rischke:2012} and therefore, the local chiral symmetry would not cause big problem. Nevertheless, the constraint-terms are conjectured to affect such QCD approaches \cite{Rischke:2012}. This well-known $\mathcal{L}_{\text{U(1)}_A}$ problem of QCD is effectively controlled by the anomaly term $c$ in the Lagrangian \cite{PC}. The squared tree-level masses of mesons $m^2$ and $m_{1}^{2}$ contain a contribution arising from the spontaneous symmetry breaking \cite{Rischke:2012}.

The introduction of scalar and vector meson nonets into the Lagrangian of P\lsm$\,$ requires redefinition for the contra-covariant derivative of the quark meson contribution represented in Eq. (\ref{derivative}), where the degrees of freedom of scalar $\Phi$ and vector $L^\mu $ and $R^\mu$ meson nonets are coupling to the electromagnetic field $A^{\mu}$. Eqs. (\ref{vector}) and (\ref{axial}) are the left-handed and right-handed field strength tensors, respectively. They represent the self interaction between the vector and axial-vector mesons with the electromagnetic field $A^{\mu}$. The local chiral invariance emerging from the globally invariant P\lsm$\,$ Lagrangian requires that $g_1=g_2=g_3=g_4=g_5=g_6=g$ \cite{Rischke:2012}
\bea
D^{\mu}\Phi & \equiv &\partial^{\mu}\Phi-ig_{1}(L^{\mu}\Phi-\Phi R^{\mu
})-ieA^{\mu}[T_{3},\Phi], \label{derivative} \\
L^{\mu\nu}  &  \equiv &\partial^{\mu}L^{\nu}-ieA^{\mu}[T_{3},L^{\nu}]-\left\{
\partial^{\nu}L^{\mu}-ieA^{\nu}[T_{3},L^{\mu}]\right\}, \label{vector} \\
R^{\mu\nu}  &  \equiv &\partial^{\mu}R^{\nu}-ieA^{\mu}[T_{3},R^{\nu}]-\left\{
\partial^{\nu}R^{\mu}-ieA^{\nu}[T_{3},R^{\mu}]\right\}.\label{axial}
\eea

It is apparent that $T_{a} = \hat{\lambda}_{a}/2$ with $a=0\dots 8$ are nine U(3) generators, where $\hat{\lambda}_{a}$ are the Gell-Mann matrices with the fields $\Phi$ of $3\times 3$ complex matrix comprising of the scalars $\sigma _{a}$ ($J^{PC}=0^{++}$), pseudoscalars  $\pi _{a}$ ($J^{PC}=0^{-+}$), $V_{a}^{\mu}$, vectors ($J^{PC}=1^{-}$) and $A_{a}^{\mu}$ axial-vectors ($J^{PC}=1^{++}$) meson states given by
\bea
\label{fieldmatrix}
\Phi &=& \sum_{a=0}^{8} T_{a}(\sigma _{a}+ i \pi _{a}),  \nn \\
L^{\mu}  &=& \sum_{a=0} ^{8} \, T_{a}\, (V_{a}^{\mu}+A_{a}^{\mu}), \\
R^{\mu}  &=& \sum_{a=0}^{8}\, T_{a}\, (V_{a}^{\mu}-A_{a}^{\mu}). \nn
\eea
${\lambda}_{0}=\sqrt{\frac{2}{3}} \, {\bf 1}$ and $T_{a}$ are  normalized such that they obey the U(3) algebra \cite{Borodulin:1995}. The chiral symmetry is explicitly broken by 
\bea
H=\sum_{a=0}^{8} T_a h_a , \qquad &&
\Delta=\sum_{a=0}^{8} T_a \delta_a.
\label{symmtery}
\eea
The symmetry breaking terms are originated by  U(3)$_{L}\times$ U(3)$_{R}=$U(3)$_{V} \times$ U(3)$_{A}$. The terms are proportional to the matrix $H$ and $\Delta$ as given in Eq. (\ref{symmtery}). This relation describes the explicit symmetry breaking due to
\begin{itemize}
\item finite quark masses in the (pseudo)-scalar and (axial)-vector sectors,  
\item breaking $U(3)_{A}$ if $H_{0},\Delta_{0}\neq0$, and 
\item breaking $U(3)_{V}\rightarrow$ SU(2)$_{V}\times U(1)_{V}$ if $H_{8},\Delta_{8}\neq0$.
\end{itemize}
For more details, the readers are referred to Ref. \cite{Lenaghan:2000ey}. It is conjectured that the spontaneous chiral symmetry breaking takes part in vacuum state. Therefore, a finite vacuum expectation value for the fields $\Phi$ and $\bar{\Phi}$ are assumed to carry the quantum numbers of the vacuum \cite{Gasiorowicz:1969}. As a result, the components of the explicit symmetry breaking term (diagonal) are $h_0$, $h_3$  and $h_8$ and $\delta_0$, $\delta_3$ and $\delta_8$ should not vanish \cite{Gasiorowicz:1969}. This leads to exacting three finite condensates $\bar{\sigma_0}$, $\bar{\sigma_3}$ and $\bar{\sigma_8}$. On the other hand, $\bar{\sigma_3}$ breaks the isospin symmetry SU(2) \cite{Gasiorowicz:1969}. To avoid this situation, we restrict ourselves to SU(3). This can be $N_f= 2+1$ \cite{Schaefer:2009} flavor pattern. Correspondingly, two degenerate light (up-quark and down-quark) and one heavier quark flavor (strange-quark), i.e. $m_u=m_d \neq m_s$ are assumed. Furthermore, the violation of the isospin symmetry is neglected. This facilitates the choice of $h_a$ ($h_0 \neq 0$, $h_3=0$ and $h_8 \neq 0$) and for $\delta _a$  ($\delta_0 \neq 0$, $\delta_3 =0$  and $\delta_8 \neq 0$). 

\begin{mathletters}
\bea 
T_{a} \, \sigma_{a} &=& \frac{1}{\sqrt{2}} \left( \begin{array}{ccc} 
        \frac{1}{\sqrt{2}} \, a_{0}^{0} + 
        \frac{1}{\sqrt{6}} \, \sigma_{8} +
        \frac{1}{\sqrt{3}} \, \sigma_{0} & a_{0}^{-} & \kappa^{-} \\
        a_{0}^{+} & -\frac{1}{\sqrt{2}} \, a_{0}^{0} +
        \frac{1}{\sqrt{6}} \, \sigma_{8} +
        \frac{1}{\sqrt{3}} \, \sigma_{0} & \bar{\kappa}^{0} \\
        \kappa^{+} & \kappa^{0} & -\sqrt{\frac{2}{3}} \, \sigma_{8} +
        \frac{1}{\sqrt{3}} \, \sigma_{0} \end{array} \right) , \\ 
T_{a} \, \pi_{a} &=& \frac{1}{\sqrt{2}} \left( \begin{array}{ccc} 
        \frac{1}{\sqrt{2}} \, \pi^{0} + \frac{1}{\sqrt{6}} \, \pi_{8} +
        \frac{1}{\sqrt{3}} \, \pi_{0} & \pi^{-} & K^{-} \\
        \pi^{+} & -\frac{1}{\sqrt{2}} \, \pi^{0} +
        \frac{1}{\sqrt{6}} \, \pi_{8} +
        \frac{1}{\sqrt{3}} \, \pi_{0} & \bar{K}^{0} \\
        K^{+} & K^{0} & -\sqrt{\frac{2}{3}} \, \pi_{8} +
        \frac{1}{\sqrt{3}} \, \pi_{0} \end{array} \right).
\eea
\end{mathletters} 
and 
\bea
T_{a}\, V_{a}^{\mu} &=& \frac{1}{\sqrt{2}}\left(
\begin{array}
[c]{ccc}%
\frac{\omega_{0}+\rho^{0}}{\sqrt{2}}&
\rho^{+} & K^{\star+}\\
\rho^{-}& \frac{\omega_{0}-\rho^{0}}{\sqrt{2}} & K^{\star0}\\
K^{\star-} & {\bar{K}}^{\star0} & \omega _8
\end{array}
\right)  ^{\mu}\text{,} \label{eq:matrix_field_R} \\
T_{a}\, A_{a}^{\mu}&=& \frac{1}{\sqrt{2}
}\left(
\begin{array}
[c]{ccc}%
\frac{f_{1_0}+a_{1}^{0}}{\sqrt{2}} &
a_{1}^{+} & K_{1}^{+}\\
a_{1}^{-} & \frac{f_{1_0}%
-a_{1}^{0}}{\sqrt{2}} & K_{1}^{0}\\
K_{1}^{-} &{\bar{K}}_{1}^{0} &f_{1_8}%
\end{array}
\right) ^{\mu}. \label{eq:matrix_field_A}%
\eea
It would be more convenient when converting the condensates $\sigma_0$ and $\sigma_8$ into a pure non-strange $\sigma_x$ and a pure strange $\sigma_y$ quark flavor \cite{Kovacs:2006} 
\bea
\label{sigms}
\left( {\begin{array}{c}
\sigma _x \\
\sigma _y
\end{array}}
\right)=\frac{1}{\sqrt{3}} 
\left({\begin{array}{cc}
\sqrt{2} & 1 \\
1 & -\sqrt{2}
\end{array}}\right) 
\left({ \begin{array}{c}
\sigma _0 \\
\sigma _8
\end{array}}
\right).
\eea
It is worthwhile to mention that $\sigma \ni (\sigma_a , \pi_a , V_a ^{\mu}, A_a ^{\mu} )$.

\subsection{Polyakov Loop Potential}
\label{PLOYAKOV}

The Lagrangian of \lsm$\,$  can be coupled to the Polyakov-loop dynamics \cite{Schaefer:2009,Mao:2010},
\begin{eqnarray}
\mathcal{L}=\mathcal{L}_{chiral}-\mathbf{\mathcal{U}}(\phi, \phi^*, T), \label{plsm}
\end{eqnarray}
The second term in Eq. (\ref{plsm}), $\mathbf{\mathcal{U}}(\phi, \phi^*, T)$, represents the effective Polyakov-loop potential \cite{Polyakov:1978vu}, which gives the dynamics of the thermal expectation value of a color-traced Wilson loop in the temporal direction \cite{Polyakov:1978vu} 
\bea
\phi (\vec{x})=\frac{1}{N_c} \langle \mathcal{P}(\vec{x})\rangle.
\eea
Then, the Polyakov-loop potential and its conjugate read 
\begin{eqnarray}
\phi = \langle\mathrm{Tr}_c \,\mathcal{P}\rangle/N_c, \qquad & &
\phi^* = \langle\mathrm{Tr}_c\,  \mathcal{P}^{\dag}\rangle/N_c, \label{phais2}
\end{eqnarray}
where $\mathcal{P}$ is the Polyakov loop, which can be expressed as a matrix in the color space \cite{Polyakov:1978vu} 
\begin{eqnarray}
 \mathcal{P}(\vec{x})=\mathcal{P}\mathrm{exp}\left[i\int_0^{\beta}d \tau A_0(\vec{x}, \tau)\right],\label{loop}
\end{eqnarray}
where $\beta=1/T$ is the inverse temperature and $A_0$ is the temporal component of Euclidean vector field \cite{Polyakov:1978vu,Susskind:1979up}.
The Polyakov-loop matrix can be re-expressed as a diagonal representation \cite{Fukushima:2003fw}, as in Eq.  (\ref{phais2}), where the gauge filed $A_{\mu}=g_s\, A_{\mu}^a\, \lambda^a/2$ with $a = 1,\dots, N_{c}^2 - 1$ and $g_s$ being the gauge coupling. 

The coupling between the Polyakov loop and the quarks is unrivalled and given by the covariant derivative $D_{\mu}=\partial_{\mu}-i\, A_{\mu}$, Eq. (\ref{plsm}), where $A_{\mu}=\delta_{\mu 0}\, A_0$ is given in the chiral limit, Eq. (\ref{plsm}) and therefore is invariant under the chiral flavor group. This is the same as the QCD Lagrangian \cite{Ratti:2005jh,Roessner:2007,Fukushima:2008wg}. In order to reproduce the thermodynamic behavior of the Polyakov loop for the pure gauge, we use  temperature-dependent potential $U(\phi, \phi^{*},T)$, which agrees with lattice QCD calculations and have $Z(3)$ center symmetry \cite{Ratti:2005jh,Roessner:2007,Schaefer:2007d,Fukushima:2008wg} as that of the pure gauge QCD Lagrangian \cite{Ratti:2005jh,Schaefer:2007d}. In case of no quarks, $\phi=\phi^{*}$ and the Polyakov loop is considered as an order parameter for the deconfinement phase-transition \cite{Ratti:2005jh,Schaefer:2007d}. In the present work, we use $U(\phi, \phi^{*},T)$ as a polynomial expansion in $\phi$ and $\phi^{*}$ \cite{Ratti:2005jh,Roessner:2007,Schaefer:2007d,Fukushima:2008wg}
\begin{eqnarray}
\frac{\mathbf{\mathcal{U}}(\phi, \phi^*, T)}{T^4}=-\frac{b_2(T)}{2}|\phi|^2-\frac{b_3
}{6}(\phi^3+\phi^{*3})+\frac{b_4}{4}(|\phi|^2)^2, \label{Uloop}
\end{eqnarray}
where $b_2(T)=a_0+a_1\left(T_0/T\right)+a_2\left(T_0/T\right)^2+a_3\left(T_0/T\right)^3$. To reproduce the pure gauge QCD thermodynamics and the behavior of the Polyakov loop as a function of  temperature, we use the parameters $a_0=6. 75$, $a_1=-1. 95$, $a_2=2. 625$, $a_3=-7. 44$, $b_3=0.75$ and $b_4=7.5$ \cite{Ratti:2005jh}.  Accordingly, the deconfinement temperature, $T_0=270~$MeV, in the pure gauge sector.

\subsection{Mean Field  Approximation}
\label{subsec:mean field} 

The partition function can be constructed, when taking into consideration a spatially uniform system in a thermal equilibrium at finite temperature $T$ and finite quark chemical potential $\mu_f$, where $f$ stands for $u, d$ and $s$ quarks. The change in particles and antiparticles is governed by the grand canonical partition function. A path integral over the quark, antiquark and meson fields leads to \cite{Schaefer:2009}
\begin{eqnarray}
\mathcal{Z}&=& \mathrm{Tr \,exp}[-(\hat{\mathcal{H}}-\sum_{f=u, d, s}
\mu_f \hat{\mathcal{N}}_f)/T] = \int\prod_a \mathcal{D} \sigma_a \mathcal{D} \pi_a \int
\mathcal{D}\psi \mathcal{D} \bar{\psi} \mathrm{exp} \left[ \int_x
(\mathcal{L}+\sum_{f=u, d, s} \mu_f \bar{\psi}_f \gamma^0 \psi_f )
\right],
\end{eqnarray}
where $\int_x\equiv i \int^{1/T}_0 dt \int_V d^3x$ and $V$ is the volume of the system. For a symmetric quark matter, the uniform blind chemical potential fulfils the conditions that $\mu_f \equiv \mu_{u}=\mu_{d}=\mu_s$ \cite{Schaefer:2007c,Schaefer:2009,blind}.  The meson fields can be replaced by their expectation values $\bar{\sigma_x}$ and $\bar{\sigma_y}$ \cite{Kapusta:2006pm}. In estimating the integration over the fermions yields, other methods were introduced  \cite{Kapusta:2006pm}. The effective mesonic potential can be deduced and the thermodynamic potential density reads  
\begin{eqnarray}
\Omega(T, \mu)=\frac{-T \mathrm{ln}
\mathcal{Z}}{V}=U(\sigma_x, \sigma_y)+\mathbf{\mathcal{U}}(\phi, \phi^*, T)+ \Omega_{\bar{q}q}(T,\mu_f). \label{potential}
\end{eqnarray}

The explicit quark contribution to the  \lsm$\,$ is given as  
\begin{equation}
  \label{qqlsm}
  \Omega_{\bar{q}q}(T,\mu_f)= \nu_c T \sum_{f=u,d,s}
  \int\limits_0^\infty \! \frac{d^3 k}{(2\pi)^3}
  \left\{ \ln (1-n_{q,f}(T,\mu_f))
  + \ln (1-n_{\bar{q},f}(T,\mu_f)) \right\},
\end{equation}
with the usual fermionic occupation numbers (for quarks) $n_{q,f}(T,\mu_{f})=\{1+\exp[(E_{f}-\mu_f)/T]\}^{-1}$.
For antiquarks $n_{\bar q,f}(T,\mu_{f}) \equiv n_{q,f} (T,-\mu_{f})$. The number of internal quark degrees of freedom is denoted by $\nu_c=2 N_{c}=6$. The flavor-dependent single-particle energies are given as $E_{f}= (k^2 + m_f^2)^{1/2}$, where $m_f$ is the flavor-dependent quark masses.  Also, the light quark sector is conjectured to decouple from strange quark sector \cite{Kovacs:2006}. Assuming degenerate light quarks, i.e. $l\equiv u, d$, then, the masses can be simplified as \cite{Kovacs:2006}
\bea
m_l = g \frac{\sigma_x}{2}, \qquad  && 
m_s = g \frac{\sigma_y}{\sqrt{2}}.  \label{sqmass}
\eea 

For PLSM, the quarks and antiquarks contributions to the potential are given as \cite{Kapusta:2006pm} 
\begin{eqnarray} \label{PloykovPLSM}
\Omega_{  \bar{q}q}(T, \mu)&=& -\nu_c \,T \sum_{f=l, s} \int_0^{\infty} \frac{d^3\vec{p}}{(2 \pi)^3} \left\{ \ln \left[ 1+3(\phi+\phi^* e^{-(E_f-\mu)/T})\times e^{-(E_f -\mu)/T}+e^{-3 (E_f-\mu)/T}\right] \right. \nonumber \\ 
&& \hspace*{32mm} \left.  +\ln \left[ 1+3(\phi^*+\phi e^{-(E_f+\mu)/T})\times e^{-(E_f+\mu)/T}+e^{-3 (E_f+\mu)/T}\right] \right\}, 
\end{eqnarray}
Based on non-strange $\sigma_x$ and strange $\sigma_y$ condensates and taking into consideration Eq. (\ref{sigms}), then the purely mesonic potential reads 
\begin{eqnarray}
U(\sigma_x, \sigma_y) &=& - h_x \sigma_x - h_y \sigma_y + \frac{m^2}{2} (\sigma^2_x+\sigma^2_y) - \frac{c}{2\sqrt{2}} \sigma^2_x \sigma_y  + \frac{\lambda_1}{2} \sigma^2_x \sigma^2_y \nn + \frac{1}{8} (2 \lambda_1
+\lambda_2) \sigma^4_x + \frac{1}{4} (\lambda_1+\lambda_2)\sigma^4_y.  \label{Upotio}
\end{eqnarray}


\section{Phase Transitions and Their Order Parameters}
\label{ChiralphaseT}

By minimizing the thermodynamic potential, Eq. (\ref{potential}), with respective to $\sigma_x$, $ \sigma_y$, $\phi$ and $\phi^*$, we obtain a set of four equations of motion $\sigma_x$, $\sigma_y$, $\phi$ and $\phi^*$. 
\bea \label{cond1}
\left.\frac{\partial \Omega}{\partial \sigma_x}= \frac{\partial
\Omega}{\partial \sigma_y}= \frac{\partial \Omega}{\partial
\phi }= \frac{\partial \Omega}{\partial \phi^*}\right. \vert_{\rm {\sigma_x=\bar{\sigma_x}, \sigma_y=\bar{\sigma_y}, \phi=\bar{\phi},\,\phi^*=\bar{\phi^*}}} =0,
\eea
meaning that $\sigma_x=\bar{\sigma_x}$, $\sigma_y=\bar{\sigma_y}$, $\phi=\bar{\phi}$ and $\phi^*=\bar{\phi^*}$ being the global minimum, where all thermodynamics quantities are related to the parameters $\sigma_x$, $\sigma_y$, $\phi$ and $\phi^*$. 

In order to determine the chiral phase-transition, $\sigma_x$ and $\sigma_y$, and the deconfinement phase-transition, $\phi$ and $\phi^*$ should be estimated. The chiral mesonic phase-structures in temperature- and density-dependence are taken as free parameters to be fitted, experimentally. These parameters are classified corresponding to scalar meson nonets $m^2$, $h_x$, $h_y$, $\lambda_1$, $\lambda_2$ and $c$ \cite{Schaefer:2009}. The vector meson nonets have the parameters $m_{1}^2$, $g_{1}$, $h_{1}$, $h_{2}$, $h_{3}$, $\delta_{x}$ and  $\delta_{y}$ \cite{Dirk Hparameters:2010}. 

\begin{figure}[hbt]
\centering{
\includegraphics[width=5.5cm,angle=-90]{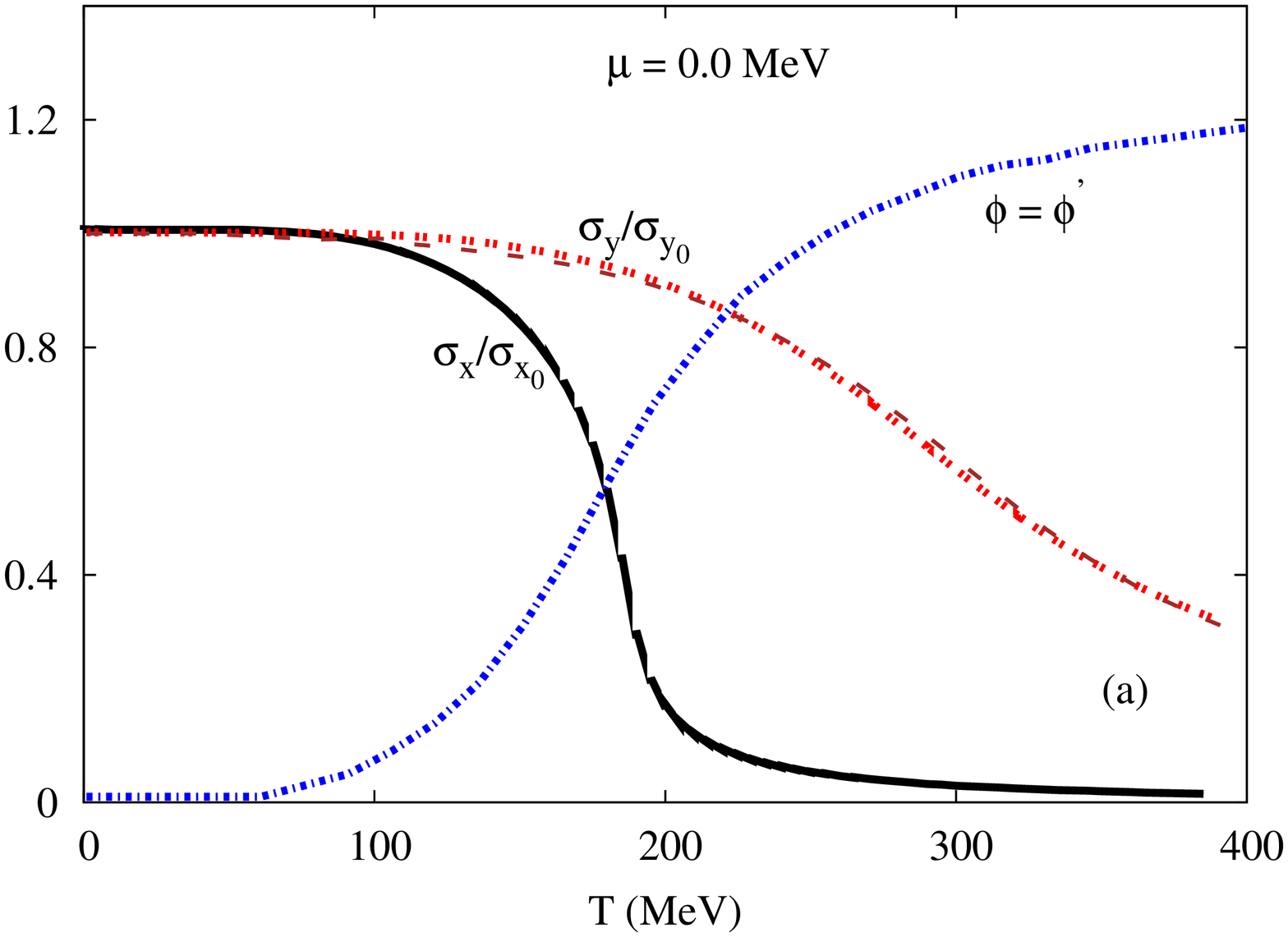}
\includegraphics[width=5.5cm,angle=-90]{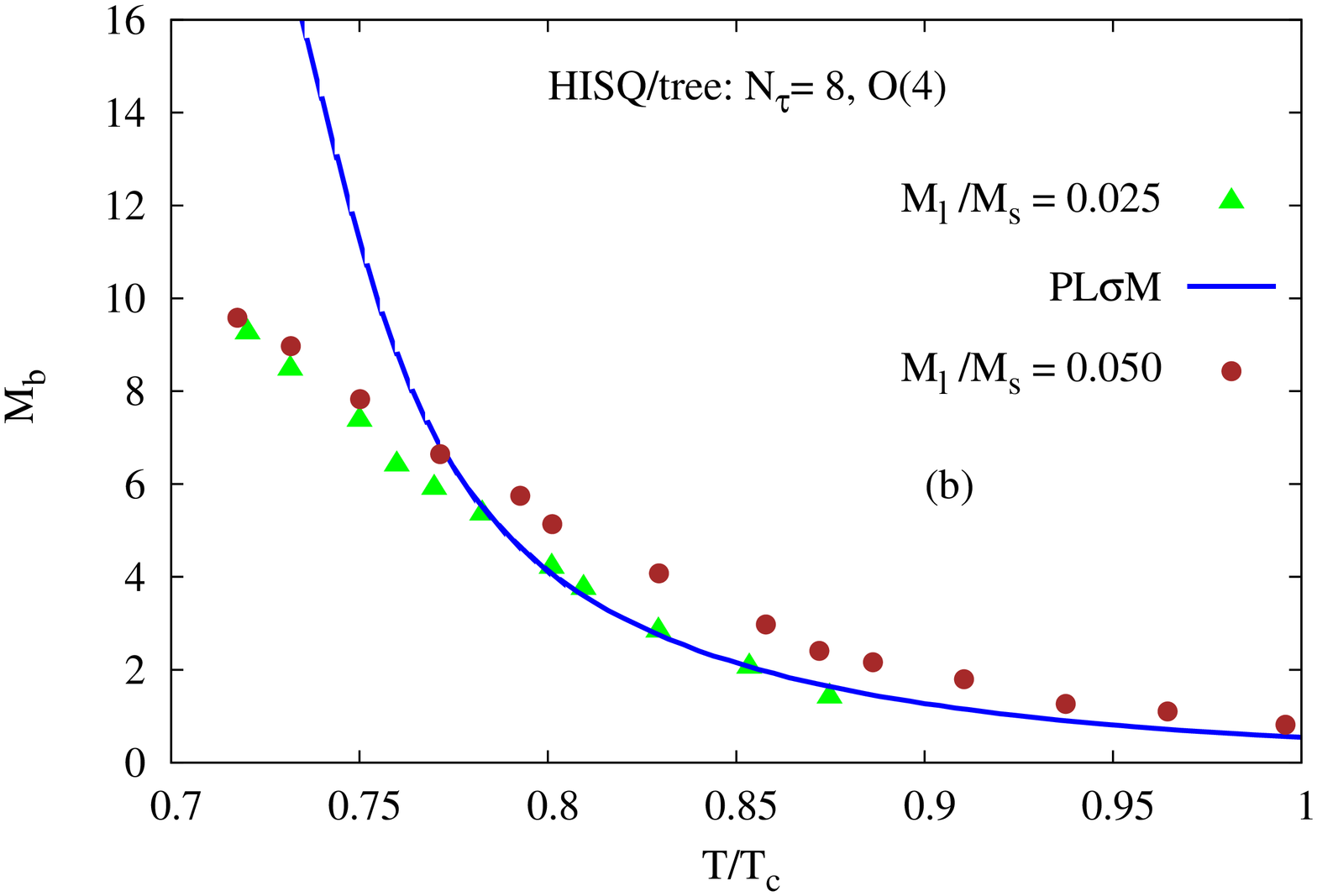}
\caption{(color online) Left-hand panel: the chiral condensates $\sigma_{x}$ and $ \sigma_{y}$ (solid and dotted curves, respectively) and the Polyakov loops $\phi$ and $\phi ^{*}$ (dashed curve at $c=0$, i.e. without anomaly) are given as functions of the temperature at vanishing baryon chemical potential. At $ \mu=0~$MeV, the two Polyakov loops are identical, i.e. $\phi$=$\phi^{*}$. Right-hand panel: the chiral condensate in $\mathcal{O}(4)~$lattices \cite{HotQCDtree} with HISQ/tree with $N_t = 8$ is compared with the P\lsm$\,$ calculations (solid curve). The rectangular symbols stand for $M_q/M_s=0.025$ and the circular ones represent $M_q/M_s=0.05$.
\label{fig:Transition}
} }
\end{figure}

In the present work, we use $\sigma=800~$MeV. At vanishing temperature, the chiral condensates for light and strange quarks are taken as $\sigma_{x_{0}}=92.4~$MeV and $\sigma_{y_{0}}=94.5~$MeV, respectively \cite{Schaefer:2009,Mao:2010}. These values are used to normalize their thermal evolution at vanishing chemical potential. In this limit, the two Polyakov loops are identical, i.e. $\langle\phi\rangle=\langle\phi^{*}\rangle$. To determining the critical temperature of the phase transition (crossover), two approaches can be implemented:
\begin{itemize}
\item The first one is the point, at which the order parameter intersects with the curve of the corresponding chiral condensate. 
\item The second one is based on the maxima/peaks of the temperature derivative of the condensates (chiral susceptibilities) for strange and nonstrange quarks. The peaks should be ordered to the critical temperatures. 
\end{itemize}
The first approach was used to derive the results depicted in Fig. \ref{fig:Transition}. Accordingly, we find that the chiral restoration of the non-strange condensate is related to $T_{c}^{q}\sim 181~$MeV, while for the strange  quark to $T_{c}^{s}\sim 270~$MeV.

The lattice QCD simulations prefer dimensionless quantities. Therefore, the chiral order parameter is expressed in the chiral condensate \cite{HotQCDtree} 
\bea
M_b= \frac{m_s \langle\bar{\sigma_x}(T, \mu)\rangle}{T^4}. \label{eq:mb}
\eea
The right-hand panel of Fig. \ref{fig:Transition} compares the chiral condensate from HISQ/tree with temporal dimensions $N_t = 8$, and two quark masses $M_q/M_s=0.025$ and $M_q/M_s=0.05$ in $\mathcal{O}(4)~$lattices \cite{HotQCDtree} with the  P\lsm$\,$  calculations for $M_b$, Eq. (\ref{eq:mb}). 

\begin{figure}[htb]
\centering{
\includegraphics[width=5.0cm,angle=-90]{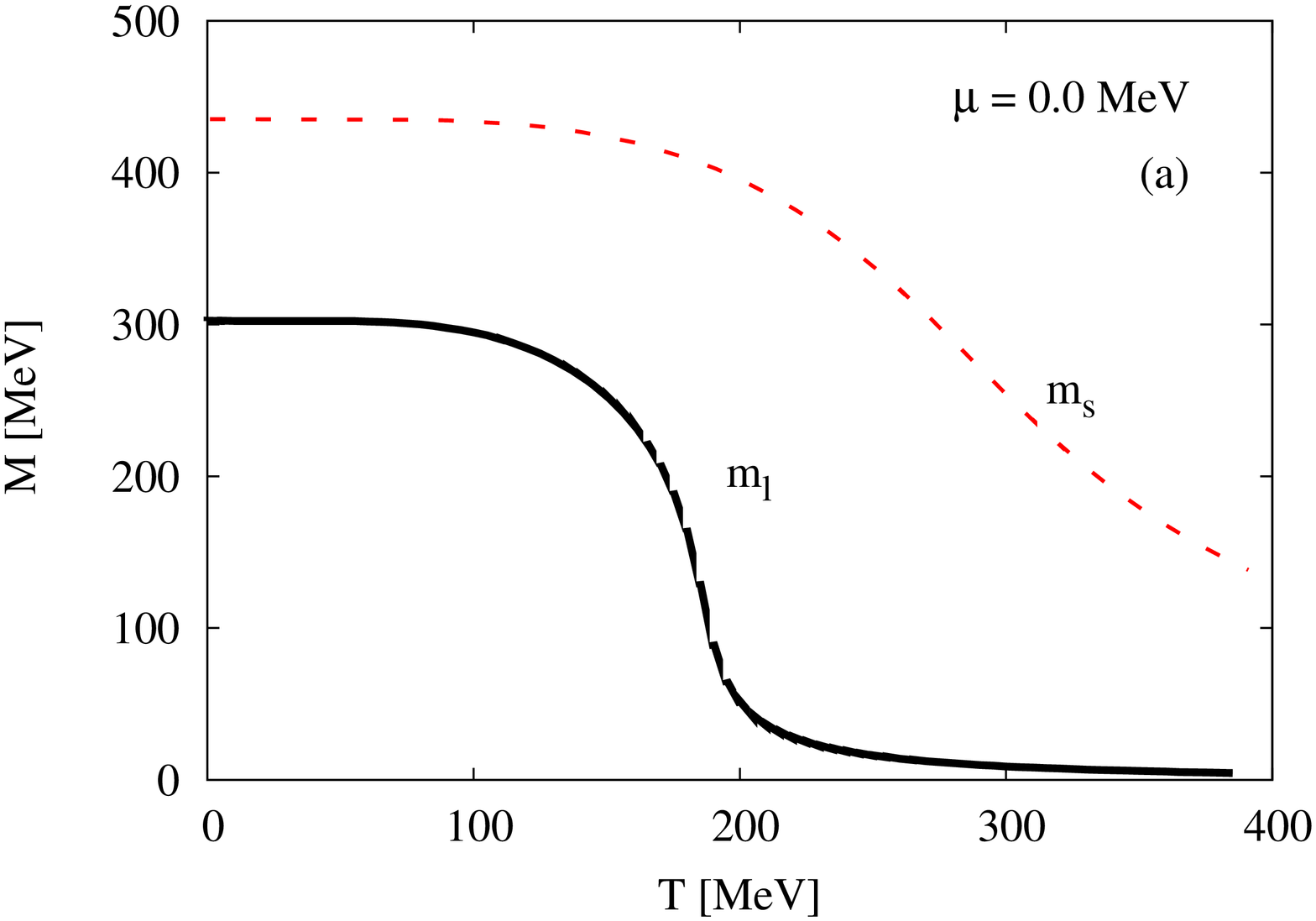}
\includegraphics[width=5.0cm,angle=-90]{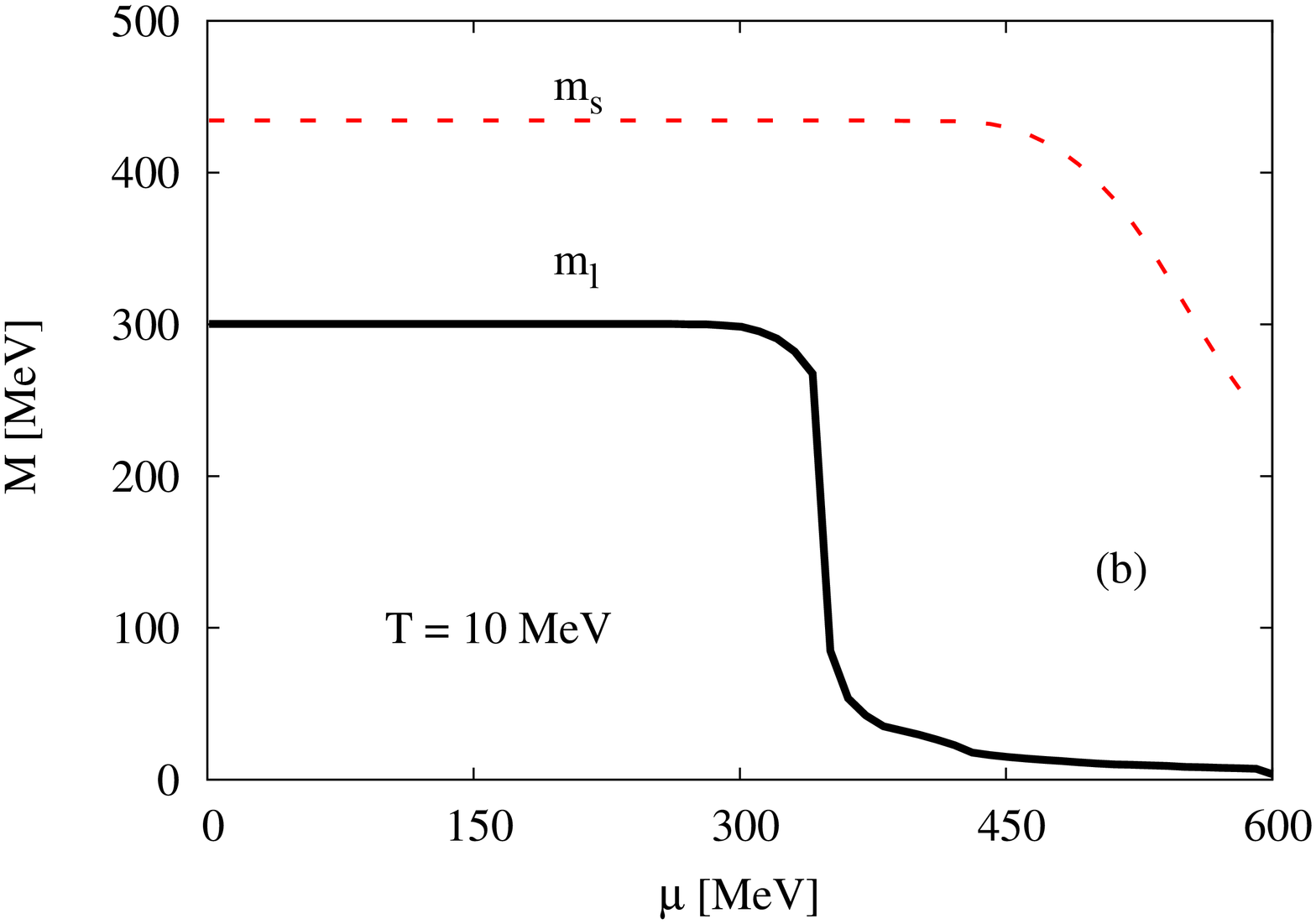}
\caption{(color online) The left-hand panel presents the thermal evolution of non-strange $m_l$ (solid curve) and strange (dashed curve) at $\mu =0~$MeV. The right-hand panel shows their dependence on the baryon chemical potential at a fixed temperature $T=10~$MeV. 
\label{fig:Massus} 
} }
\end{figure}

When the light constituent quark mass takes the value $m_l = 300~$MeV, the coupling $g=6.5$ and the strange constituent quark mass reads $m_s \sim 433~$MeV. These are normalized to the values at zero temperature $T$ and vanishing baryon-chemical potential $\mu$. In cases of finite $T$ and vanishing $\mu$ and vanishing $T$ and finite $\mu$, the chiral phase-transition is determined by non-strange and strange quarks fields, Eq. (\ref{sqmass})  as shown in Fig. \ref{fig:Massus}. The left-hand panel of Fig. \ref{fig:Massus} shows the thermal evolution of non-strange and strange quarks at vanishing $\mu$. The right-hand panel shows their density dependence at $T=10~$MeV. The contribution of finite quark mass seems to have a considerable effect the chiral phase-transition. To this end, the normalized condensates are studied in $T$- and $\mu$-dependence. 

\begin{figure}[htb]
\includegraphics[width=5cm,angle=-90]{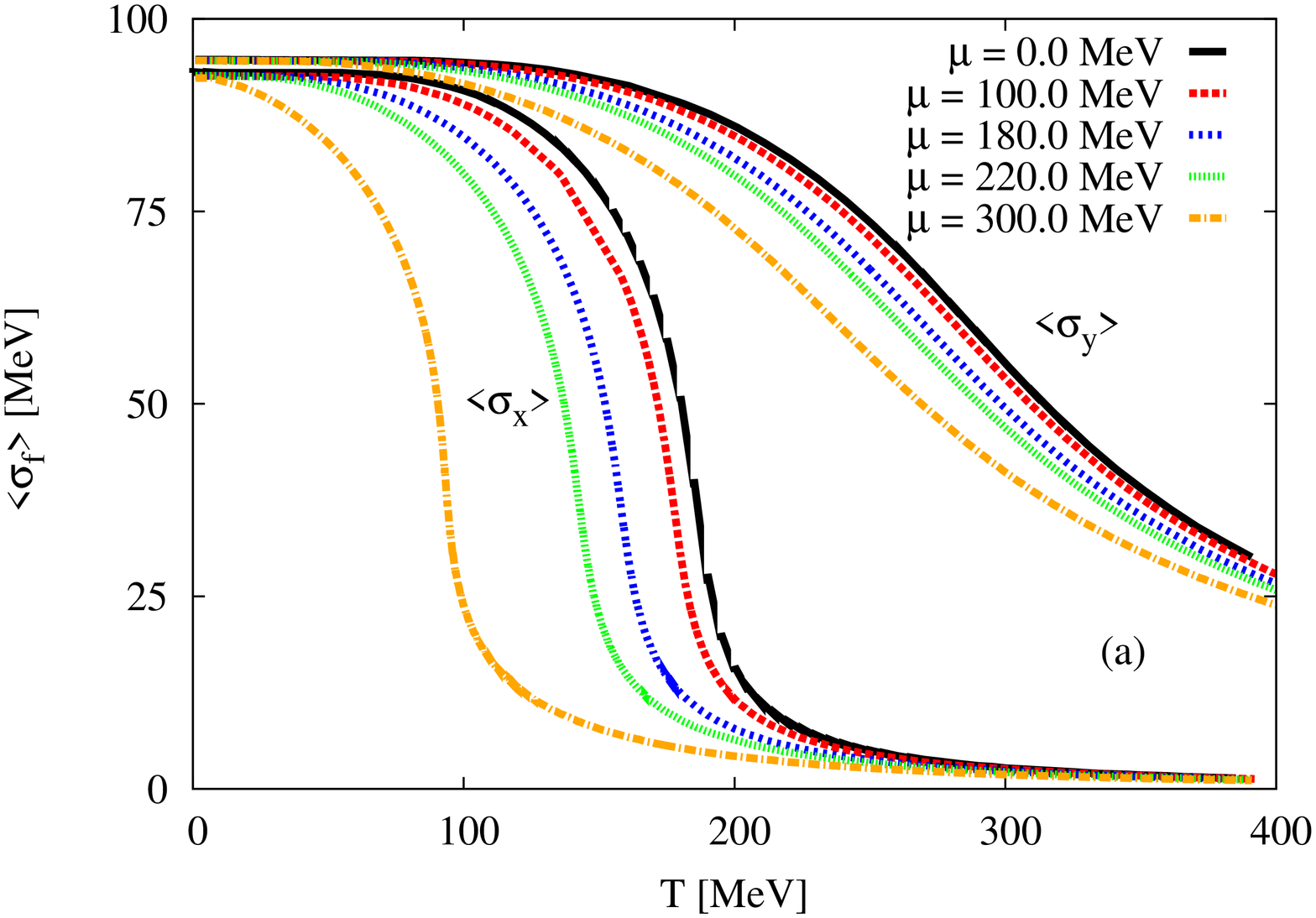}
\includegraphics[width=5cm,angle=-90]{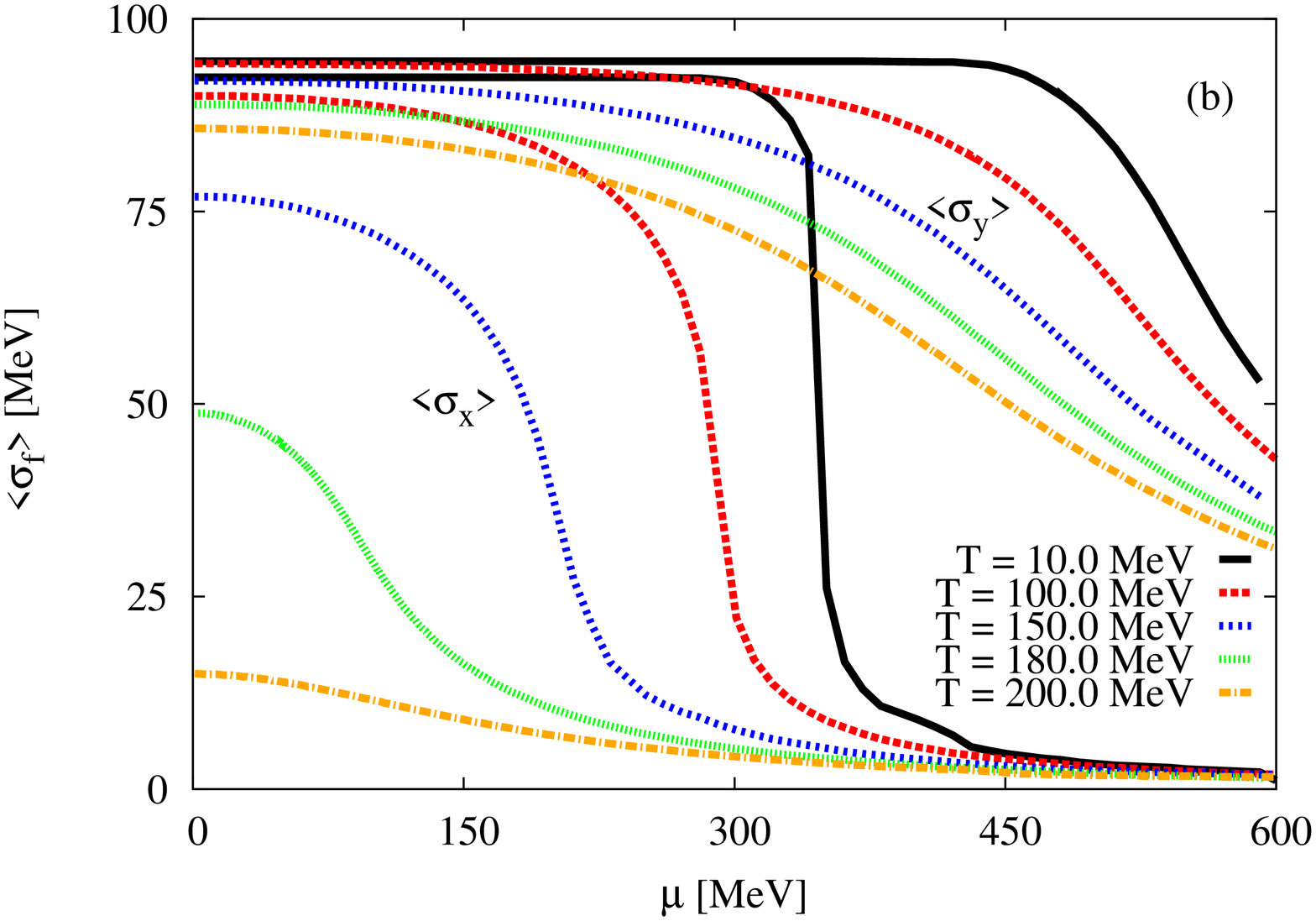}
\caption{(Color online) Left-hand panel: the averaged chiral condensate, $\langle\sigma_x\rangle$, is given as functions of temperature at different chemical potentials, $\mu=0$, $100$, $180$, $200$ and $300~$MeV (solid curves from on top, then downwards to bottom, respectively). For the chiral condensate $\langle\sigma_y\rangle$, we fix the same values of $\mu$. The right-hand panel presents the dependence on $\mu$, where the temperatures are fixed at the given values $T=10$, $100$, $150$, $180$ and $200~$MeV (solid curves from on top, then downwards to bottom, respectively).
\label{sxsyDiif} 
}
\end{figure}

For the in-medium thermal and dense effects on the mesonic masses, we present in the left-hand panel of Fig. \ref{sxsyDiif} the chiral condensates at varying temperatures and fixed baryon chemical potentials. In doing this, we take into consideration the thermal and dense dependences of the chiral condensates. For instance, we present the chiral condensates at different temperatures and  chemical potentials. At these temperatures and chemical potentials, we should estimate the thermal and dense dependences of the mesonic states. We notice that the values of $\sigma_{x}$ and $\sigma_{y}$ decrease with increasing $T$. There is a rapid decrease within a narrow range of temperatures. The light quarks are more sensitive than the strange quarks. This likely describes the characteristics of the chiral phase-transition. 

There is a similar decrease in both quantities with increasing hadronic dense-medium (baryon-chemical potential), right-hand panel of Fig. \ref{sxsyDiif}. We notice that the sudden decrease around the chiral phase-transition is sharper than the one in the left-hand panel. This would indicate that the chiral phase-transition at large density and low temperature (very near to the abscissa of the QCD phase diagram \cite{Tawfik:2004sw}) is much prompt than the one at low chemical potential and high temperature. The earlier would likely be characterized as a first-order phase-transition, while the latter as a moderate phase-transition (crossover) \cite{Tawfik:2004sw,FodorCO}.

We also notice that the fast decrease of $\sigma_x$ takes place earlier and faster than that of $\sigma_y$.  For instance, in the left-hand panel of fig. \ref{fig:Massus}, we find that $T_{c}^q=181~$MeV at vanishing density, and the decreases are smooth, while at finite baryon-chemical density and fixed $T=10~$MeV, the critical value $\mu =360~$MeV. This would be interpreted as a {\it smooth} phase-transition know as crossover \cite{lQCDco}. Thus, in presence of the Polyakov loop-potential, UA(1) of the symmetry breaking term is kept constant throughout the chiral and deconfinement phase-transition.

\begin{figure}[htb]
\includegraphics[width=5cm,angle=-90]{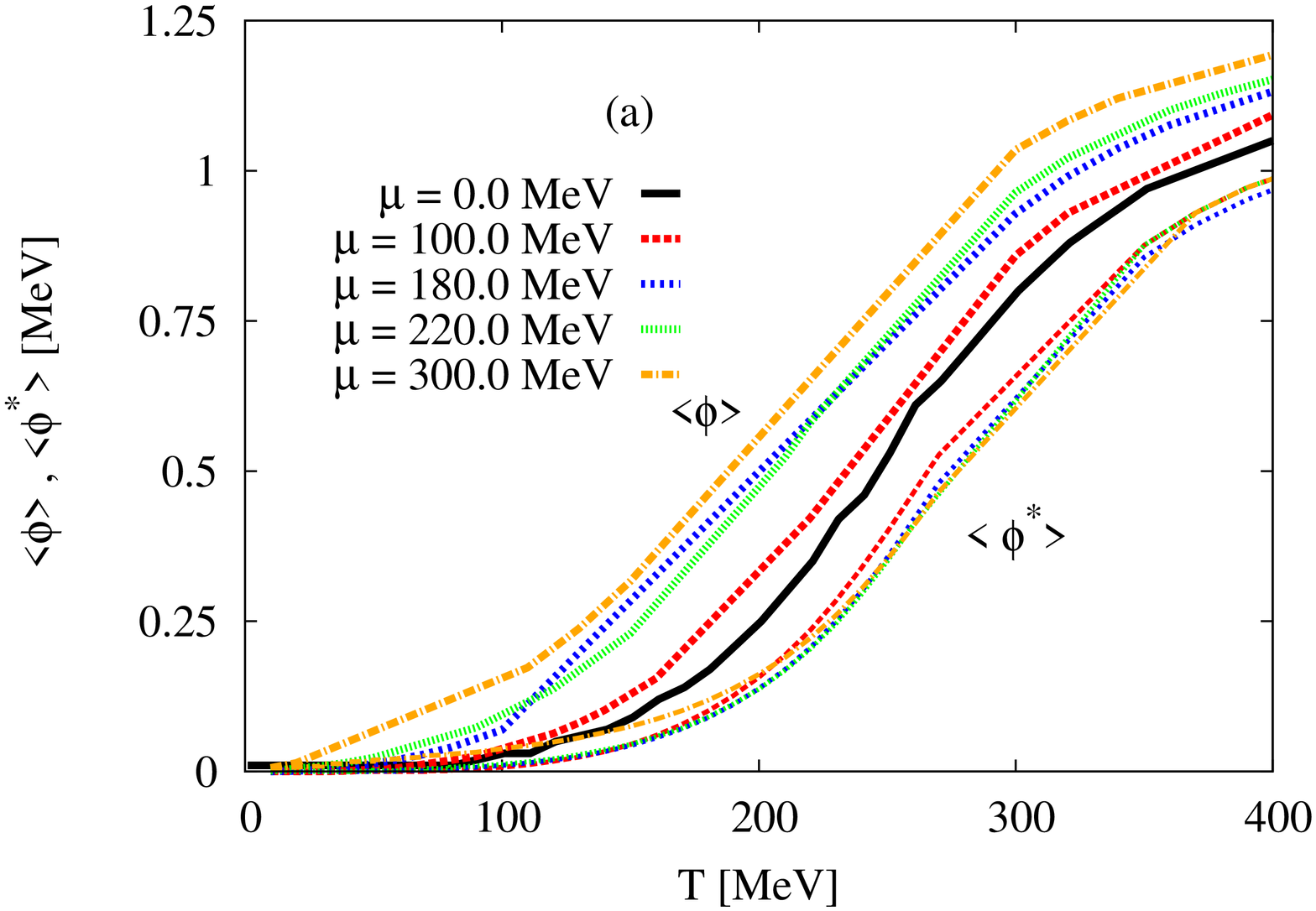}
\includegraphics[width=5cm,angle=-90]{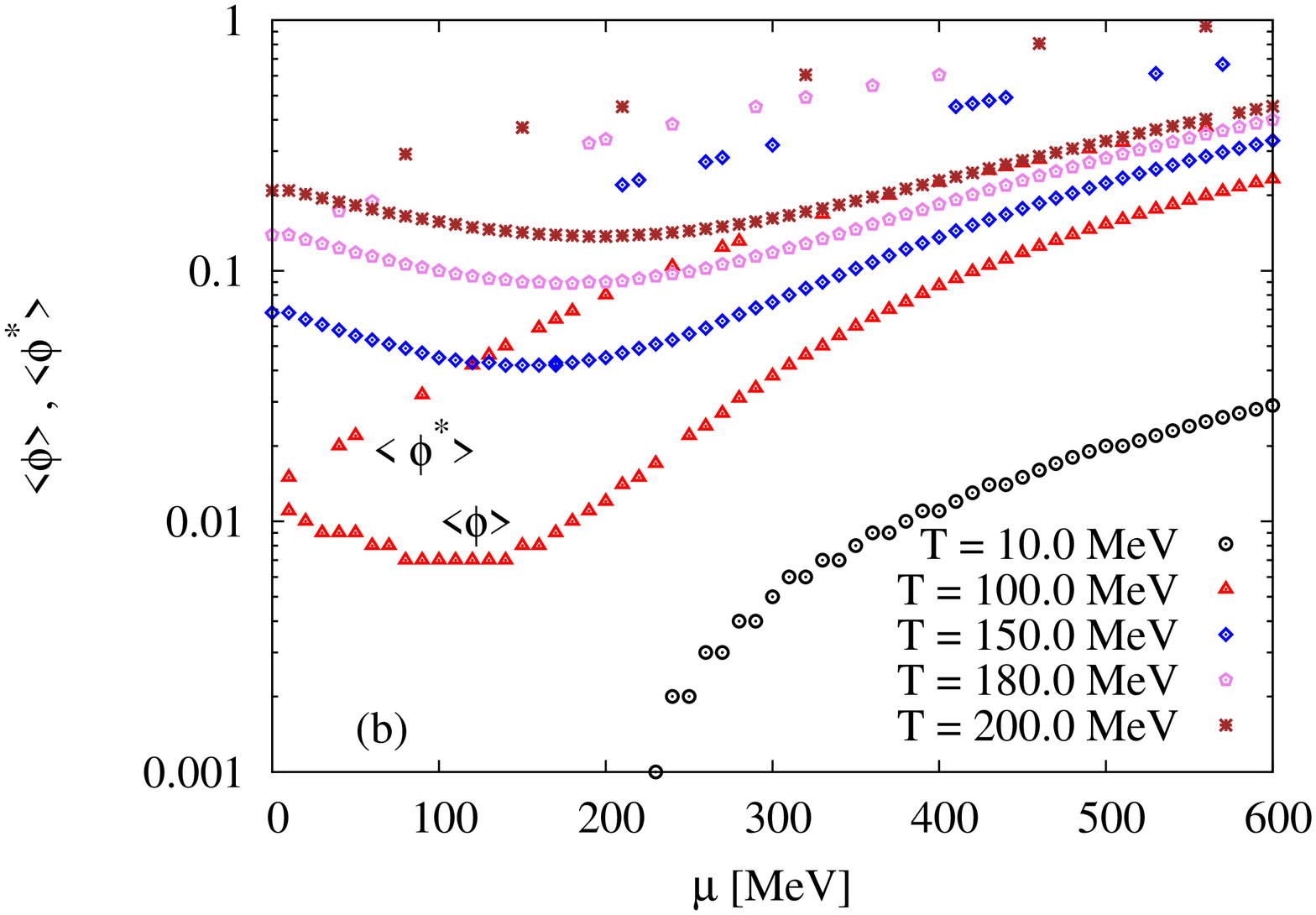}
\caption{(Color online) The expectation values of the Polyakov-loop fields, $\phi$ and $\phi ^{*}$, (left-hand panel) are given as function of temperature at different chemical potentials, $\mu=0$, $100$, $180$, $200$ and $300~$MeV for $\phi ^*$ (solid curve, from the forward to backwards, respectively) and for $\phi $ (solid curve, from the backward to forward, respectively). The chiral condensates, $\phi$ and $\phi ^{*}$, (right-hand panel) are given as functions of the chemical potential numerically in log-scale at different temperatures $T=10$, $100$, $150$, $180$ and $200~$MeV as points (from backward to forward, respectively) $\phi$ (top curves) and $\phi^{*}$ (bottom curves). 
\label{phiDiif} 
}
\end{figure}

For the results depicted in Fig. \ref{phiDiif}, we analyse the deconfinement phase-transition at varying baryon-chemical potentials and temperatures and include  the Ploykov-loop corrections to the meson masses at fixed five different temperatures and five different chemical potentials. The thermal effects of the hadronic medium on the evolution of $\phi$ seems to be very smooth. In hadronic dense-medium, the slope of $\phi (\mu)$ seems to depend on the temperature. It is always positive and increases rapidly with $\mu$, while $\phi (\mu)^{*}$ decreases slowly comparing to $\phi (\mu)$. Both quantities intersect at a characteristic value of $\mu$ depending on value of the temperature $T$.

\section{Masses of Sixteen Mesonic States}
\label{Meson}

\subsection{Inclusion of Anomalous Terms}

It is assumed that the contribution of the quark potential to the Lagrangian vanishes in the vacuum. Therefore, the meson potential determines the mass matrix, entirely. In other words, the meson masses do not receive any contribution from quark/antiquark in vacuum. Thus, the meson masses are governed by the meson potential \cite{Lenaghan:2000ey,Schaefer:2009}. 

The masses are defined by the second derivative of the grand potential $\Omega (T,\mu_f)$, Eq. (\ref{potential}), evaluated at its minimum Eq. (\ref{eq:meson_masses}), with respect to the corresponding fields. In the present calculations, the minima are estimated by vanishing expectation values of all scalar, pseudoscalar, vector and axial-vector fields. The pure strange $\bar \sigma_x$ and non-strange $\bar \sigma_y$ condensates are finite
\bea
  m^2_{i,{ab}} = \left. \frac{\partial^2 \Omega(T,\mu_f )}{\partial  \zeta_{i,a} \partial  \zeta_{i,b}}  \right\vert_{\rm min}, \label{eq:meson_masses}
\eea
where $i$ stands for scalar, pseudoscalar, vector and axial-vector mesons and $a$ and $b$ range from $0,\dots,8$. In vacuum, the mesonic sectors are formulated in the non-strange and strange basis:
\begin{itemize}
\label{massesALL}
\item Scalar meson masses are given as  
\bea
  m^2_{a_0} &=& m^2 + \lambda_1 \le(\bsig_x^2 + \bsig_y^2\ri) + \frac{3
    \lambda_2}{2} \bsig_x^2 +\frac{\sqrt{2} c}{2} \bsig_y ,\\
  m^2_{\kappa} &=& m^2 + \lambda_1 \le(\bsig_x^2 + \bsig_y^2\ri) +
  \frac{\lambda_2}{2} \le(\bsig_x^2 + \sqrt{2} \bsig_x \bsig_y +
    2 \bsig_y^2\ri) + \frac{c}{2}\bsig_x ,\\
  m^2_\sigma &= &m^2_{s,{00}} \cos^{2}\theta _s +
  m^2_{s,{88}}\sin^{2}\theta_s + 2  m^2_{s,{08}} \sin \theta_s \cos
  \theta_s ,\\
  m^2_{f_0}  &=& m^2_{s,{00}} \sin^{2}\theta _s +
  m^2_{s,{88}}\cos^{2}\theta_s - 2  m^2_{s,{08}} \sin \theta_s \cos \theta_s , 
\eea
with 
\bea 
m^2_{s,{00}} &=&m^2 + \frac{\lambda_1}{3} \le(7 \bsig_x^2 + 4
  \sqrt{2}\bsig_{x}\bsig_{y} + 5 \bsig_y^2\ri) + \lambda_2\le(\bsig_x^2 +
  \bsig_y^2\ri) -\frac{\sqrt{2}c}{3}\le(\sqrt{2} \bsig_x +
  \bsig_y\ri),\nonumber \\
 m^2_{s,{88}} &=& m^2 + \frac{\lambda_1}{3} \le(5 \bsig_x^2 - 4
  \sqrt{2}\bsig_{x}\bsig_{y} + 7 \bsig_y^2\ri) +
  \lambda_2\le(\frac{\bsig_x^2}{2} + 2\bsig_y^2\ri)
  +\frac{\sqrt{2}c}{3}\left(\sqrt{2} \bsig_x -
  \frac{\bsig_y}{2}\right),\nonumber  \\
  m^2_{s,{08}} &=& \frac{2 \lambda_1}{3}\le(\sqrt{2}\bsig_x^2 - \bsig_x \bsig_y - \sqrt{2}\bsig_y^2 \ri)
  + \sqrt{2}\lambda_2 \le(\frac{\bsig_x^2}{2} - \bsig_y^2 \ri)
  + \frac{c}{3\sqrt{2}} \le(\bsig_x - \sqrt{2}\bsig_y\ri). \nn
\eea
and $s$ (scalar) refers to $i$ in Eq. (\ref{eq:meson_masses}).

\item Pseudoscalar meson masses read
\bea
  m^2_\pi &=& m^2 + \lambda_1 \le(\bsig_x^2 + \bsig_y^2\ri) +
  \frac{\lambda_2}{2} \bsig_x^2 -\frac{\sqrt{2} c}{2} \bsig_y, \\
  m^2_K &=& m^2 + \lambda_1 \le(\bsig_x^2 + \bsig_y^2\ri) +
  \frac{\lambda_2}{2} \le(\bsig_x^2 - \sqrt{2} \bsig_x \bsig_y +
    2 \bsig_y^2\ri) - \frac{c}{2} \bsig_x, \\
    m^2_{\eta'} &=& m^2_{p,{00}} \cos^{2}\theta _p + m^2_{p,{88}}\sin^{2}\theta _p + 2 m^2_{p,{08}} \sin \theta _p \cos \theta_p, \\
  m^2_{\eta} &=& m^2_{p,{00}} \sin^{2}\theta_p + m^2_{p,{88}}\cos^{2}\theta_p - 2 m^2_{p,{08}} \sin \theta_p \cos \theta_p,
\eea
with
\bea
m^2_{p,{00}} &=& m^2 + \lambda_1 \le(\bsig_x^2 + \bsig_y^2\ri)
  + \frac{\lambda_2}{3}\le(\bsig_x^2 + \bsig_y^2\ri) +\frac{c}{3}\le(2
  \bsig_x + \sqrt{2} \bsig_y\ri), \nonumber \\
  m^2_{p,{88}} &=& m^2 + \lambda_1 \le(\bsig_x^2 + \bsig_y^2\ri)
  + \frac{\lambda_2}{6}\le(\bsig_x^2 + 4 \bsig_y^2\ri) -\frac{c}{6}\le(4
  \bsig_x - \sqrt{2}\bsig_y\ri), \nonumber \\
  m^2_{p,{08}} &=& \frac{\sqrt{2} \lambda_2}{6} \le(\bsig_x^2 - 2
  \bsig_y^2\ri) - \frac{c}{6}\le(\sqrt{2}\bsig_x - 2 \bsig_y\ri), \nonumber
\eea
the mixing angles are given by
\bea
  \tan 2\theta_i &=& \frac{2 m^2_{i,{08}}}{m^2_{i,{00}}-m^2_{i,{88}}}\
  ,\  i=s,p\ .
\eea
and $p$ (pseudoscalar) refers to $i$ in Eq. (\ref{eq:meson_masses}).

\item Vector meson  masses are given as
\bea
m_{\rho}^{2}  &=&m_{1}^{2}+\frac{1}{2} \le(h_{1}+h_{2}+h_{3}\ri) \bsig_x^{2} 
+\frac{h_{1}}{2}\bsig_y^{2}+2\delta_{x}\;,\label{m_rho}\\
m_{K^{\star}}^{2}  &=&m_{1}^{2}+\frac{\bsig_x^{2}}{4} \le(g_{1}^{2}+2h_{1}
+h_{2}\ri) +\frac{\bsig_x \bsig_y}{\sqrt{2}}(h_{3}-g_{1}^{2})+\frac{\bsig_y^{2}}{2}\le(g_{1}^{2}+h_{1}+h_{2}\ri)+\delta_{x}+\delta_{y}\;,\label{m_K_star}\\
m_{\omega_{x}}^{2}  &=&m_{\rho}^{2}\;,\\
m_{\omega_{y}}^{2}  &=&m_{1}^{2}+\frac{h_{1}}{2}\bsig_x^{2} + \le(
\frac{h_{1}}{2}+h_{2}+h_{3}\ri) \bsig_y^{2}+2\delta_{y}\;,
\label{V}%
\eea
and vectors $V^{\mu}$ refer to $i$ in Eq. (\ref{eq:meson_masses}).

\item And finally the axial-vectors masses are
\bea
m_{a_{1}}^{2}  &=&m_{1}^{2}+\frac{1}{2} \le(2g_{1}^{2}+h_{1}+h_{2}-h_{3}\ri) \bsig_x^{2}+\frac{h_{1}}{2}\bsig_y^{2}+2\delta_{x} ,\label{m_a_1}\\
m_{K_{1}}^{2}  &=&m_{1}^{2}+\frac{1}{4}\le(  g_{1}^{2}+2h_{1}+h_{2}\ri)
\bsig_x^{2}-\frac{1}{\sqrt{2}}\bsig_x \bsig_y \le(h_{3}-g_{1}^{2}\ri) + \frac{1}{2}\le(g_{1}^{2}+h_{1}+h_{2}\ri)  \bsig_y^{2}  + \delta_{x}+\delta_{y}, \label{m_K_1}\\
m_{f_{1x}}^{2}  &=& m_{a_{1}}^{2}, \\
m_{f_{1y}}^{2}  &=& m_{1}^{2}+\frac{\bsig_x^{2}}{2} h_{1}+\le(2g_{1}^{2}+\frac{h_{1}}{2}+h_{2}-h_{3}\ri)  \bsig_y^{2}+2\delta_{y}.
\label{AV}%
\eea
and axialvector $A^{\mu}$ refer to $i$ in Eq. (\ref{eq:meson_masses}).
\end{itemize}

The evolution of masses of (pseudo)-scalar states depends on the anomaly term of  $\mathcal{L}_{\text{U(1)}_A}$. This term causes anomaly in $c$-term. The way of choosing the anomaly term defines/describes of the structure of the hadronic states \cite{Wolf:anomaly}. The anomaly term, which we have implemented here, agrees with the calculation of Refs. \cite{Schaefer:2009,eta,V. Tiwari:2009} but differs from Ref. \cite{Rischke:2012}. Moreover, the estimated masses of (axial)-vector states are not affected by the anomaly term \cite{Rischke:2012}.

The quantum and thermal fluctuations of the mesonic fields are neglected. It is worthwhile to mention that the integration over the mesonic fields is renounced. Furthermore, the mesonic fields are replaced by their expectation values, $\sigma _0 \, and \,\sigma _8$, resulting in the mesonic potential $U(\sigma_0, \sigma_8)$. The quarks are treated as quantum fields. The integration over the quark fields yields determinant, which can be rewritten as a trace over a logarithm defined by Eq. (\ref{qqlsm}) for \lsm$\,$ and Eq. (\ref{PloykovPLSM}) for P\lsm$\,$. The Matsubara formalism \cite{Kapusta:1989} gives an estimation for the quark contribution to the meson masses, section \ref{QGP}.

In order to include the quark contribution to the grand potential, the meson masses should be modified due to the in-medium effects. In calculating the second derivative, Eq. (\ref{eq:meson_masses}), we take into account Eq. (\ref{qqlsm}) and diagonalize the resulting quark mass matrix. Then, we can deduce an expression for the modification in the meson masses \cite{Schaefer:2009}. 
\bea
 m^2_{i,{ab}} = \left. \frac{\partial^2 \Omega(T,\mu_f )}{\partial  \zeta_{i,a} \partial  \zeta_{i,b}}  \right\vert_{\rm min} &=& \nu_{c}\sum_{f=l,s} \int \frac{d^3 p}{(2\pi)^3} \frac{1}{2E_{q,f}} \biggl[\left(n_{q,f} + n_{\bar{q},f}
  \right) \biggl( m^{2}_{f,a b} - \frac{m^{2}_{f,a}
 m^{2}_{f, b}}{2 E_{q,f}^{2}} \biggr)  - \left( b_{q,f}+ b_{\bar{q},f}\right) \biggl(\frac{m^{2}_{f,a}  m^{2}_{f, b}}{2 E_{q,f} T}
\biggr) \biggr]. \hspace*{8mm} \label{eq:ftlsmass}
\eea 
The quark mass derivative with respect to the meson fields $\zeta_{i,a}$, $m^2_{f,a} \equiv \partial m^2_f/\partial \zeta_{i,a}$  and that with respect to the meson fields $\zeta_{i,a} \partial  \zeta_{i,b}$, $m^2_{f,{ab}} \equiv \partial m^2_f/\partial \zeta_{i,a} \partial  \zeta_{i,b}$ are listed in Tab. \ref{tab:qmassd}. Correspondingly, the antiquark function $b_{\bar{q},f}(T,\mu_f) = b_{q,f}(T,-\mu_f)$, where
\begin{equation}
b_{q,f}(T,\mu_f)= n_{q,f}(T,\mu_f) (1-n_{q,f}(T,\mu_f)).
\end{equation}

A expression for the meson mass modification can be estimated from P\lsm, Eq. (\ref{PloykovPLSM}), and the diagonalization of  the resulting quark mass matrix \cite{V. Tiwari:2009},  
\bea
 m^2_{i,{ab}} = \left. \frac{\partial^2 \Omega(T,\mu_f )}{\partial  \zeta_{i,a} \partial  \zeta_{i,b}}  \right\vert_{\rm min} &=& \nu_{c}\sum_{f=l,s} \int \frac{d^3 p}{(2\pi)^3} \frac{1}{2E_{q,f}} \biggl[ (N_{q,f} + N_{\bar{q},f} ) \biggl( m^{2}_{f,a b} - \frac{m^{2}_{f,a}
 m^{2}_{f, b}}{2 E_{q,f}^{2}} \biggr)  + (B_{q,f} + B_{\bar{q},f}) \biggl(\frac{m^{2}_{f,a}  m^{2}_{f, b}}{2 E_{q,f} T}
\biggr) \biggr].   \hspace*{8mm} \label{eq:ftmass}
\eea 
In estimating $m^2_{i,{ab}}$, the definitions $E_{q,f}(T,\mu)=E_{q,f}(T,-\mu)$ and 
\bea
N_{q,f} &=& \frac{\Phi e^{-\,E_{q,f}/T} + 2 \Phi^* e^{- 2\, E_{q,f}/T} + e^{-3\,E_{q,f}/T}}{1+3(\phi+\phi^* e^{-E_{q,f}/T}) e^{-E_{q,f}/T}+e^{-3E_{\bar{q},f}/T}}, \\
N_{\bar{q},f} &=& \frac{\Phi^* e^{-E_{\bar{q},f}/T} + 2 \Phi e^{- 2 E_{\bar{q},f}/T} + e^{-3E_{\bar{q},f}/T}}{ 1+3(\phi^*+\phi e^{-E_{\bar{q},f}/T}) e^{-E_{\bar{q},f}/T}+e^{-3E_{\bar{q},f}/T}},
\eea
are implemented \cite{V. Tiwari:2009}. Furthermore, for quark $B_{q,f}=3(N_{{q},f} )^2 - C_{{{q}},f} $ and for antiquark $B_{\bar{q},f}=3(N_{{{\bar{q}},f} )^2 - C_{{\bar{q}}},f}$,  where 
\bea
C_{{q},f} &=& \frac{\Phi e^{-\,E_{q,f}/T} +4 \Phi^* e^{-2\,E_{q,f}/T} +3 e^{-3\,E_{q,f}/T}}{1+3(\phi+\phi^* e^{-E_{q,f}/T})\, e^{-E_{q,f}/T}+e^{-3E_{\bar{q},f}/T}}, \\
C_{\bar{q},f} &=&  \frac{\Phi^* e^{-E_{\bar{q},f}/T} + 4 \Phi e^{-2E_{\bar{q},f}/T} +3 e^{-3E_{\bar{q},f}/T}}{ 1+3(\phi^*+\phi e^{-E_{\bar{q},f}/T})\, e^{-E_{\bar{q},f}/T}+e^{-3E_{\bar{q},f}/T}},
\eea
are defined \cite{V. Tiwari:2009}.

The quark masses has to be taken into account and accordingly same isospin of light quarks $m_u=m_d$, but different for $m_s$. The first and second derivatives of squared quark mass in non-strange and strange basis with respect to meson fields are evaluated at minimum \cite{Schaefer:2009}. In Tab. \ref{tab:qmassd}, the summation over the two light flavors denoted by symbol $l$ are in given in the first two columns present the first and second derivatives of squared light quark masses, respectively. The last two columns are devoted to the strange quark mass. In spite of the consideration of  SU(2) isospin symmetry, the derivatives the first and second derivatives of squared light quark masses are different for the $u$- and $d$-quark, where their summation is cancelled out \cite{Schaefer:2009}.

\begin{table}[htb]
\begin{tabular}{c c c c c c }
\hline
&& $m^{2}_{l,a} m^{2}_{q,b}/g^4$ & $m^{2}_{l,ab}/g^2$& $m^{2}_{s,a} m^{2}_{s,b}/g^4$ & $m^{2}_{s,ab}/g^2$\\
\hline \hline
$\sigma_0$ & $\sigma_0$ &$\frac{1}{3}\sigma_{x}^{2}$& $\frac{2}{3}$& $\frac{1}{3}
\sigma_{y}^{2}$& $\frac{1}{3}$ \\
$\sigma_1$ & $\sigma_1$ &$\frac{1}{2}\sigma_{x}^{2}$& $ 1$ & $0$ & $0$\\
$\sigma_4$ & $\sigma_4$ &$ 0$ &$\sigma_x \frac{\sigma_x + \sqrt{2} \sigma_y}
{\sigma_{x}^{2} -2 \sigma_{y}^{2}}$ & $0$ & $\sigma_y \frac{ \sqrt{2} \sigma_x +2 \sigma_y}{2 \sigma_{y}^{2} - \sigma_{x}^{2}}$ \\
$\sigma_8$ & $\sigma_8$ &$ \frac{1}{6} \sigma_{x}^{2}$ & $\frac{1}{3}$ & 
$ \frac{2}{3}\sigma_{y}^{2} $ & $\frac{2}{3} $ \\
$\sigma_0$ & $\sigma_8$ & $ \frac{\sqrt{2} }{6} \sigma_{x}^{2}$ & $ \frac{\sqrt{2}}{3}$ & 
$- \frac{\sqrt{2}}{3} \sigma_{y}^{2}$ & $- \frac{\sqrt{2}}{3}$\\ \hline  
$\pi_0$ & $\pi_0$ & $0$ & $\frac{2}{3}$ & $0$ & $\frac{1}{3}$ \\
$\pi_1$ & $\pi_1$ & $0$ & $1$ & $0$ & $0$ \\
$\pi_4$ & $\pi_4$ & $0$ & $\sigma_x \frac{\sigma_x -\sqrt{2}\sigma_y}{\sigma_{x}^{2} 
- 2 \sigma_{y}^{2}}$ & $0$ & $\sigma_y \frac{\sqrt{2} \sigma_x -2 \sigma_y}
{\sigma_{x}^{2} -2 \sigma_{y}^{2}}$ \\
$\pi_8$ & $\pi_8$ & $0$ & $\frac{1}{3}$ & $0$ & $\frac{2}{3}$\\
$\pi_0$ & $\pi_8$ & $0$ & $\frac{\sqrt{2}}{3}$ & $0$ & $-\frac{\sqrt{2}}{3}$ \\ 
\hline
\end{tabular}
\caption{The first and second derivatives of squared quark masses in non-strange (first two columns) and strange (last two columns) basis with respect to the meson fields are evaluated at minima \cite{Schaefer:2009}.
\label{tab:qmassd} 
}
\end{table}


\begin{table}[htb]
\begin{tabular}{ c c c c c c c}
\hline
& Sector & Symbol & PDG \cite{PDG:2012} & P\lsm &\begin{tabular}{c} PNJL \cite{NJL:2013,P. Costa:PNJL} \end{tabular} &\begin{tabular}{c|c}\multicolumn{2}{ c }{Lattice QCD} \\
\hline
Hot QCD\cite{HotQCD} & PACS-CS \cite{PACS-CS} \\ \end{tabular} \\
\hline \hline
&\begin{tabular}{c}
Scalar \\$J^{PC}=0^{++}$
\end{tabular}
&\begin{tabular}{c}
$a_{0}$~\\ ~$\kappa $~\\~$\sigma$~\\~$ f_{0}$~\\
\end{tabular} 
&\begin{tabular}{c}
$a_{0}(980^{\pm 20})$~\\ $K_{0}^* (1425^ {\pm 50}) $ \\$\sigma(400-1200)$\\$ f_{0}(1200-1500 )$~\\
\end{tabular}  
&\begin{tabular}{c}
$1026 $~\\ ~$ 1115 $~\\~$ 800 $~\\~$ 1284$~\\
\end{tabular} 
&\begin{tabular}{c}$837 $~\\ ~$ 1013 $~\\~$ 700 $~\\~$ 1169$~\\ \end{tabular} & \\  
\hline
&\begin{tabular}{c}\\
Pseudoscalar \\
$J^{PC}=0^{-+}$
\end{tabular} 
&\begin{tabular}{c}
$\pi$~\\ ~$K $~\\~ $\eta$~\\~$\eta ^{'}$~\\
\end{tabular} 
&\begin{tabular}{c}
$\pi ^0 (134.97^{\pm 6.9} )$~\\ $K^0$ ($497.614^{\pm 24.8 }  $) \\ $\eta(547.853^{\pm 27.4} )$~\\$\eta ^{'}(957.78^{ \pm 60})$\\
\end{tabular}  
&\begin{tabular}{c}
$120$~\\ ~$\,509$~\\~ $ 553 $~\\~$ 965 $~\\
\end{tabular}  &\begin{tabular}{c}$126$~\\ ~$\, 490$ \\ $505$ \\ $949$\\ \end{tabular} &
\begin{tabular}{l}$134^{\pm 6} ~~~~~~~~~~135.4^{\pm 6.2}$ \\ $422.6^{\pm 11.3}\, \,~~~498^{\pm 22} $\\ $579^{\pm 7.3}\, \,~~~~~~~688^{\pm 32} $\\$-\, \,~~~~~~~~~- $~\\ \end{tabular}
 \\  
\hline 
&\begin{tabular}{c}
Vector\\ $J^{PC}=1^{-}$
\end{tabular}
&\begin{tabular}{c}
~$\rho$~\\ ~$\omega _X$~\\ ~ $K^*$ ~\\ ~$ \omega _y$~\\
\end{tabular}
 &\begin{tabular}{c}
$\rho (775.49^{\pm 38.8})$\\ ~$\omega (782.65^{\pm 44.7})$~\\ ~ $K^* (891.66^{\pm 26})$ ~\\ ~$ \phi (1019.455^{\pm 51})$~\\
\end{tabular}
&\begin{tabular}{c}
~$745$~\\ ~$745$~\\~ $894$ ~\\ ~$ 1005$~\\
\end{tabular}
 &\begin{tabular}{c}$ - $~\\ ~$ - $~\\~ $ - $~\\~$ - $~\\ \end{tabular} &\begin{tabular}{l} $\,\; 756.2^{\pm 36}~~~~~~597^{\pm 86} $\\ $\; 884^{\pm 18}\, ~~~~~~~~861^{\pm 23} $\\ $\, 1005^{\pm 93}\, ~~~~~~~1010.2^{\pm 77} $\\$-\, \,~~~~~~~~~- $~\\ \end{tabular}
 \\
\hline 
&\begin{tabular}{c}
Axial-Vector\\ 
$J^{PC}=1^{++}$
\end{tabular}
&\begin{tabular}{c}
~$a_1$~\\ ~$f_{1x}$~\\ ~ $K_{1}^{*}$ ~\\ ~$ f_{1y}$~\\
\end{tabular}
 &\begin{tabular}{c}
$a_1 (1030 - 1260)$\\ ~$f_{1}(1281^{ \pm 60})$~\\ ~ $K_{1}^{*} (1270^{\pm 7} )$ ~\\ $ f_{1}(1420^{\pm 71.3})$\\
\end{tabular}
 &\begin{tabular}{c}
~$980 $~\\ ~$980$~\\ ~ $1135 $ ~\\ ~$1315$~\\
\end{tabular}
&\begin{tabular}{c}$ - $~\\ ~$ - $~\\~ $ - $~\\~$ - $~\\ \end{tabular} 
  &\begin{tabular}{c}
\end{tabular} \\  
\hline
\end{tabular}
\caption{A comparison between (pseudo)-scalar and (axial)-vector meson sectors in P\lsm ~(present work) and the corresponding results from PNJL \cite{P. Costa:PNJL}. Both are compared with the experimental measurements, PDG \cite{PDG:2012} and the lattice QCD simulations \cite{HotQCD,PACS-CS}.
\label{masscomp}
}
 \end{table} 
 
Tab. \ref{masscomp} presents  a comparison between the different scalar and vector meson nonets in various effective thermal models, like P\lsm ~(present work) and PNJL \cite{P. Costa:PNJL} confronted to PDG \cite{PDG:2012} and lattice QCD calculations \cite{HotQCD,PACS-CS}. Some remarks are now in order.  The errors are deduced from the fitting for the parameters used in calculating the equation of states and other thermodynamics quantities. The fitting requires information from the experimental inputs about (axial)-vector and (pseudo)-scalar states. The output results are very precise for some light hadrons described by the present model, the P\lsm. We aim to describe hadron vacuum phenomenology with such an extreme precision and not only to describe the hadron spectrum in both thermal- and hadronic dense-medium. We show the effects of the chiral condensate and deconfinement phase-transition in order to characterize the chiral phase-structure of many hadrons. The PNJL model is limited to study (pseudo)-scalar meson states. Only pseudoscalar and vector meson masses are available in the lattice QCD calculations (HotQCD Collaboration) \cite{HotQCD}and (PACS-CS Collaboration) \cite{PACS-CS}.  

The estimation of the meson masses seems to agree well with Refs. \cite{Schaefer:2009,eta,V. Tiwari:2009,Rischke:2012}. But for mixing strange with nonstrange scalar states, one state  $<1~$GeV and another one $>1~$GeV were obtained in Ref. \cite{Rischke:2012}. To this end the authors needed to implement Gyuri fit to correct this \cite{PC}.

\subsubsection{Temperature Dependence}
\label{T_dependence}
 
\begin{figure}[htb]
\centering{
\includegraphics[width=3.5cm,angle=-90]{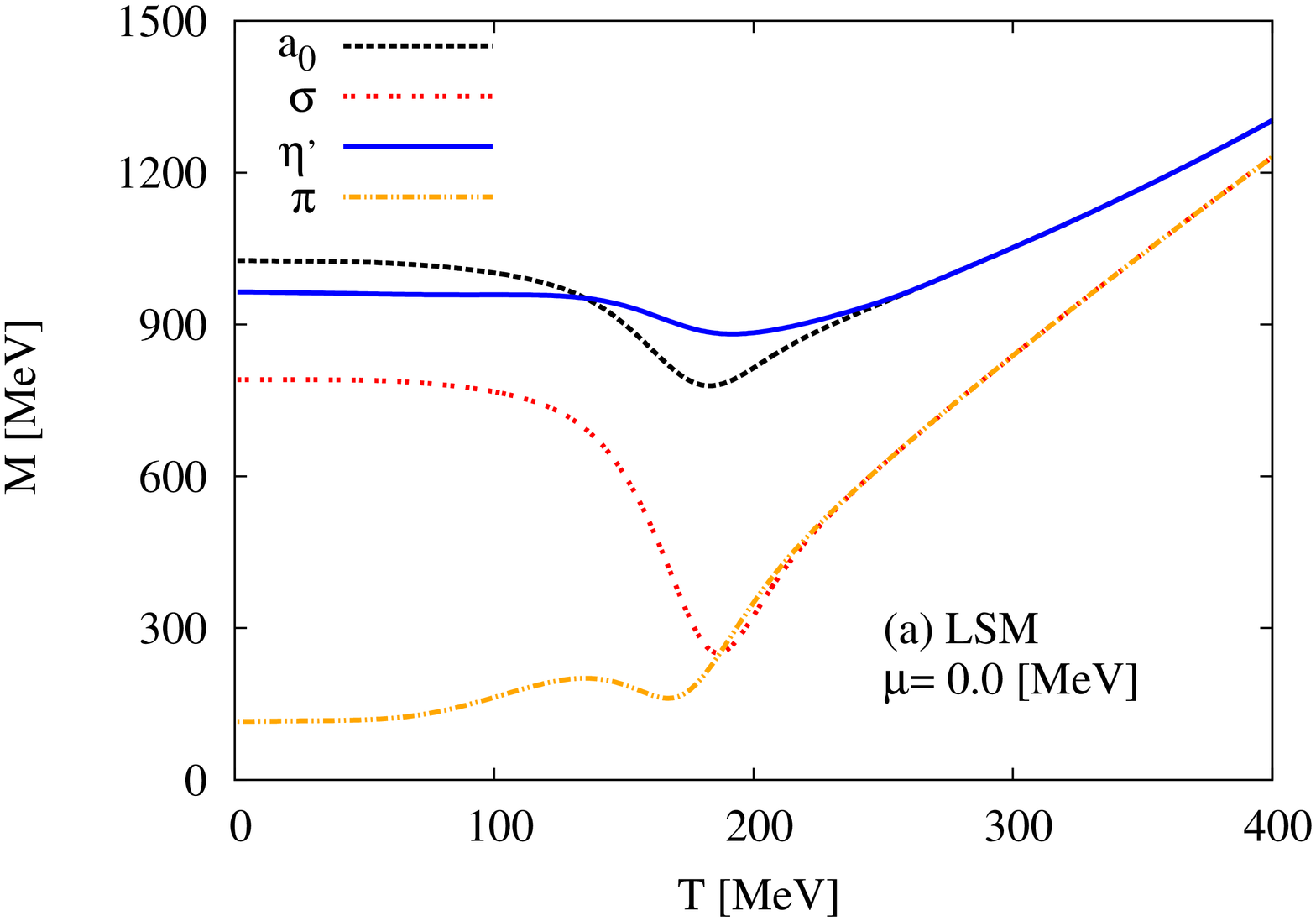}
\includegraphics[width=3.5cm,angle=-90]{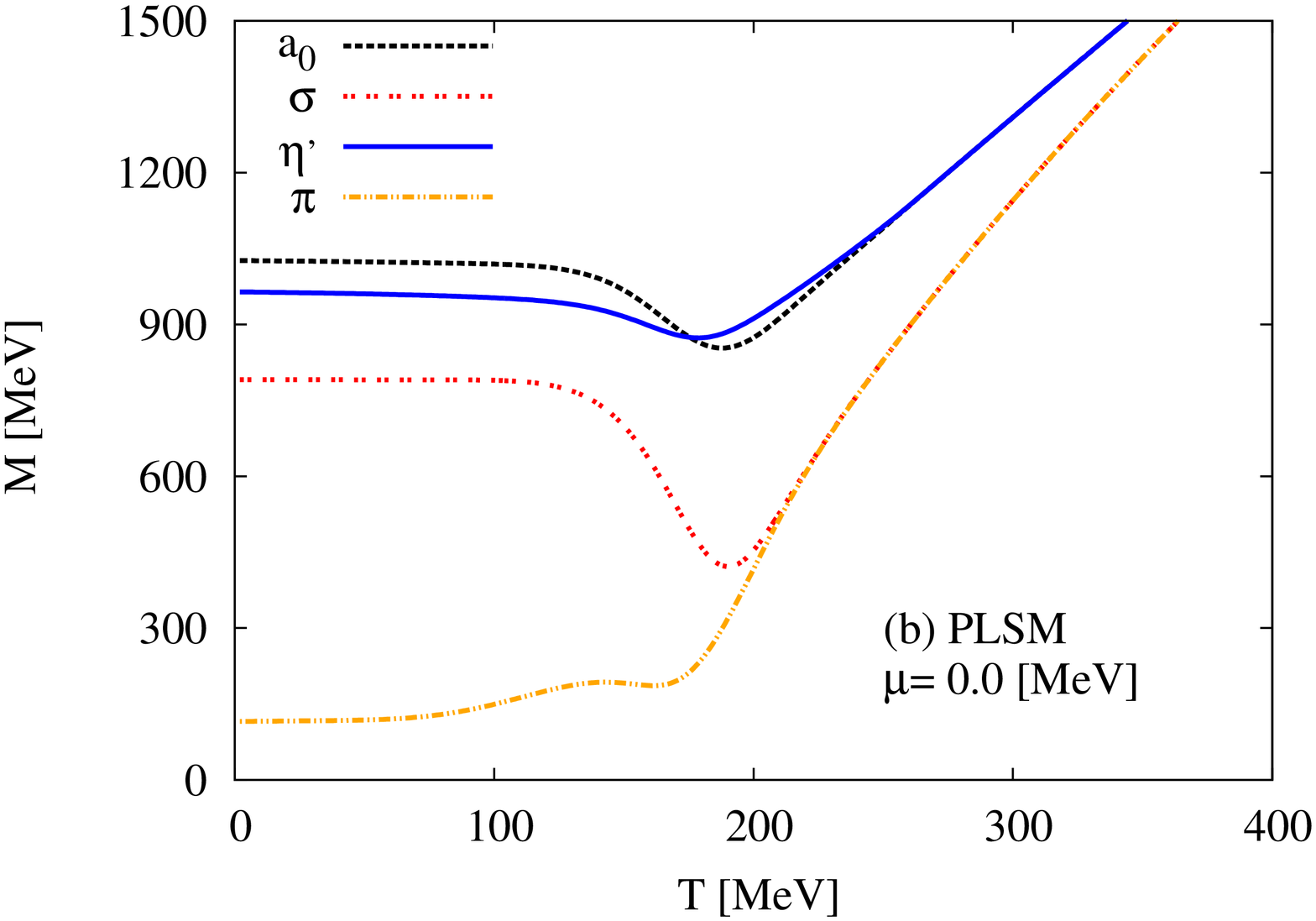}\\
\includegraphics[width=3.5cm,angle=-90]{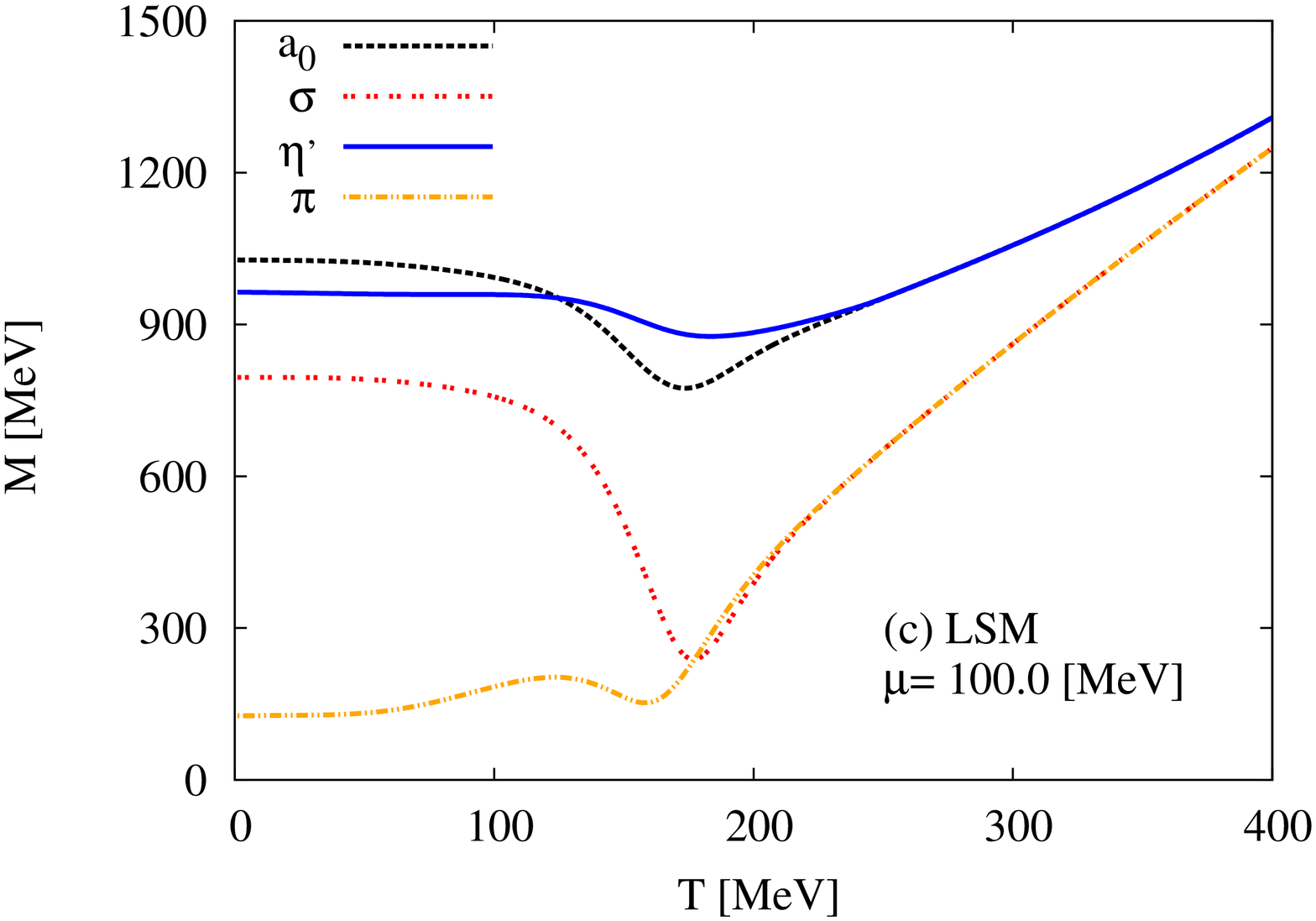}
\includegraphics[width=3.5cm,angle=-90]{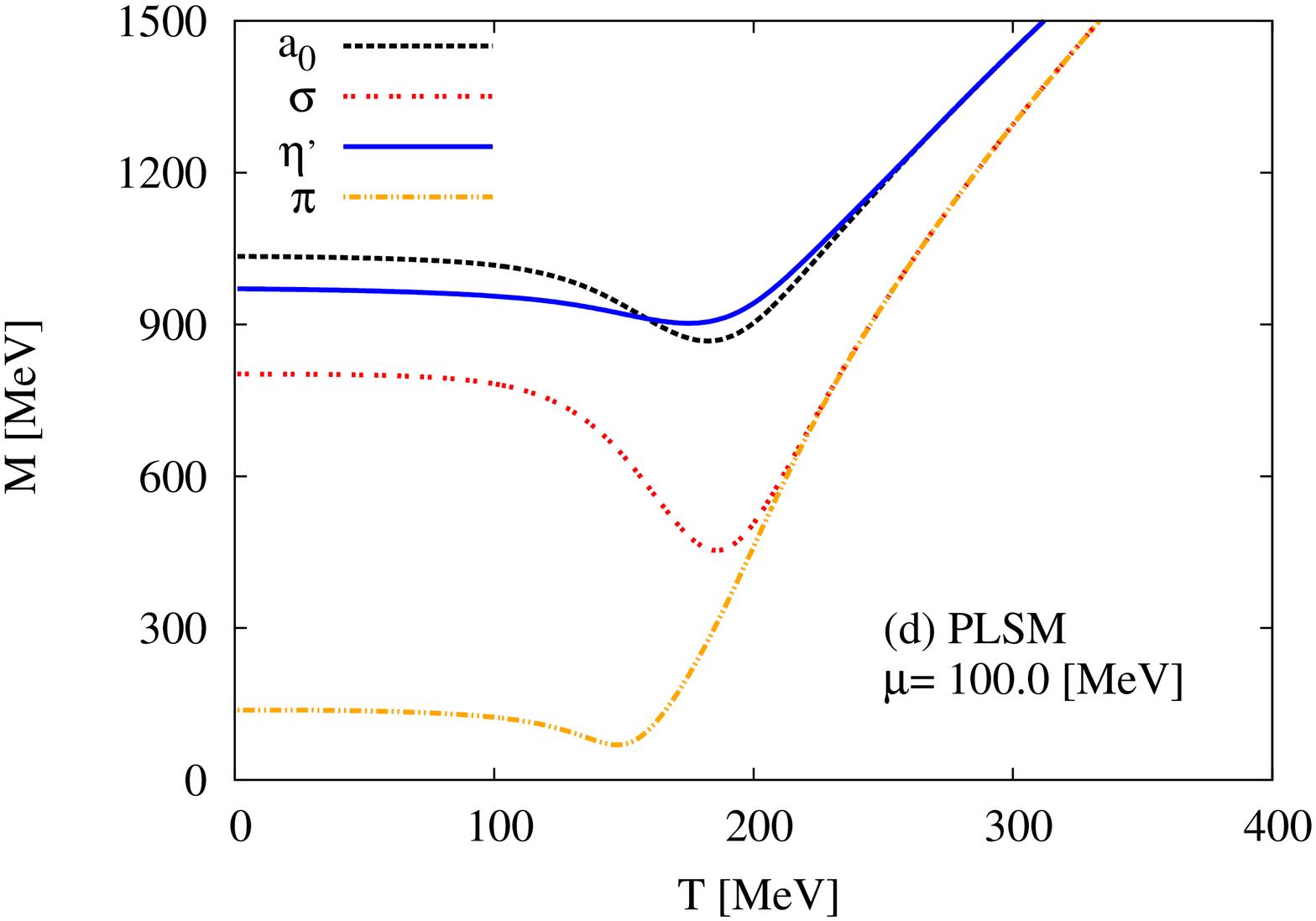}\\
\includegraphics[width=3.5cm,angle=-90]{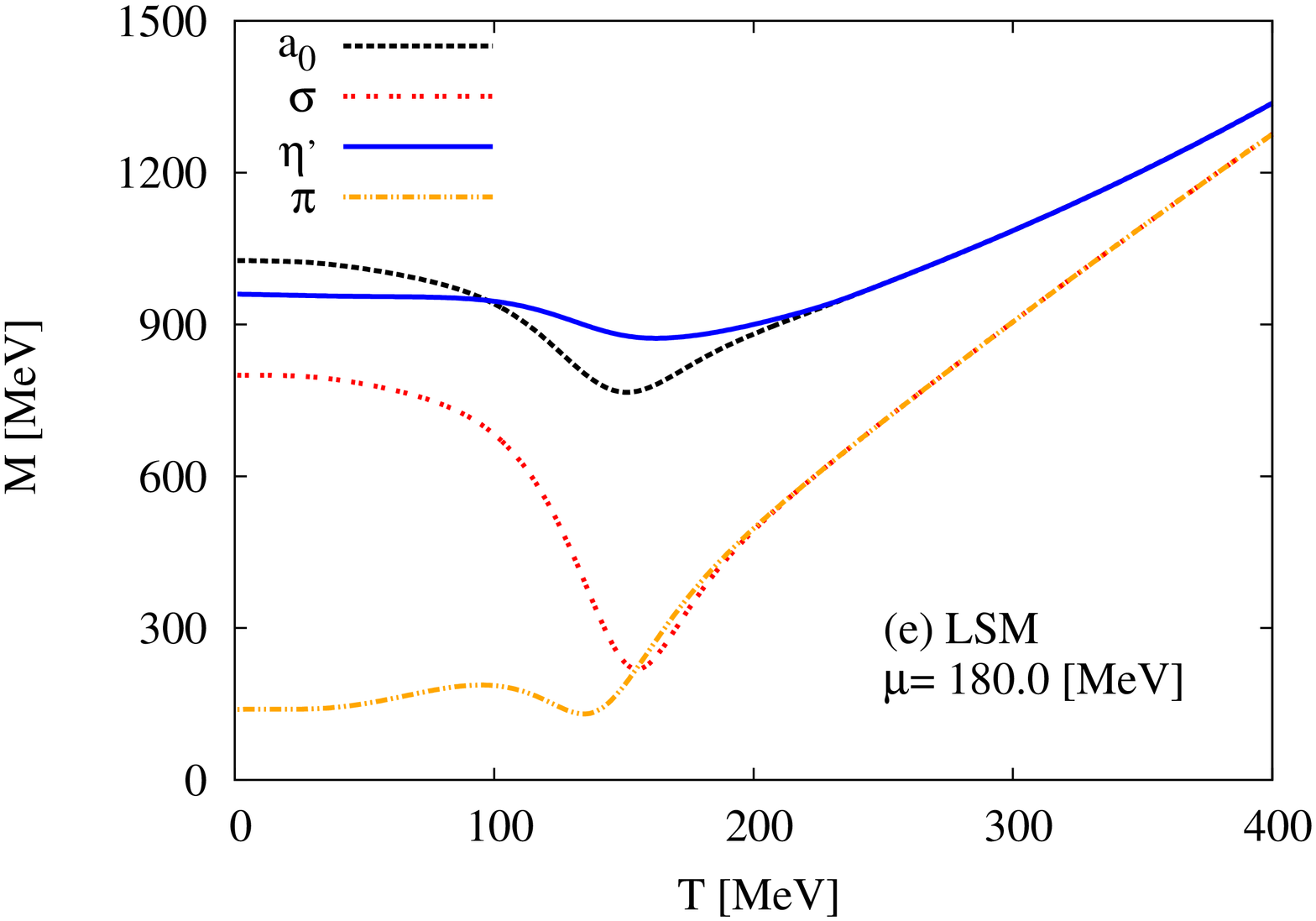}
\includegraphics[width=3.5cm,angle=-90]{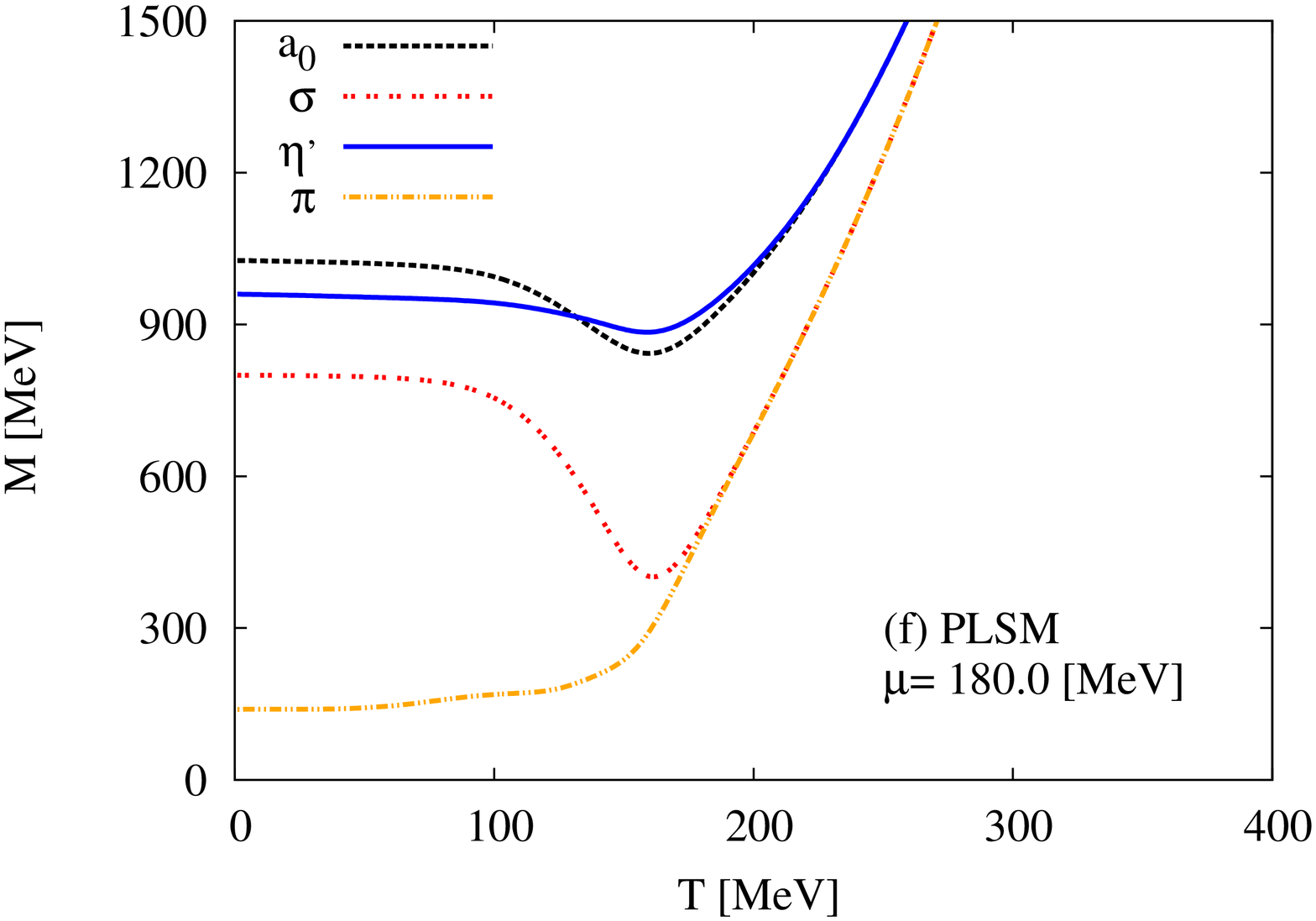}\\
\includegraphics[width=3.5cm,angle=-90]{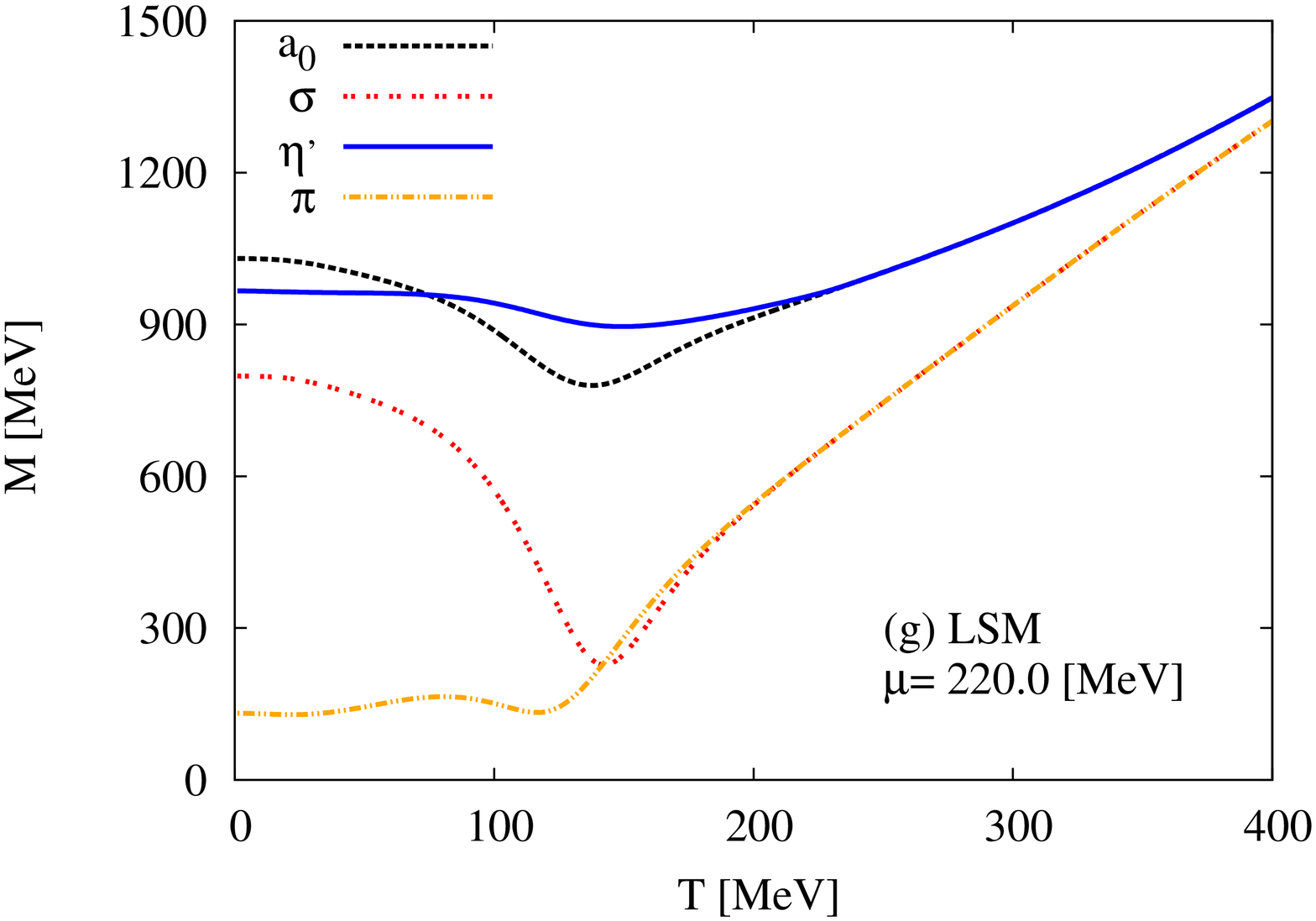}
\includegraphics[width=3.5cm,angle=-90]{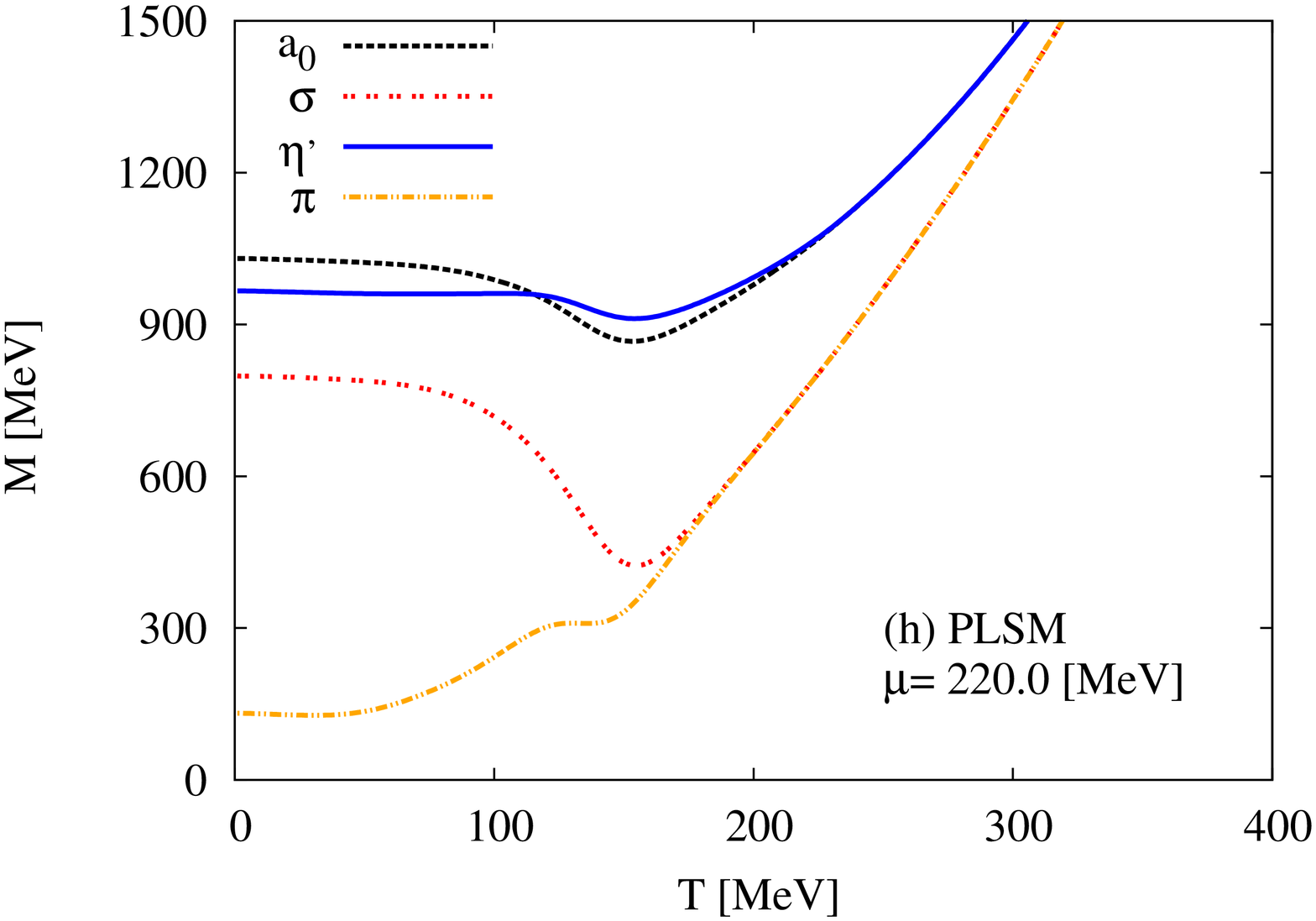}\\
\includegraphics[width=3.5cm,angle=-90]{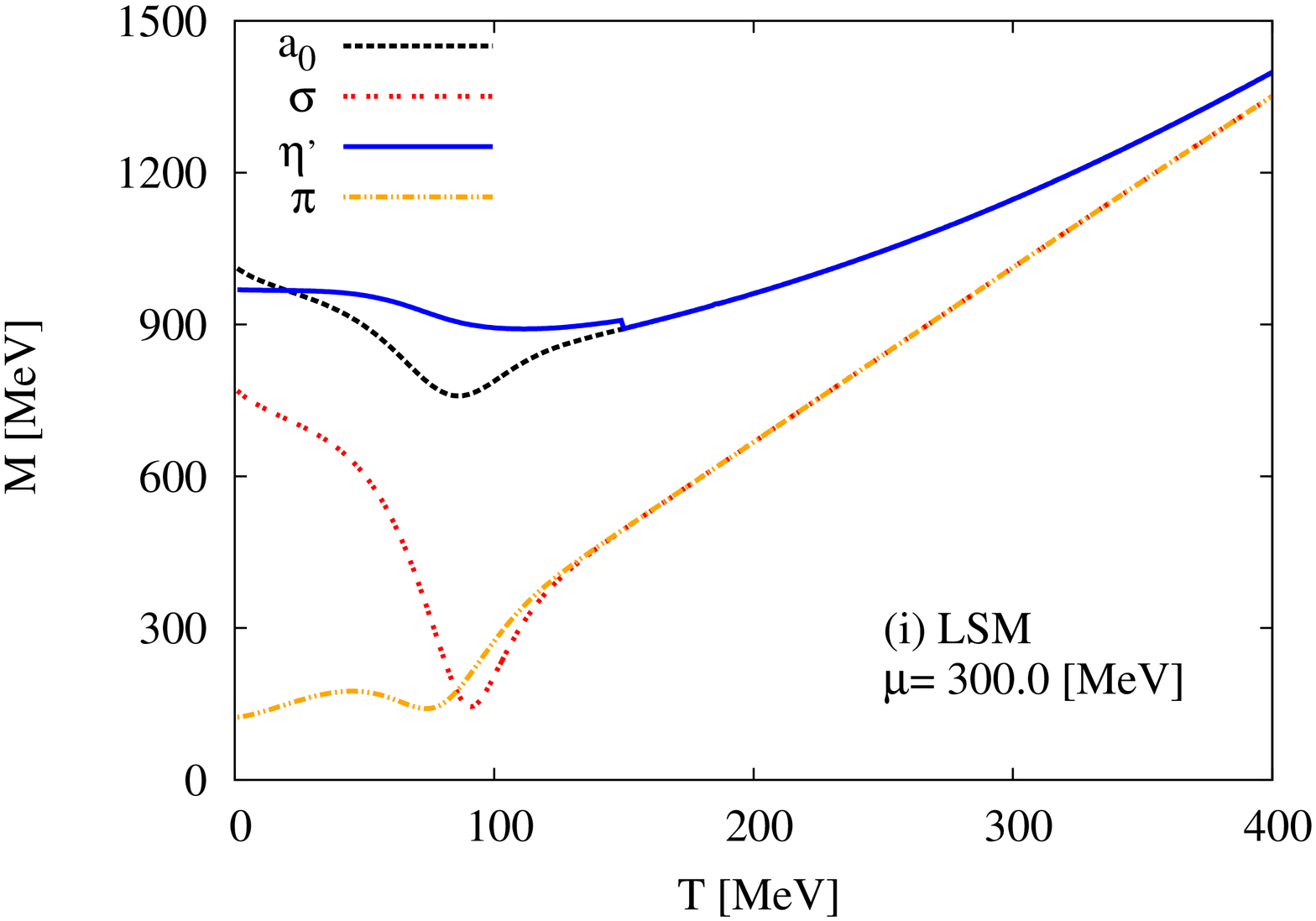}
\includegraphics[width=3.5cm,angle=-90]{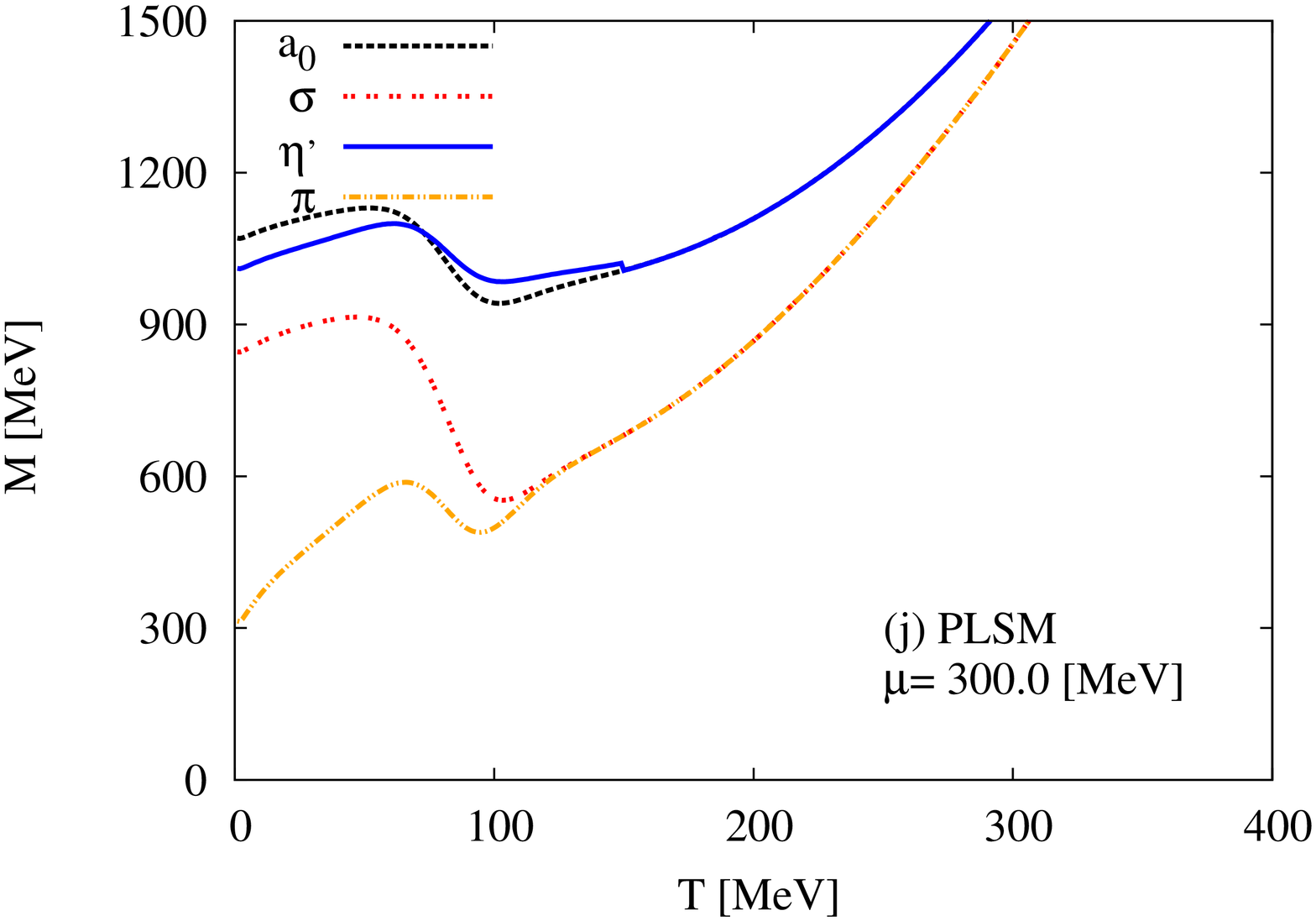}
\label{1SP_diff_MU}
\caption{(Color online) Scalar $a_0$ (dashed curve) and $\sigma$ (dotted curve)  and pseudoscalar states $\eta^{,}$ (solid curve)  and $\pi$  (dashed-dotted curve) are given as function of temperature at different baryon-chemical potentials $\mu=0$, $100$, $180$, $220$ and $300~$MeV. The left-hand panel shows \lsm$\,$ results. The P\lsm$\,$ are presented in the right-hand panel. 
\label{SPT1}
}}
\end{figure}

\begin{figure}[htb]
\centering{
\includegraphics[width=3.5cm,angle=-90]{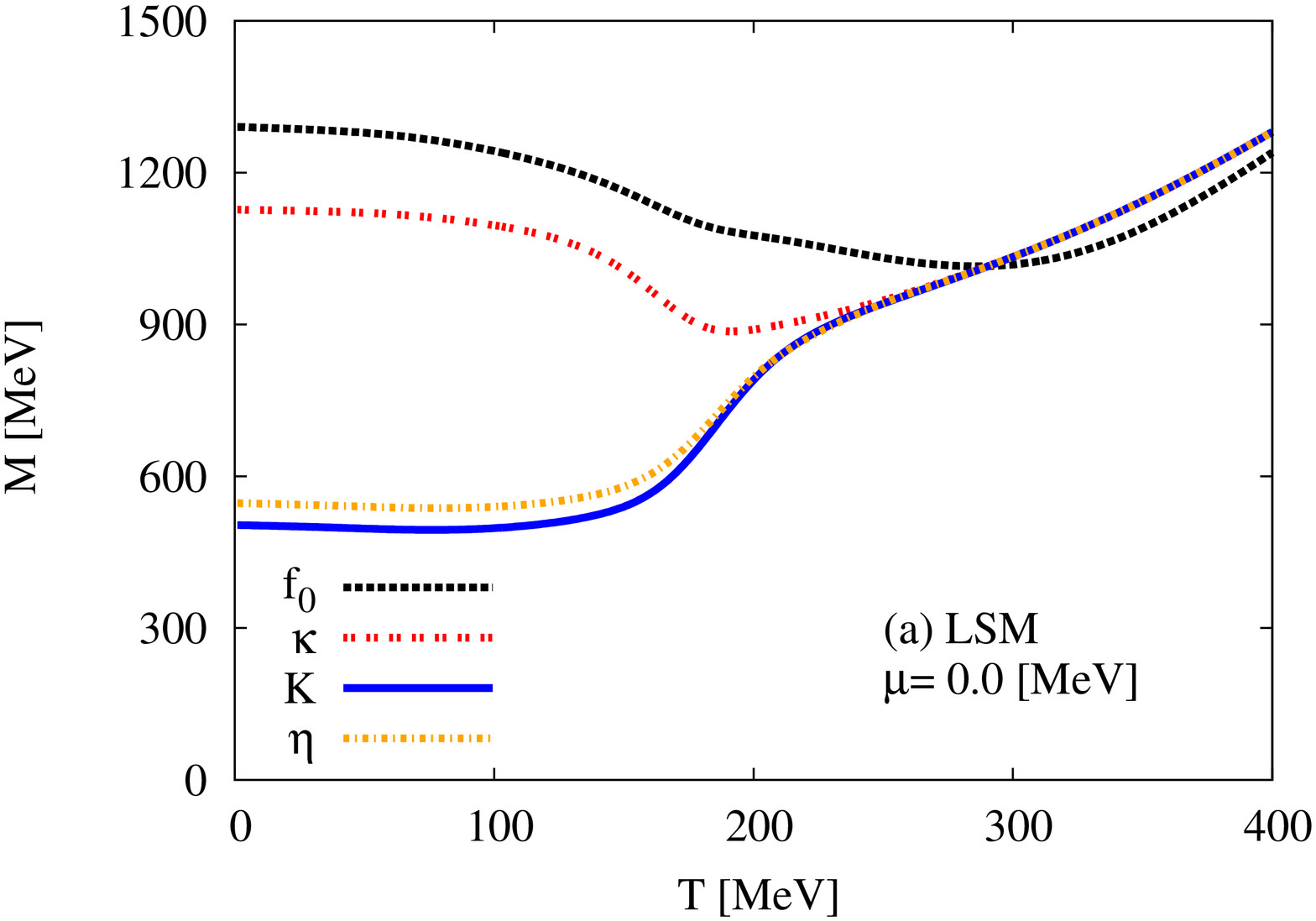}
\includegraphics[width=3.5cm,angle=-90]{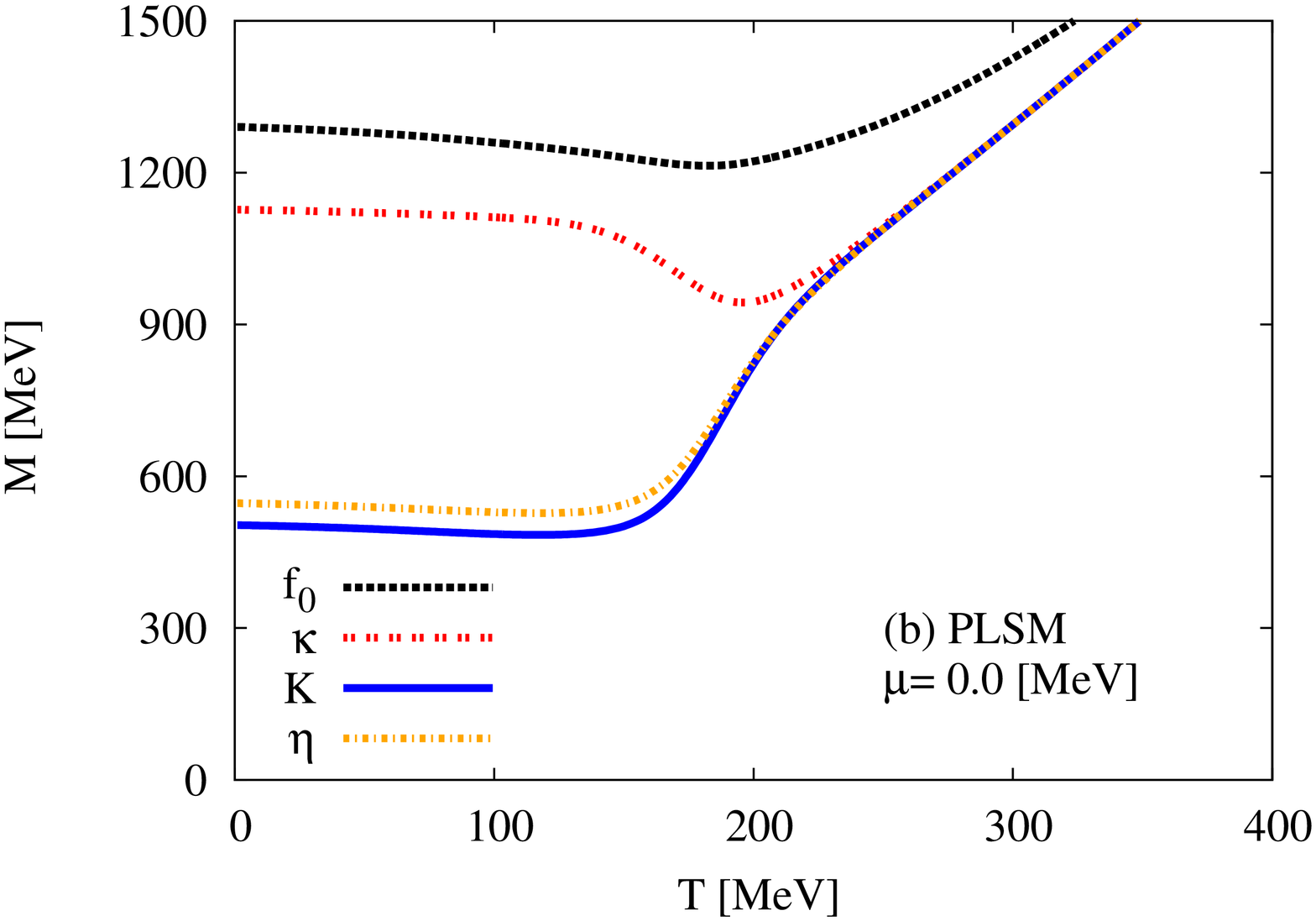}\\
\includegraphics[width=3.5cm,angle=-90]{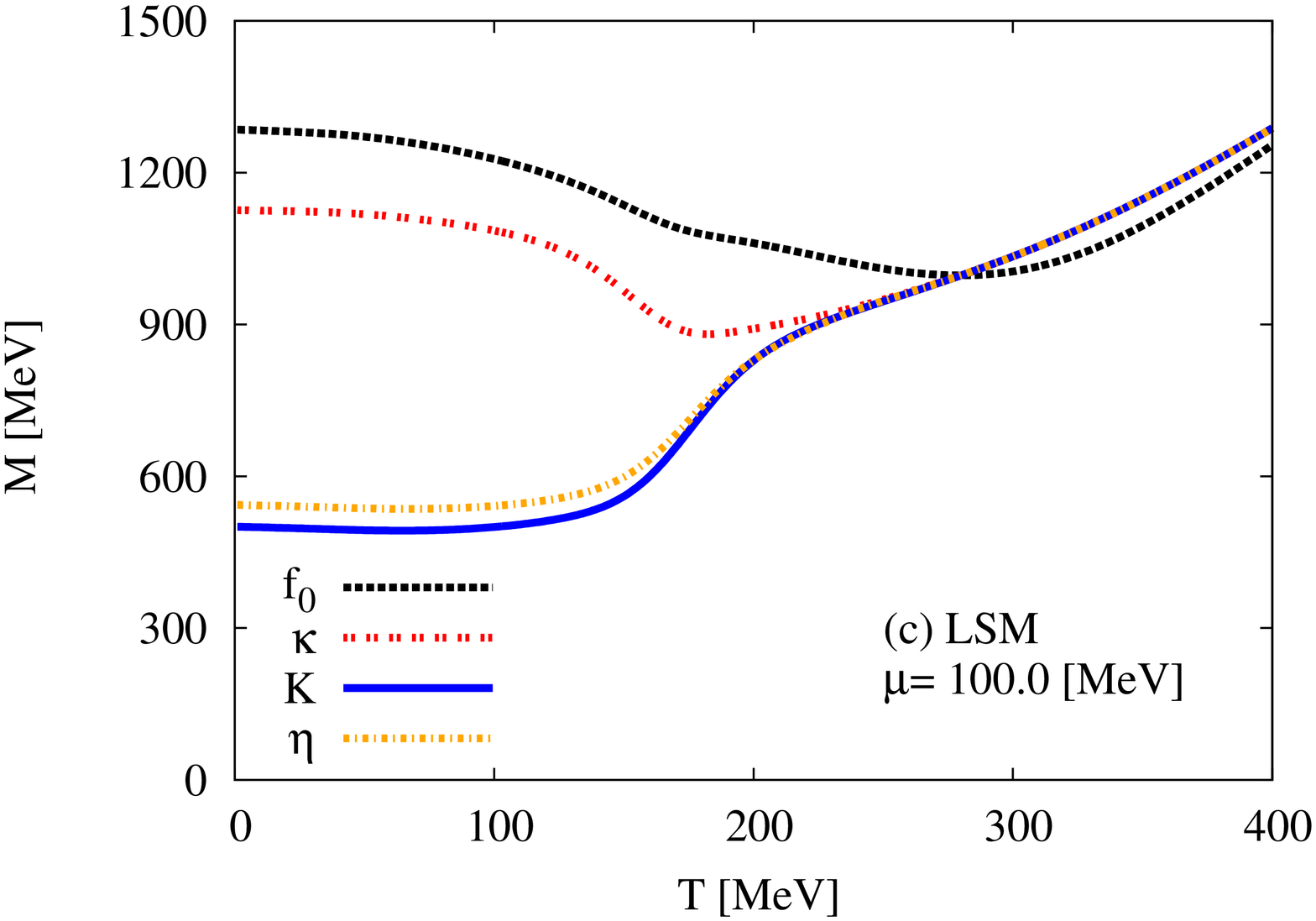}
\includegraphics[width=3.5cm,angle=-90]{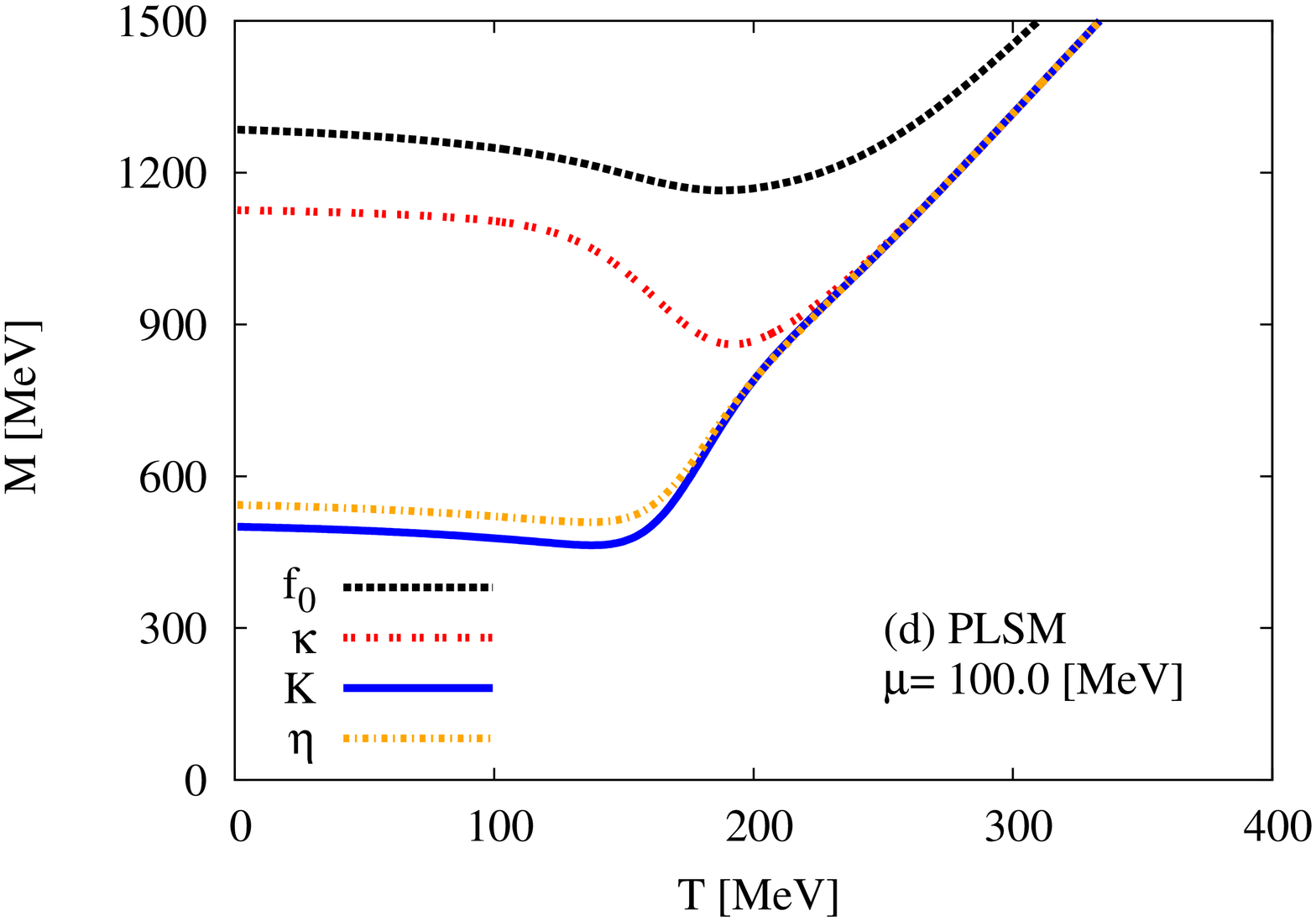}\\
\includegraphics[width=3.5cm,angle=-90]{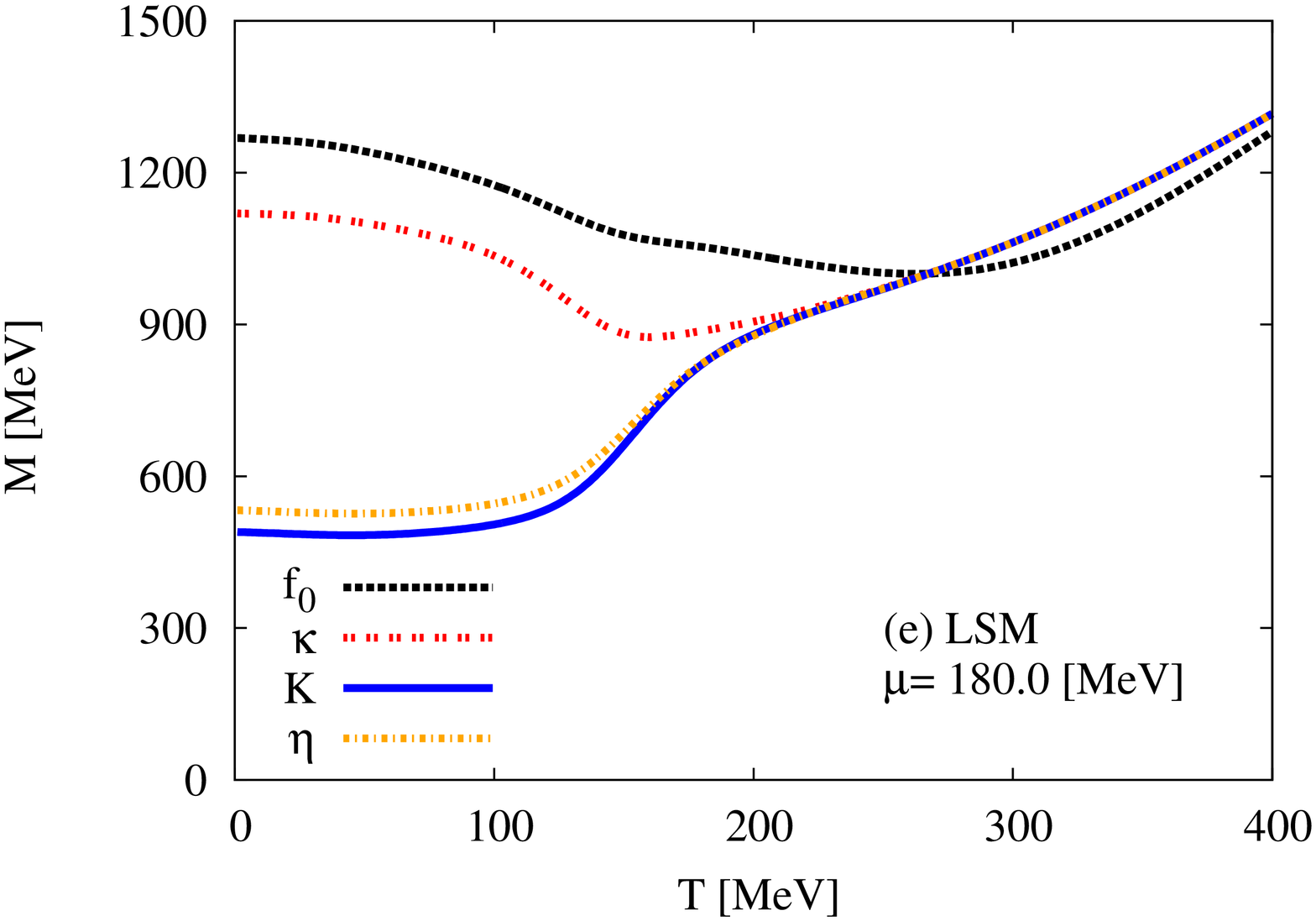}
\includegraphics[width=3.5cm,angle=-90]{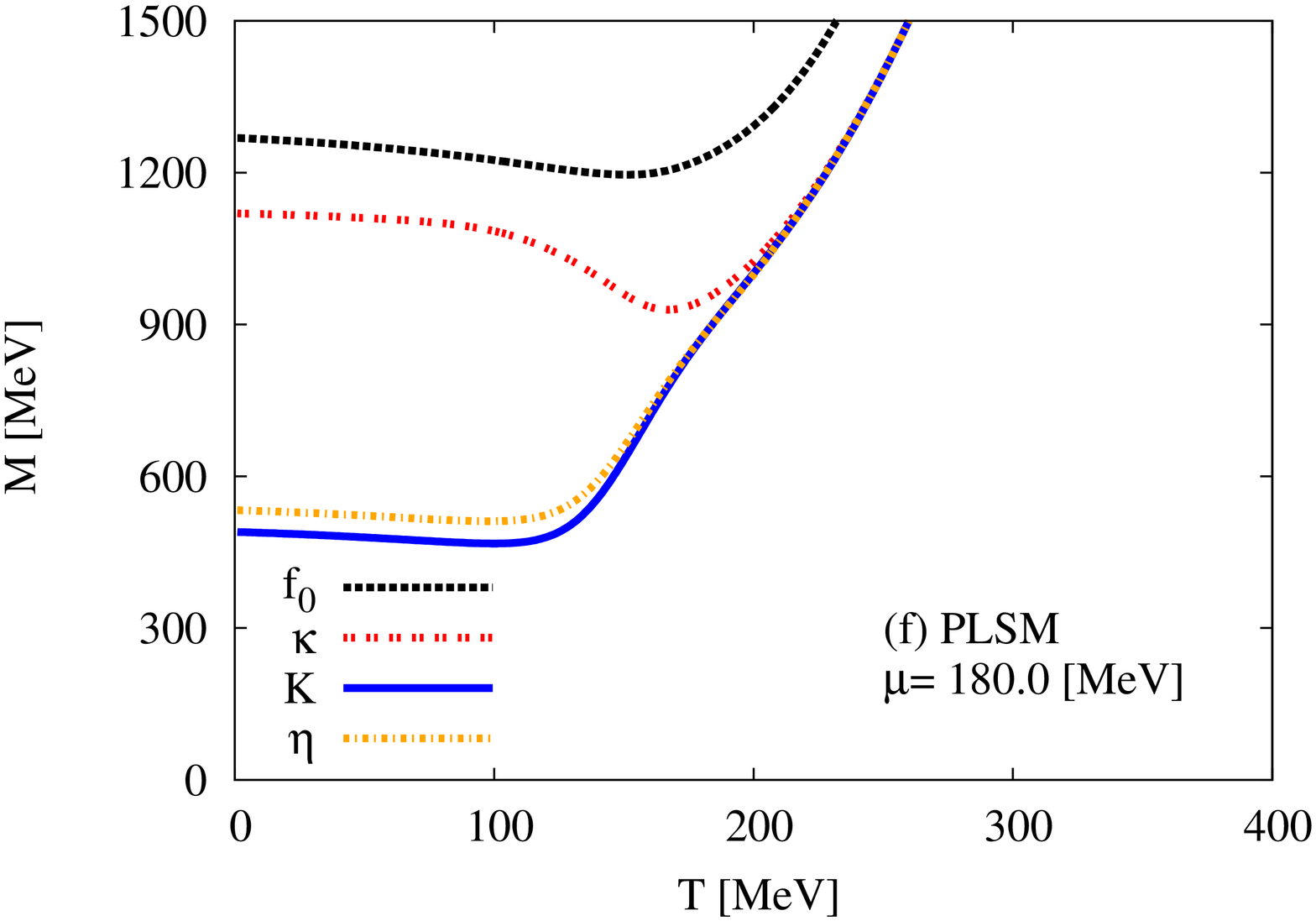}\\
\includegraphics[width=3.5cm,angle=-90]{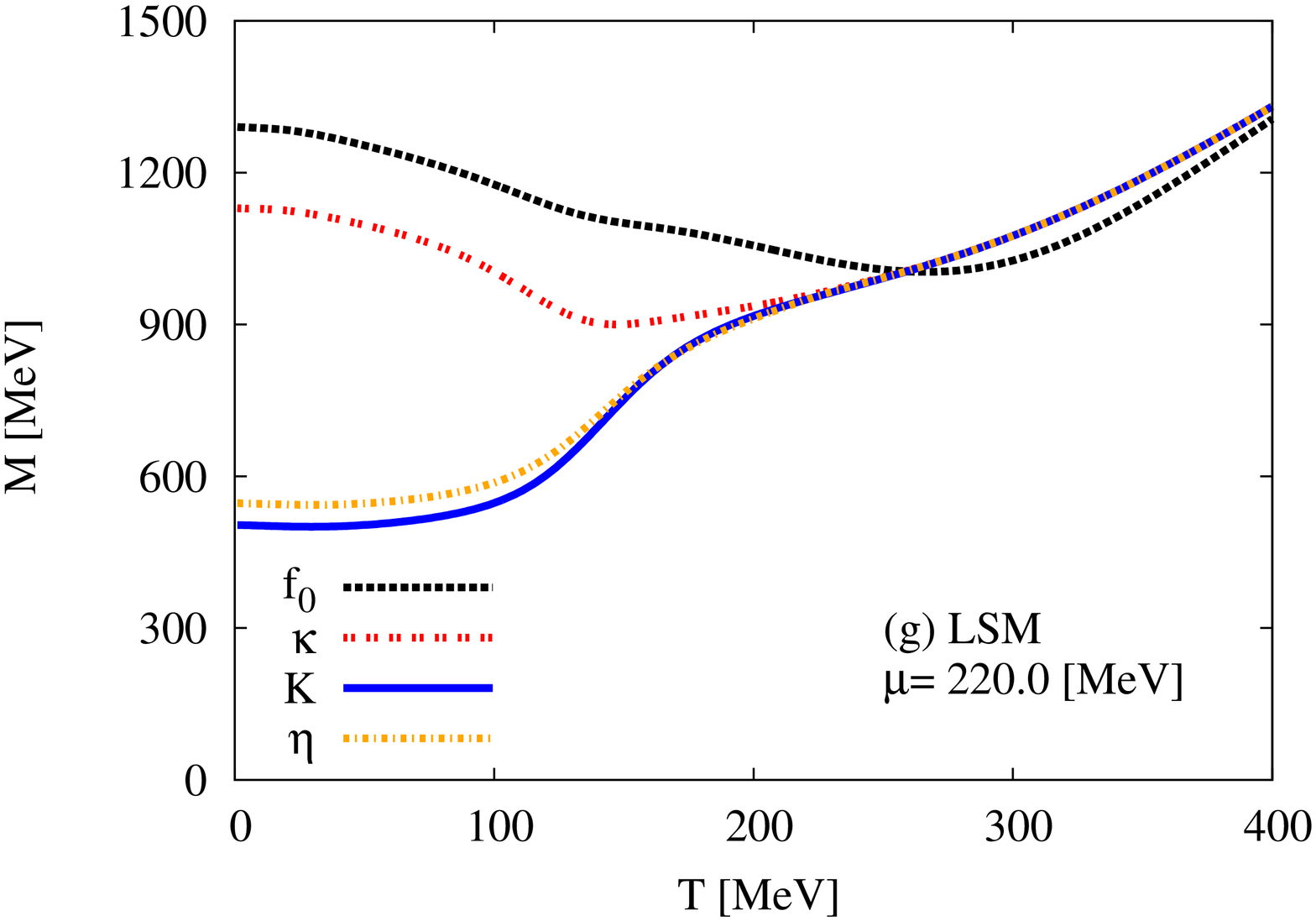}
\includegraphics[width=3.5cm,angle=-90]{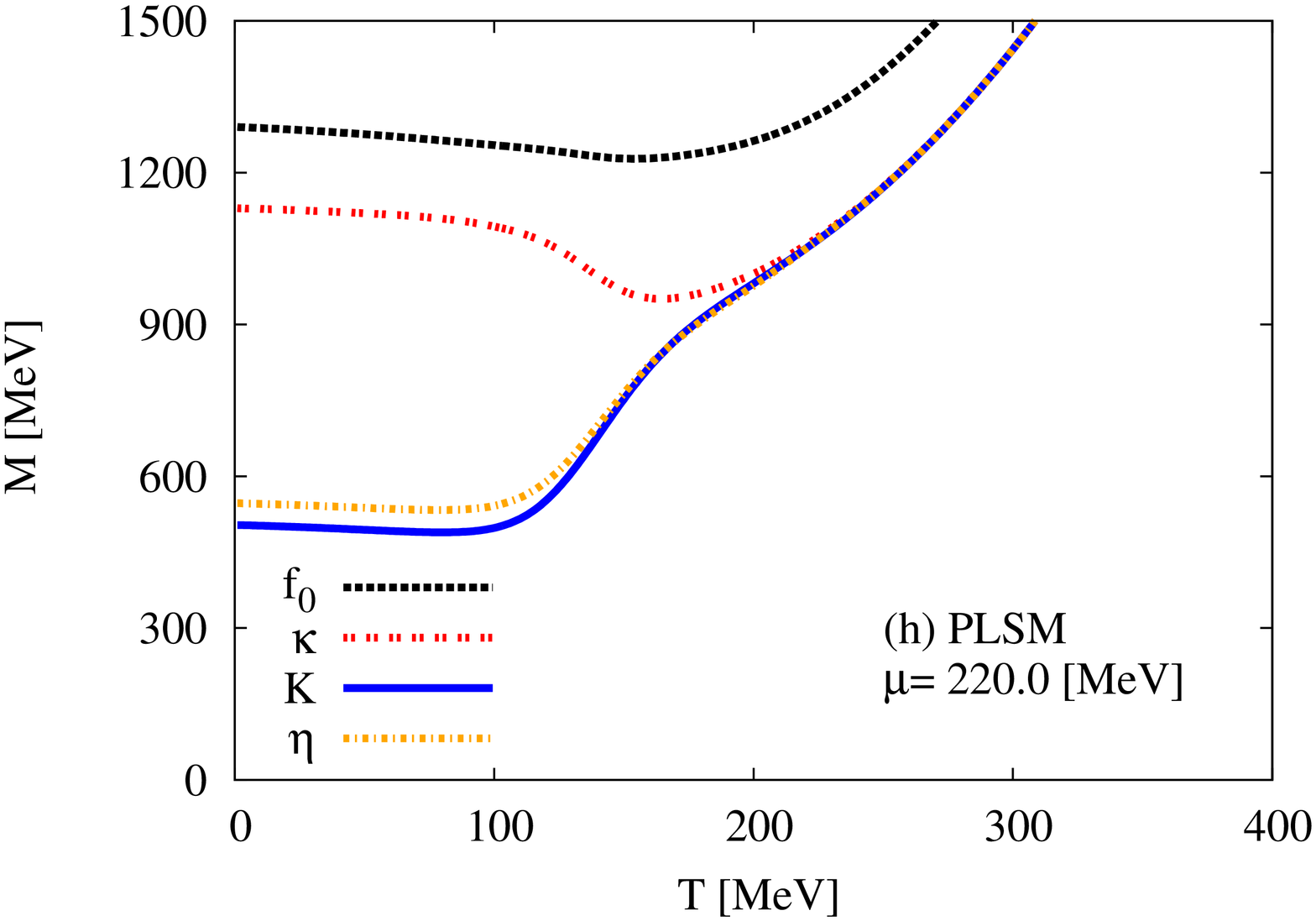}\\
\includegraphics[width=3.5cm,angle=-90]{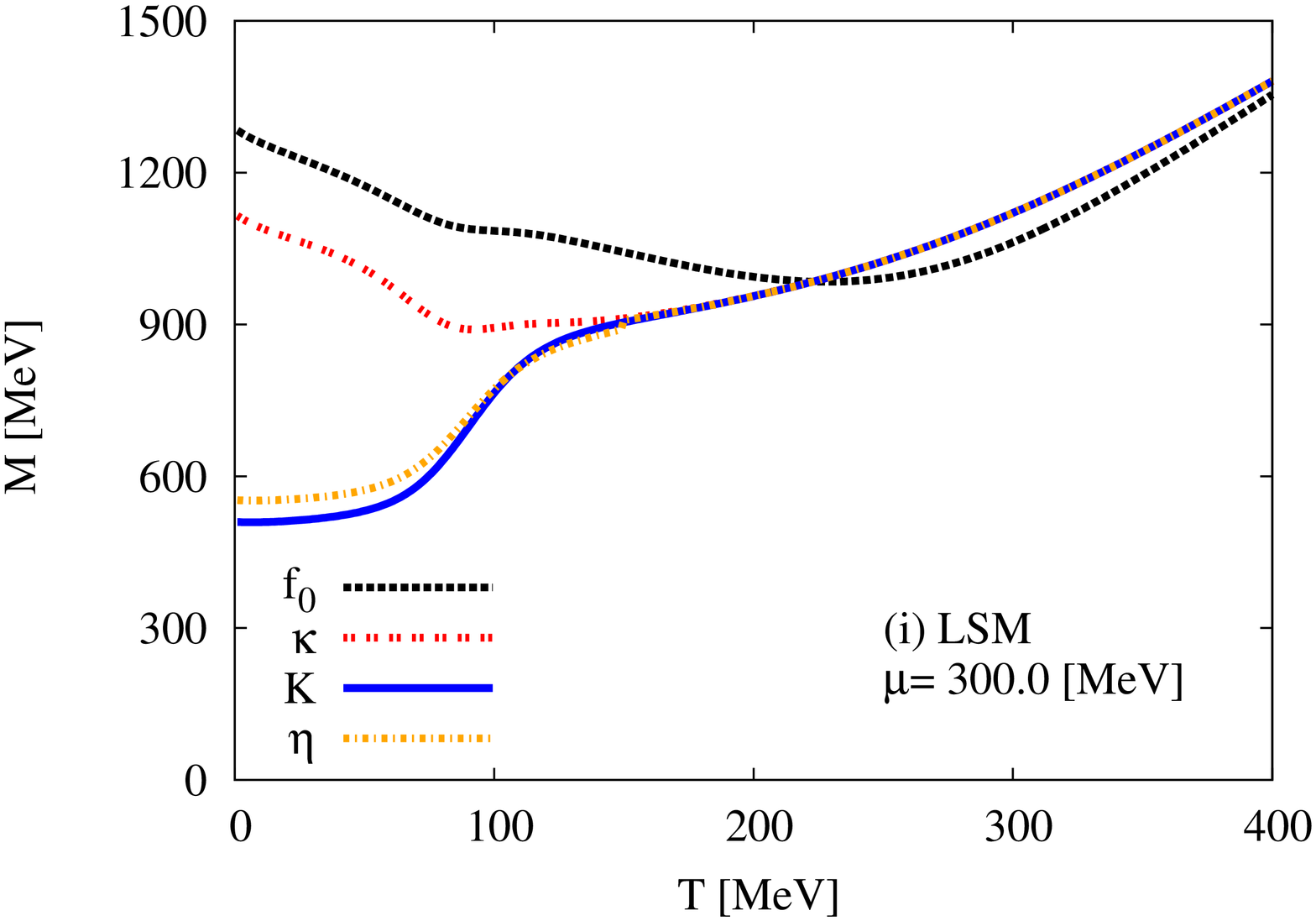}
\includegraphics[width=3.5cm,angle=-90]{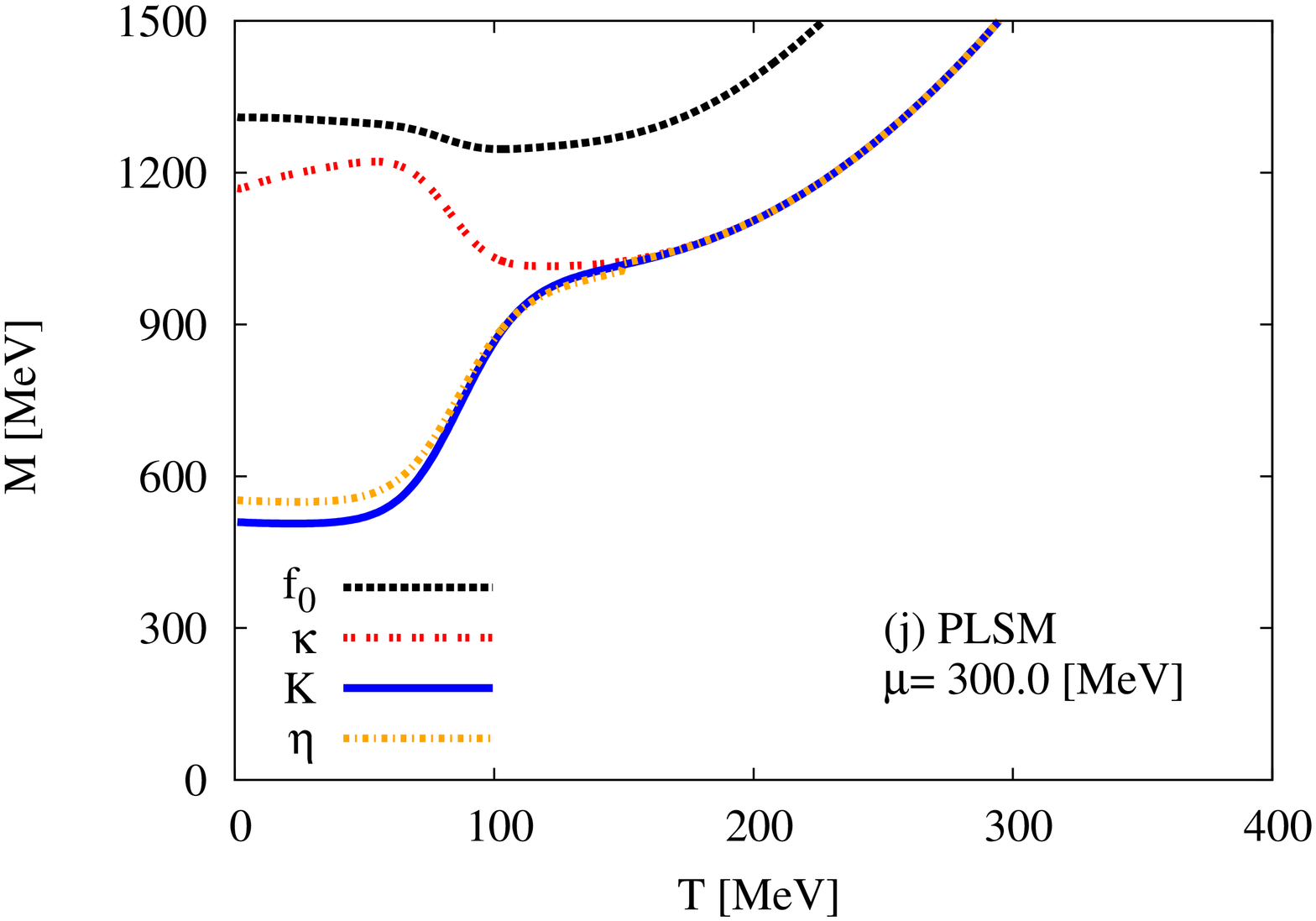}
\caption{(Color online) Left-hand panel (\lsm) and right-hand panel (P\lsm): scalars $f_0$ (horizontal dashed curve) and $\kappa$ (vertical dashed curve) and pseudoscalars $\eta $ (dotted curve)  and $K$ (solid curve) are given in dependence on the temperature at different baryon-chemical potentials $\mu=0$, $100$, $180$, $220$ and $300~$MeV.
\label{SPT2}
}}
\end{figure}

\begin{figure}[htb]
\centering{
\includegraphics[width=3.5cm,angle=-90]{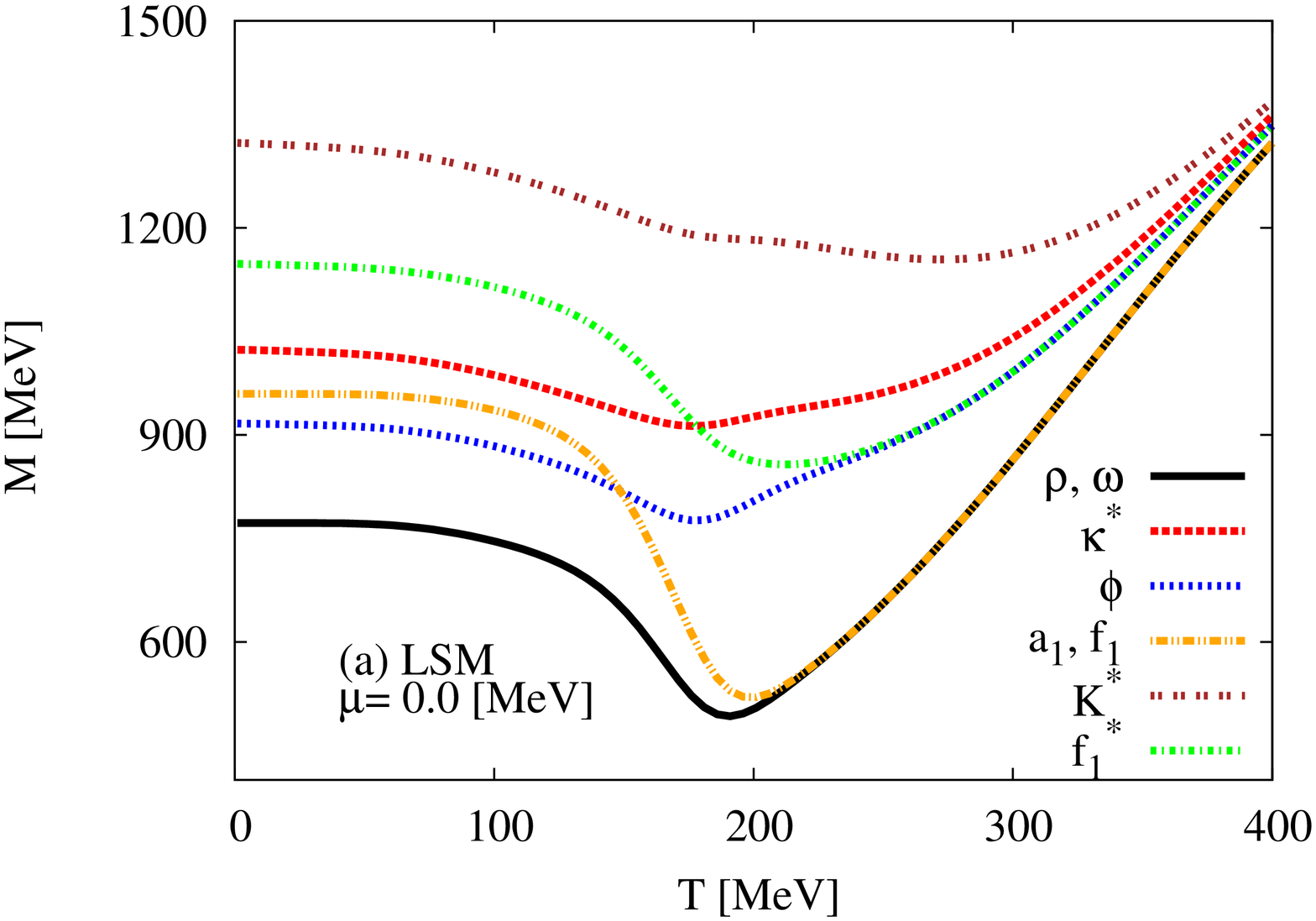}
\includegraphics[width=3.5cm,angle=-90]{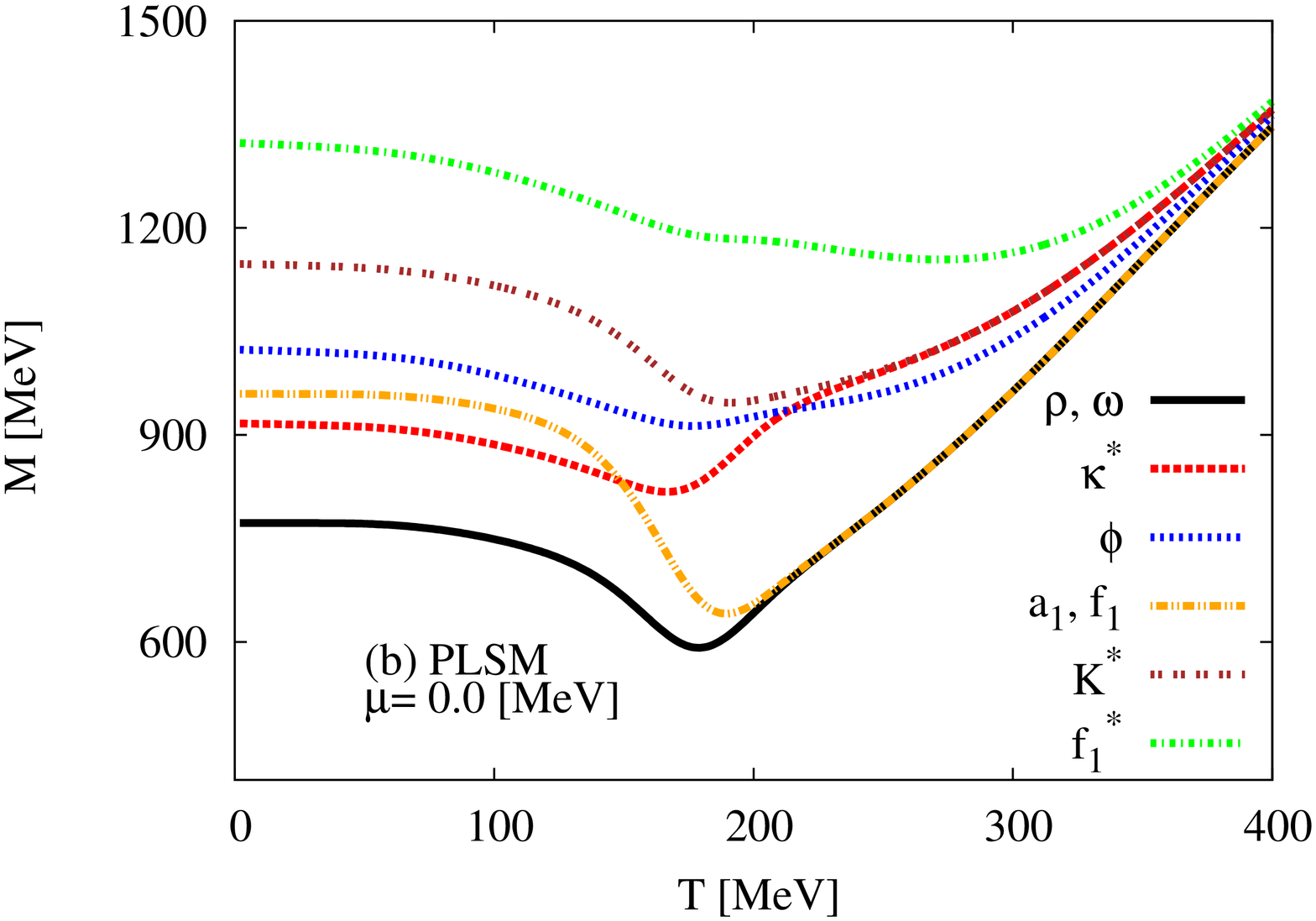}\\
\includegraphics[width=3.5cm,angle=-90]{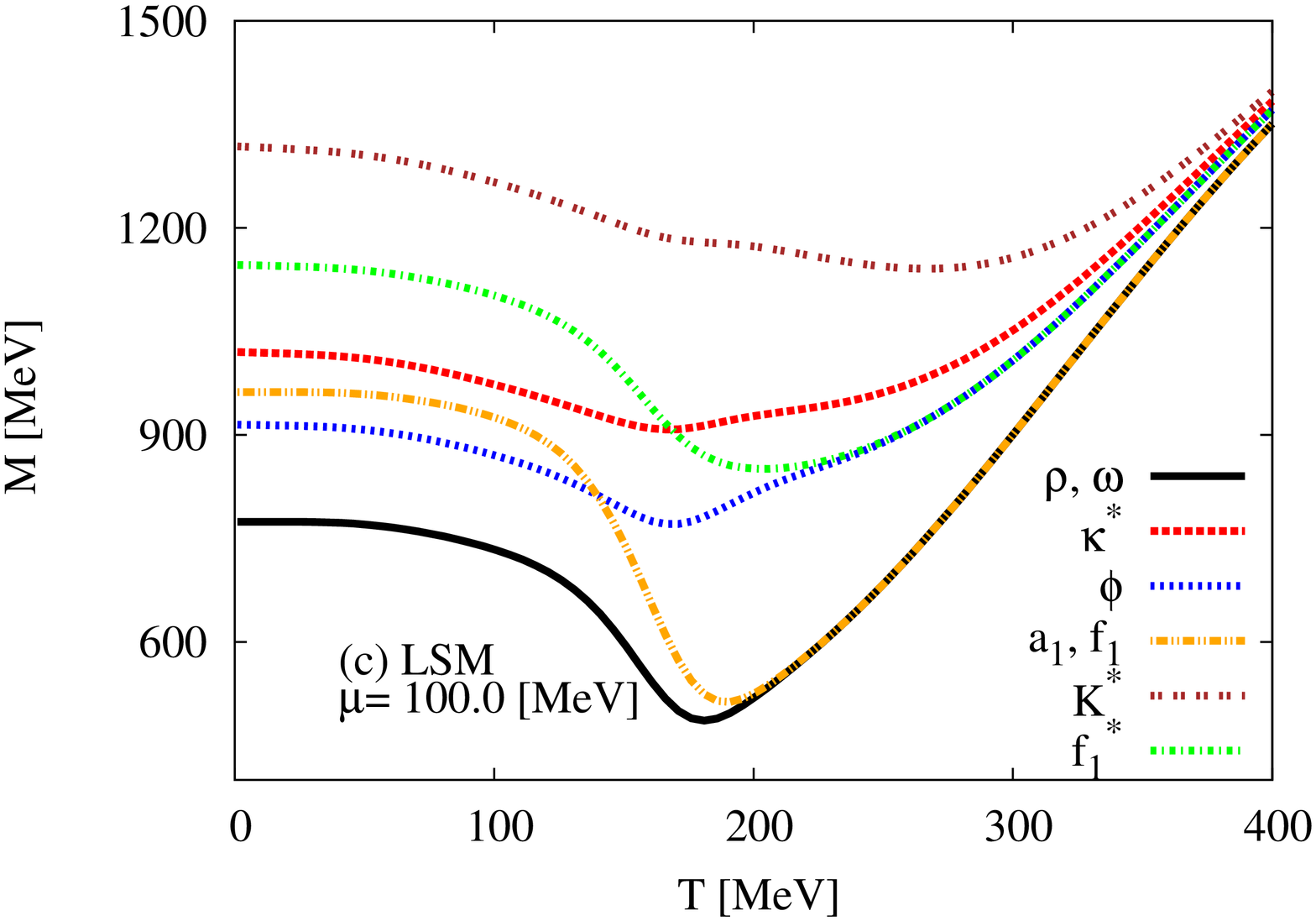}
\includegraphics[width=3.5cm,angle=-90]{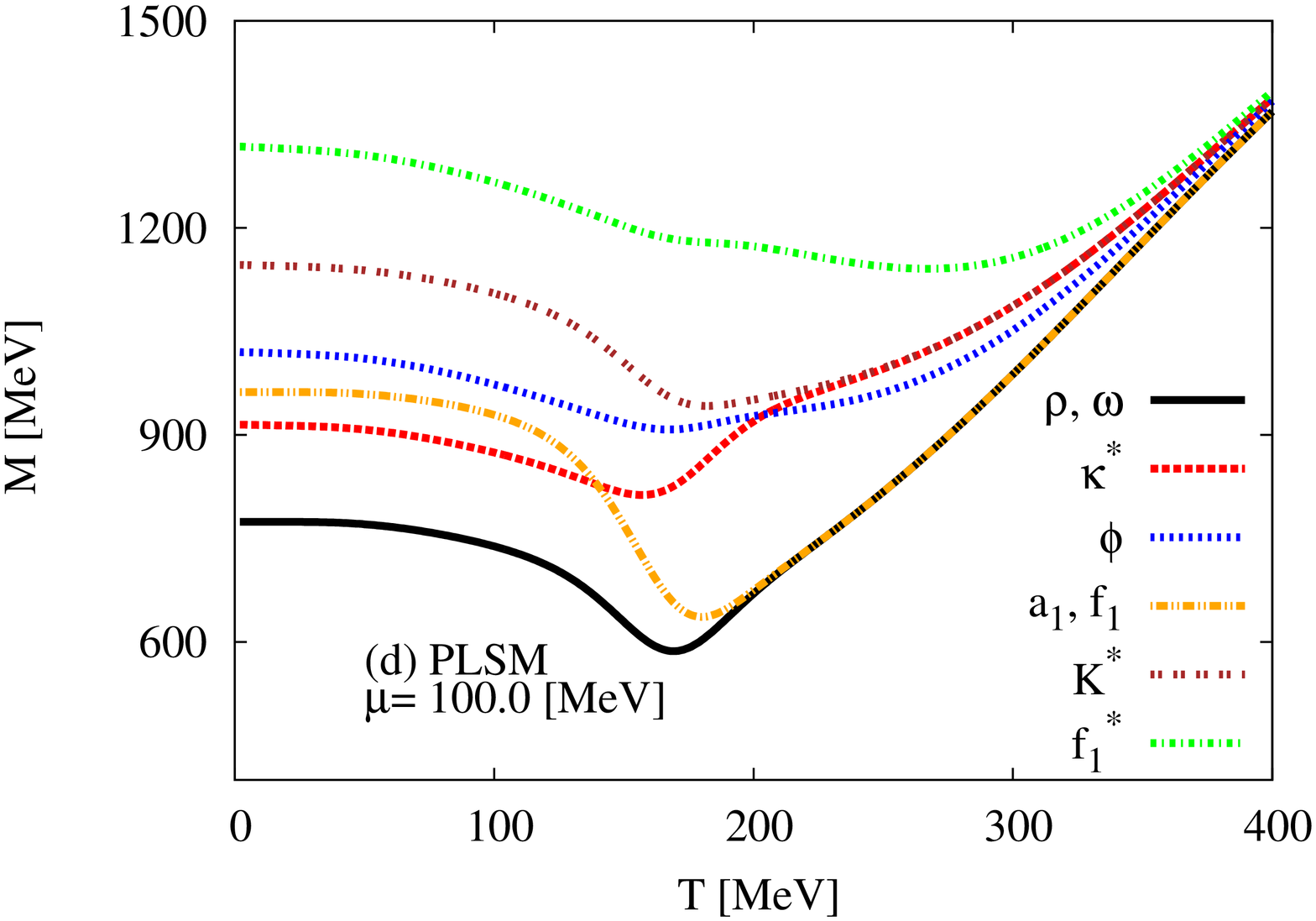}\\
\includegraphics[width=3.5cm,angle=-90]{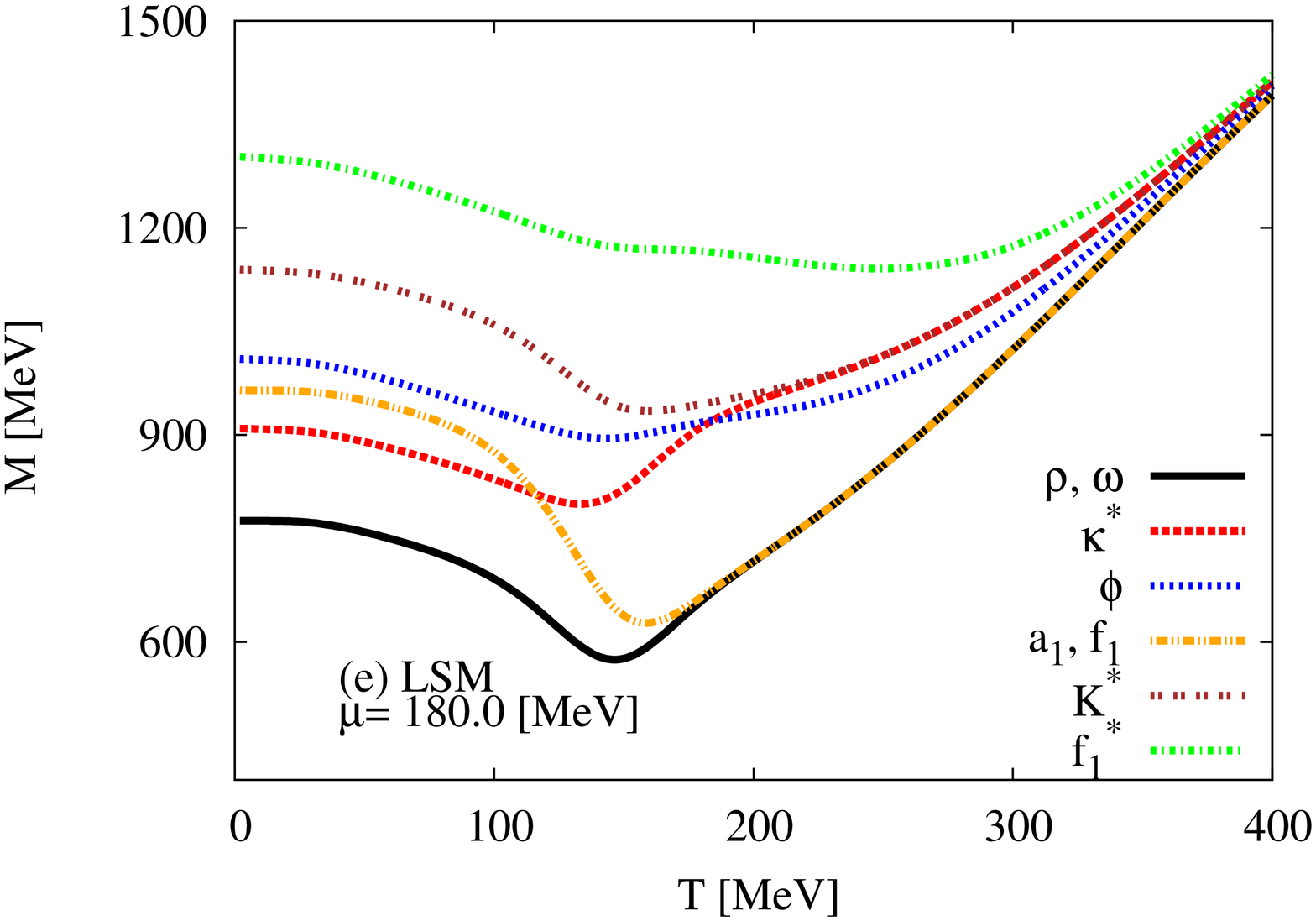}
\includegraphics[width=3.5cm,angle=-90]{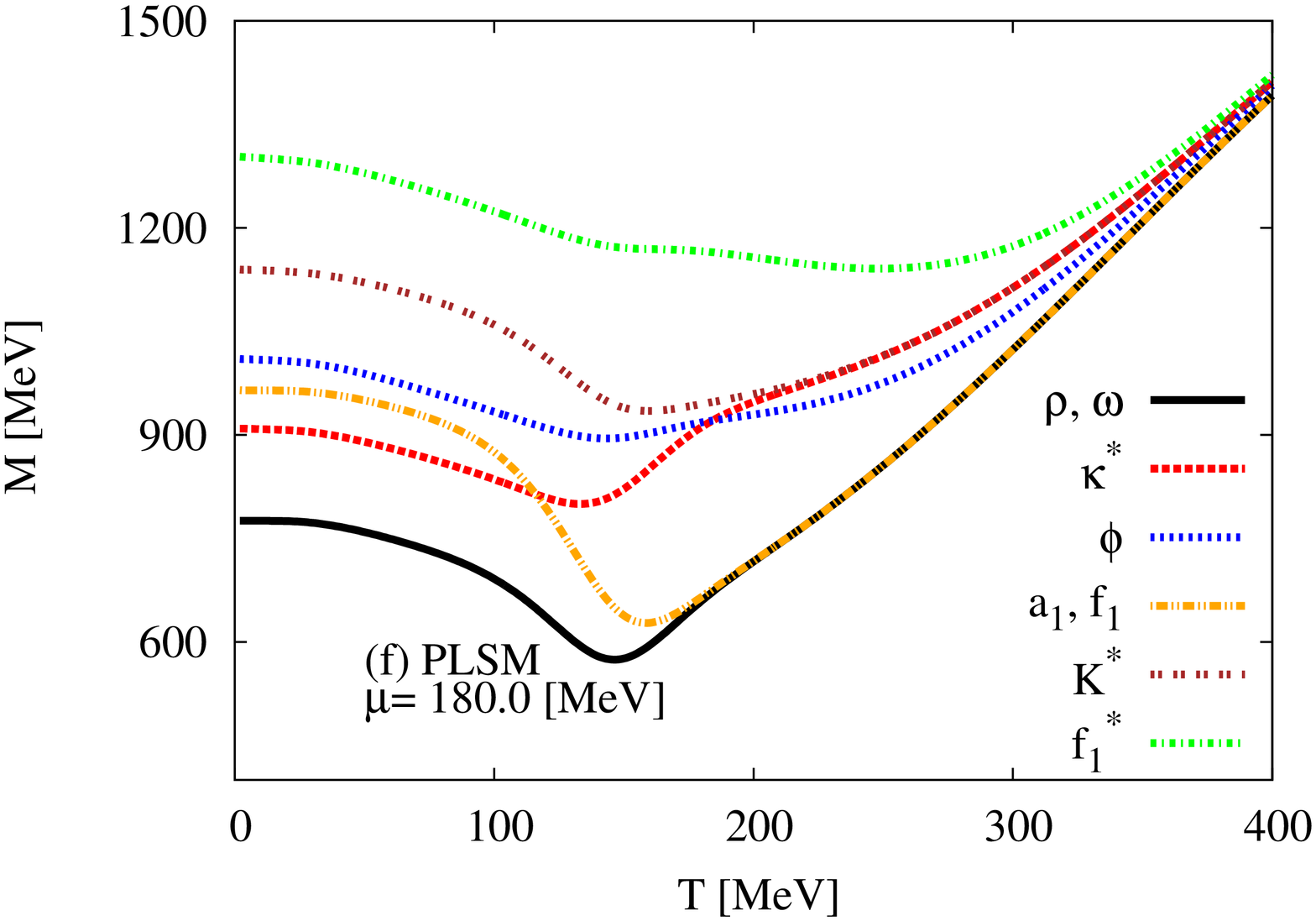}\\
\includegraphics[width=3.5cm,angle=-90]{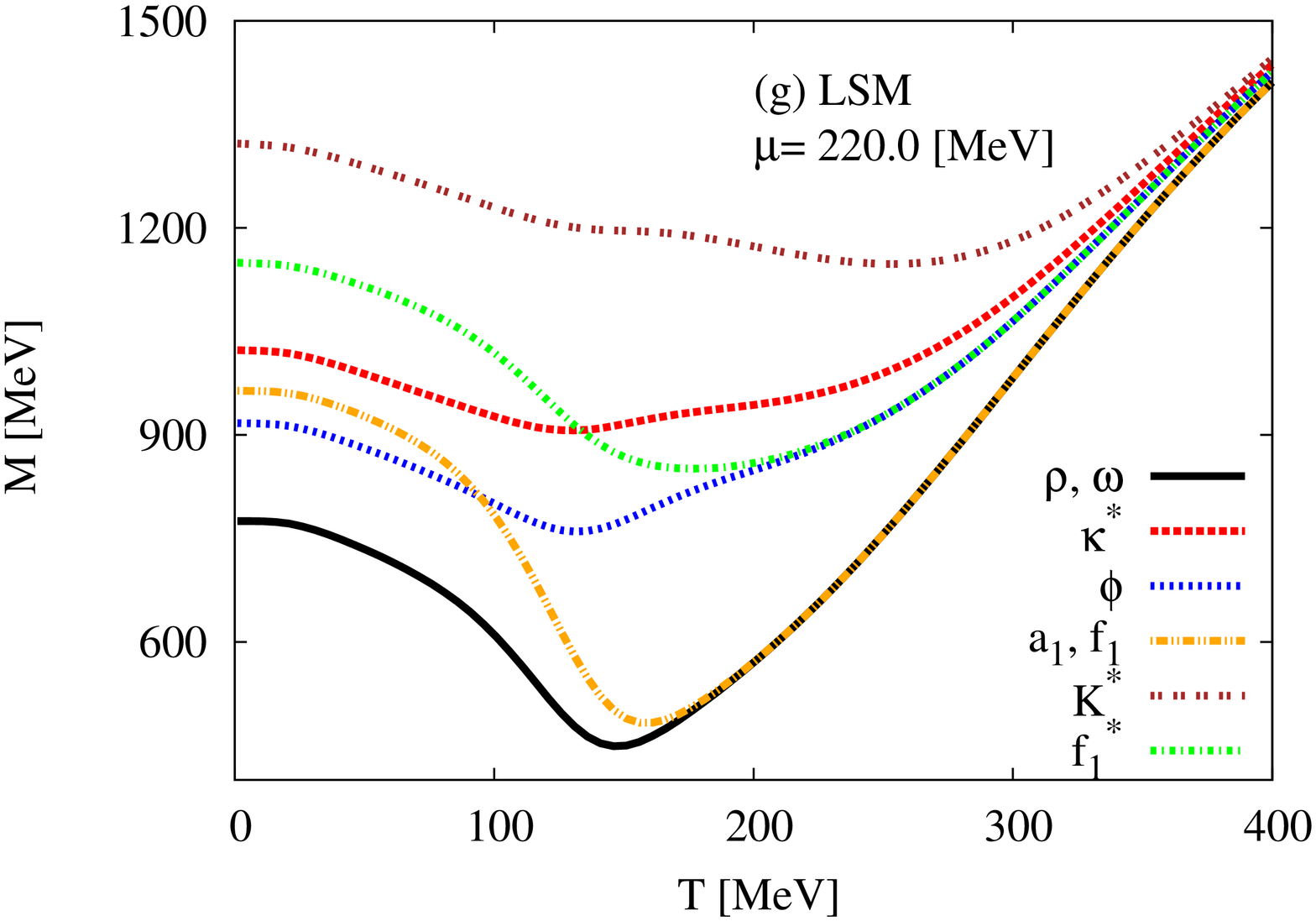}
\includegraphics[width=3.5cm,angle=-90]{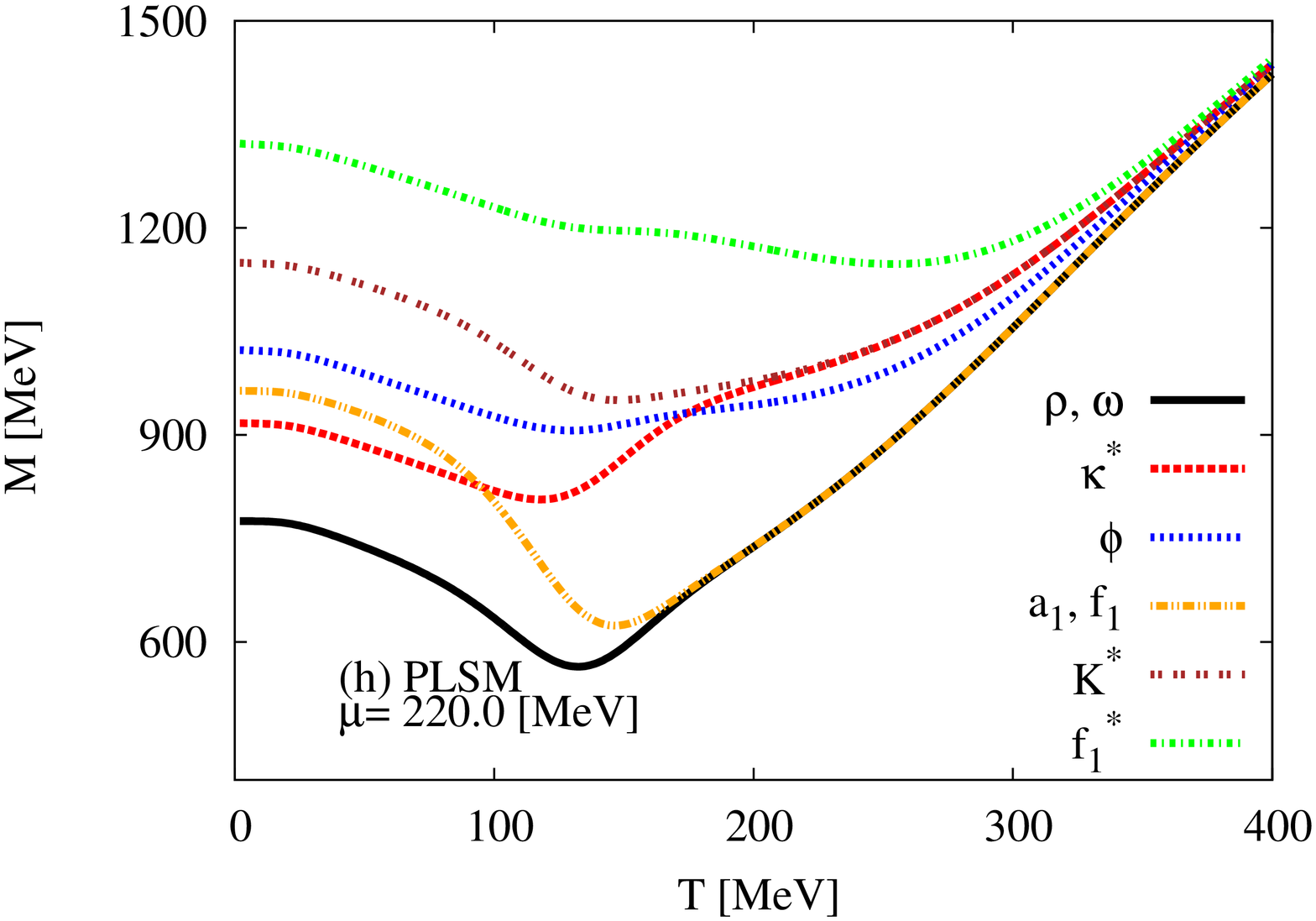}\\
\includegraphics[width=3.5cm,angle=-90]{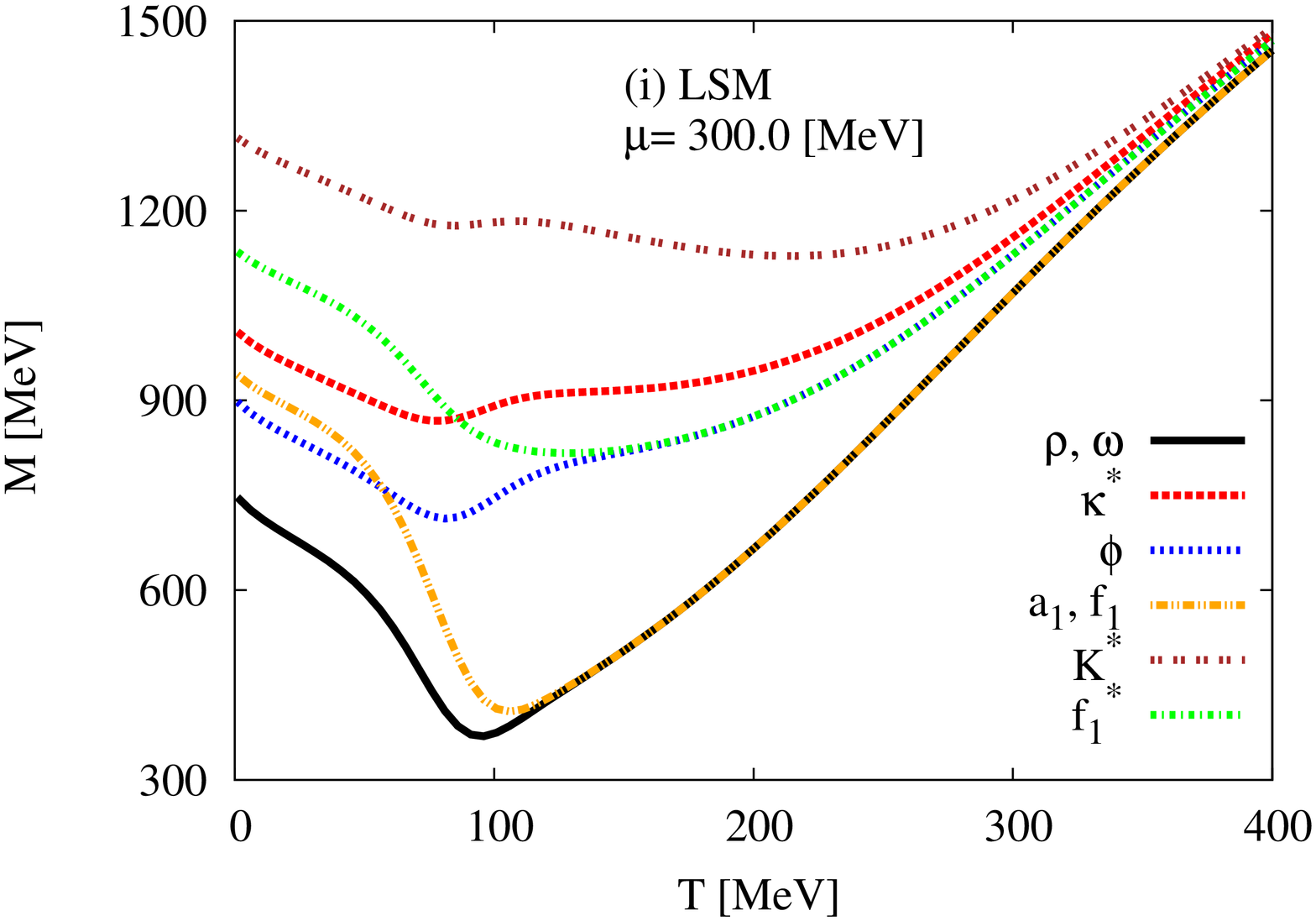}
\includegraphics[width=3.5cm,angle=-90]{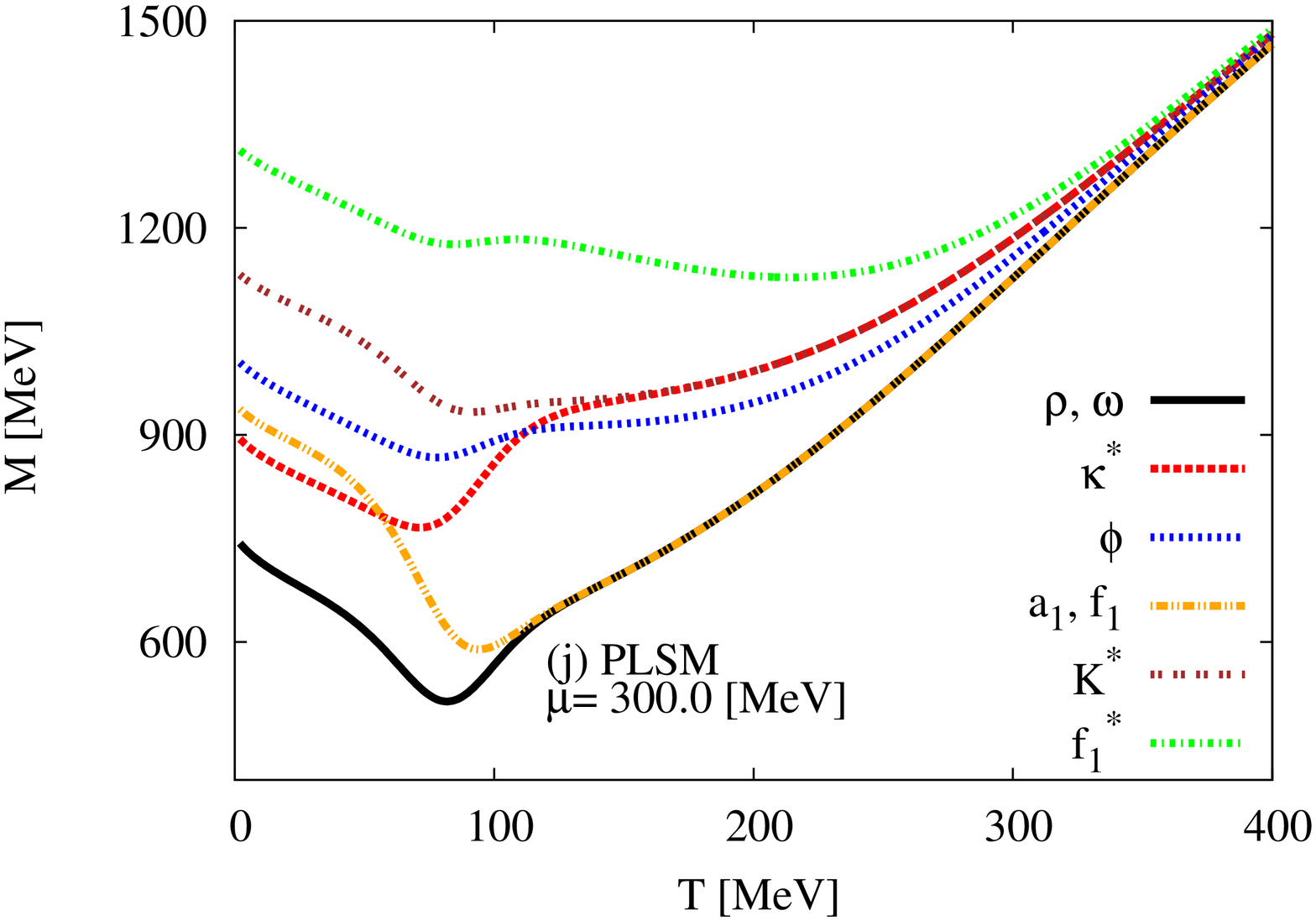}
\caption{(Color online) Left-hand panel (\lsm) and right-hand panel (P\lsm): vector mesons $\rho$ and $\omega$ (solid  curve), $\kappa^{*}$ (long-dotted curve) and $\phi$ (dotted curve)  and axial-vector mesons $a_1=f_1 $ (dashed-dotted curve) , $K^*$  (dotted curve) and $f_{1}^*$ (short dashed-dotted curve) are given in $T$-dependence at different baryon-chemical potentials $\mu=0$, $100$, $180$, $220$ and $300~$MeV.
\label{AVT1} 
}}
\end{figure}

In the presence of chiral symmetry breaking and the correction of Polyakov-loop potential, we present different scalar and vector meson nonets in thermal- and hadronic dense-medium and estimate the corresponding meson spectrum. We start with meson masses at finite temperature and varying baryon-chemical potential in both \lsm$\,$ and P\lsm. The thermal evolution for scalar and pseudoscalar are shown in Figs. \ref{SPT1} and \ref{SPT2}, respectively. The vector and axial-vector are presented in Fig. \ref{AVT1}. In the same way, the mass spectrum at nonzero chemical potential in both \lsm$\,$ and P\lsm$\,$  in dense-medium are shown in Figs. \ref{SPMU1} and \ref{SPMU2} for scalar and pseudoscalar mesons and in Fig. \ref{AVMU1} for both vector and axial-vector mesons.

The temperature variations of mesonic masses can be understood as the in-medium thermal effects on the mesonic states. As shown in Figs. \ref{SPT1} and \ref{SPT2}, respectively, the bosonic thermal contributions to the mesonic masses decrease with increasing the temperature, while the fermionic contributions increase at high temperatures. The fermionic (quark) contributions are negligible at small  temperatures. At high temperatures, the bosonic thermal contributions dominates. This leads to degeneration in the mesonic masses, which in turn leads to a natural change in chiral and deconfinement phase-transition with increasing temperature.

In Fig. \ref{SPT1}, the left-hand panel shows the two scalar meson sectors, $a_0$, $\sigma$ and the two pseudoscalar meson sectors, $\eta ^{,}$ and $\pi$, in thermal hadronic medium at vanishing baryon-chemical potentials $\mu$ in the presence of U(1)$_A$ symmetry breaking. The U(1)$_A$ symmetry breaking gets effectively restored and repeals the mass gap between the chiral partners \cite{Schaefer:2009}, where at very large temperatures comparable to the strange quark mass, the difference between the strange and non-strange mesons becomes negligible, Fig. \ref{fig:Massus}. Accordingly, all mesonic masses will degenerate. Since at very high temperature, the major effect takes place in the strange mass, such as $a_0$ and $\eta^{,}$ the masses of $\sigma$ and $\pi$ degenerate in close vicinity of reduced temperature. This result is compatible with the result reported in Ref. \cite{Schaefer:2009}. The masses of $a_0$ and $\eta^{,}\sim 250\,$MeV and  masses of $\sigma$ and $\pi\sim 181~$MeV. The term with U(1)$_A$ symmetry breaking appears in the meson masses through the anomaly breaking term, $c$. It is strongly related to the strange condensate $\sigma_y$. In the right-hand panel, the Ployakov-loop correction is introduced. This correction seems to enhance the quark dynamics and raise the mass degeneration in a sharp and fast way.

In Fig. \ref{SPT1}, the different panels present an systematic study for the effects of the chemical potentials on the sixteen mesonic states. We find that increasing the baryon-chemical potential (from top to bottom panels) enhances the degeneration of the mesonic masses. For example, at $\mu=100\,$MeV, four meson states $a_0$, $\eta^{,}$ become degenerate at $\sim 240\,$MeV, $\sigma$ and $\pi$ at $\sim 180$, while at $\mu=220~$MeV, the four states $a_0$, $\eta^{,}$ degenerate at $\sim 170\,$MeV, $\sigma$ and $\pi$ at $\sim 125\,$MeV. This has a close relationship with the chiral condensate and the deconfinement phase-transition. In Fig. \ref{sxsyDiif}, the chiral condensates $\bar{\sigma}_x$ and $\bar{\sigma}_y$ and deconfinement phase-transition $\phi$ and $\phi^*$ vary with $T$ and $\mu$. The contributions from the non-strange quarks to the rapid crossover in the non-strange sector are different and affect the contributions of the mesonic masses, very strongly \cite{Schaefer:2009}.

Fig. \ref{SPT2} presents the thermal evolution of the scalars $f_0$ (horizontal dashed curve) and $\kappa$ (vertical dashed curve)  and pseudoscalars $\eta$ (dotted curve)  and $K$  (solid curve)  at different baryon-chemical potentials $\mu=0$, $100$, $180$, $220$ and $300~$MeV. We find that the masses of these states degenerate at $T\sim 240\,$MeV, especially in \lsm. In the same way as shown in Fig. \ref{SPT1} for example, at $\mu=100\,$MeV, the temperatures at which the three mesonic states $\kappa$ ,$K$ and $\eta$, become degenerate $T\sim 240$. Strength of the stability state at low temperatures delays as the density increases.

The left-hand panel of Fig. \ref{AVT1} gives the thermal evolution of $\rho$, $\omega$, $a_1$, $f_1$, $\kappa^*$, $K^*$ and $\phi$ calculated in the \lsm$\,$. We find that the masses of these states degenerate at $T\sim 200\,$MeV, while $\kappa^*$, $K^*$ and $\phi$ at $T\sim 240\,$MeV. At high temperatures, it is obvious that the effects of the non-strange mass vanish. This makes the differences between the various masses disappear. Increasing the baryon-chemical potential reduces the temperatures, at which the masses degenerate. This can be understood on the basis of the thermal evolution of the chiral condensates and the deconfinement phase-transition, shown in Fig. \ref{sxsyDiif}. At $\mu=300\,$MeV, $\rho$, $\omega$, $a_1$ and $f_1$ degenerate at $T\sim 110\,$MeV, while $\kappa^*$, $K^*$ and $\phi$ degenerate at $T\sim 140\, MeV$.  The right-hand panel presents the same results but calculated in P\lsm$\,$.  The Ployakov-loop correction causes a sharp and fast mass-degeneration.

The mass degeneration can be interpreted as an effect of the fermionic vacuum fluctuations on the chiral symmetry restoration \cite{Schaefer:2009}, especially on the condensate $\sigma _y$. The effect seems to melt the strange condensate faster than the non-strange one $\sigma _x$, Fig. (\ref{fig:Transition}). At very high temperature, the mass gap between mesons seems to disappear and decrease with the melting strange condensate $\sigma_y$. This mass gap appears at low temperatures, where the non-strange condensate remains finite. At temperatures higher than  the critical value only strange condensate remains finite. This thermal effects is strongly related to the degeneration of the meson masses.
 
\subsubsection{Density dependence}
\label{Mu_dependence}

\begin{figure}[hbt]
\centering{
\includegraphics[width=3.5cm,angle=-90]{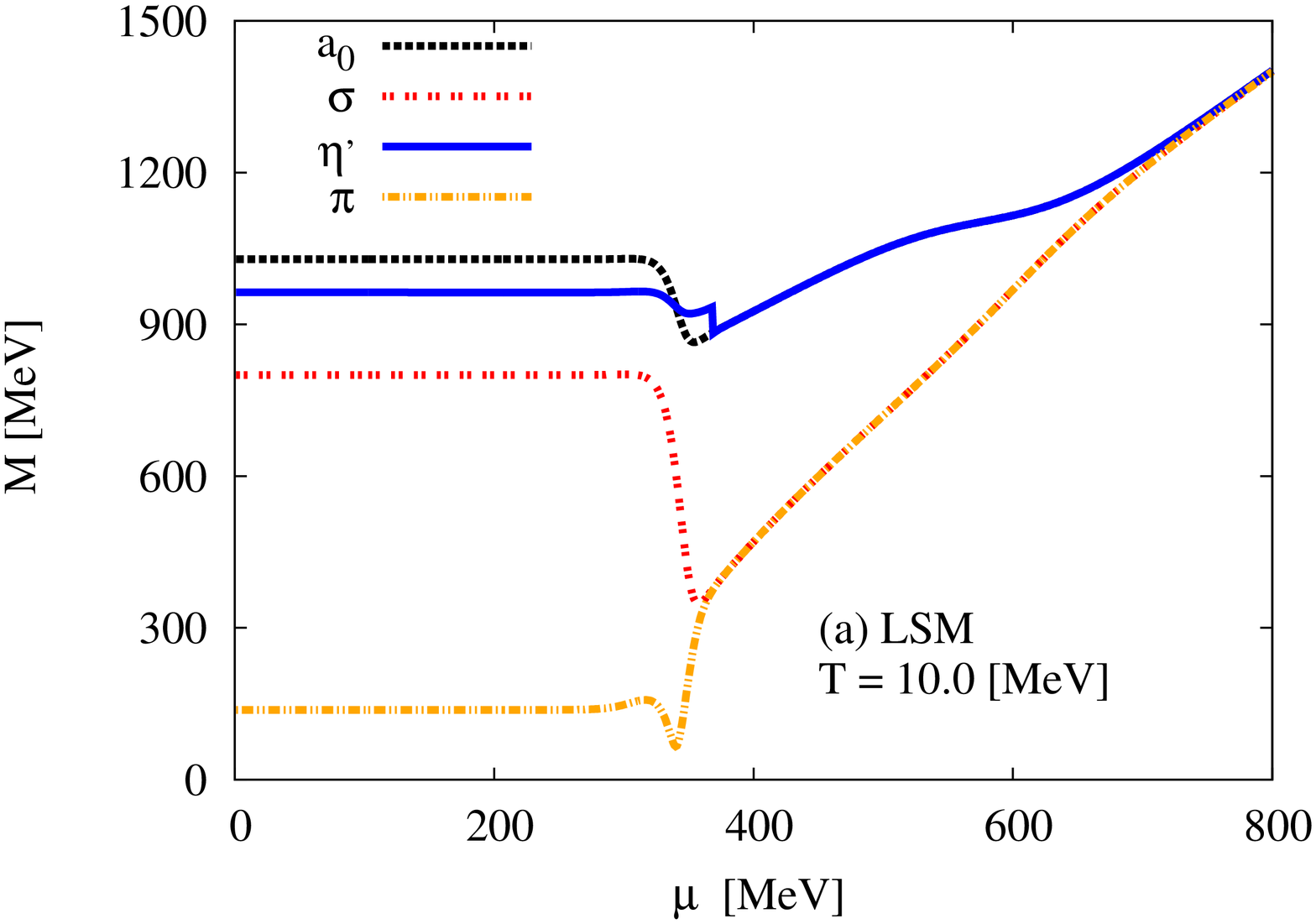}
\includegraphics[width=3.5cm,angle=-90]{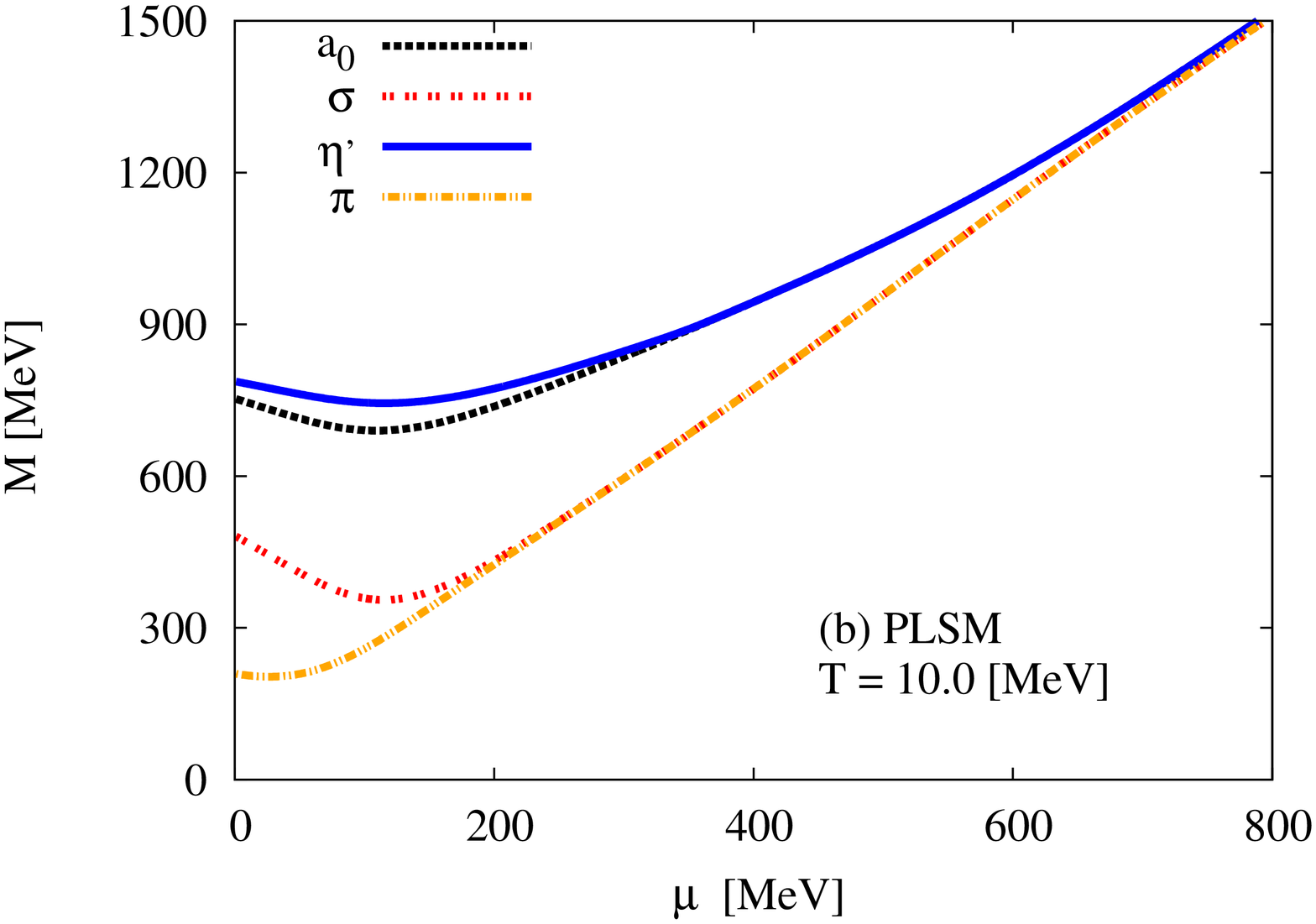}\\
\includegraphics[width=3.5cm,angle=-90]{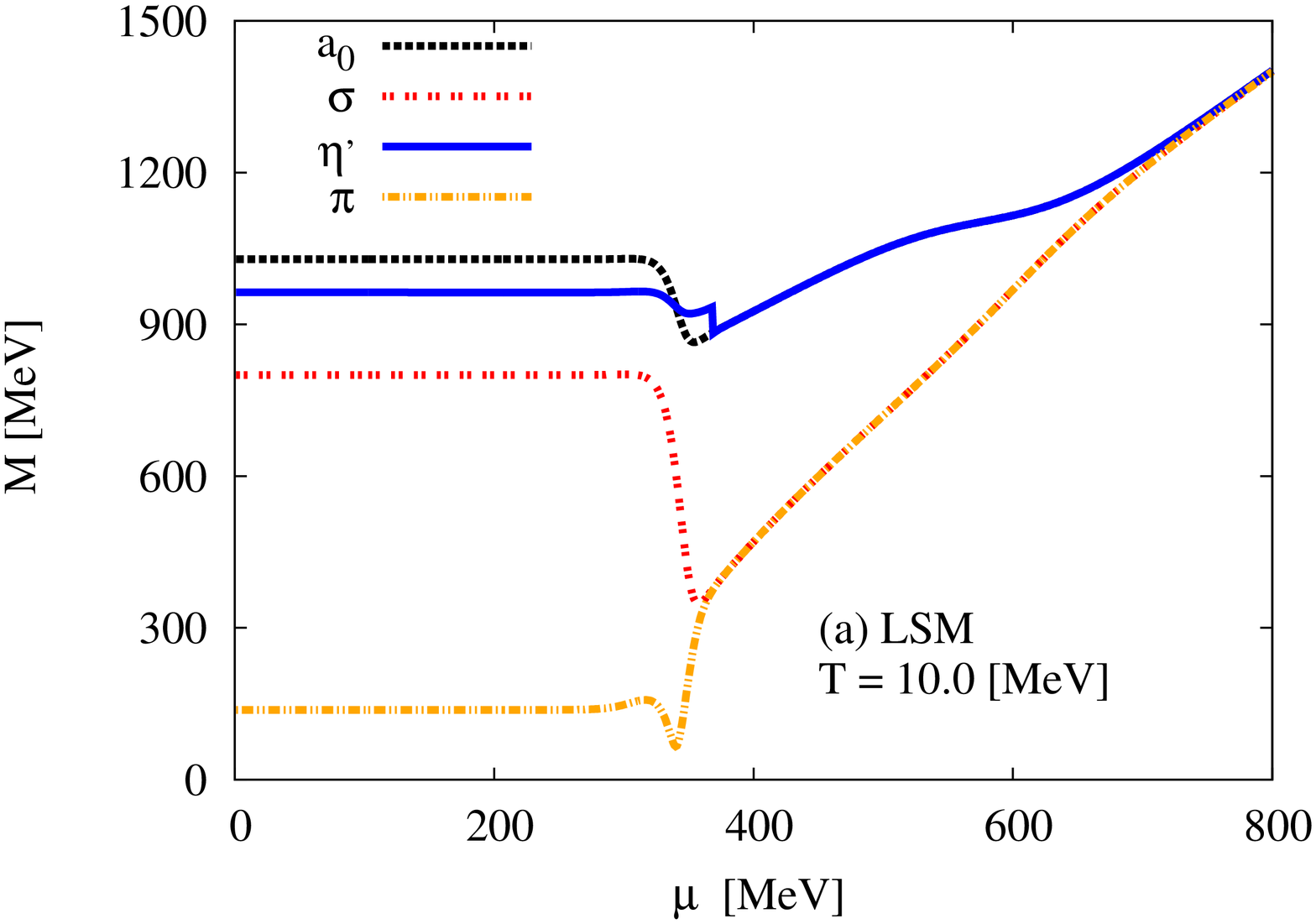}
\includegraphics[width=3.5cm,angle=-90]{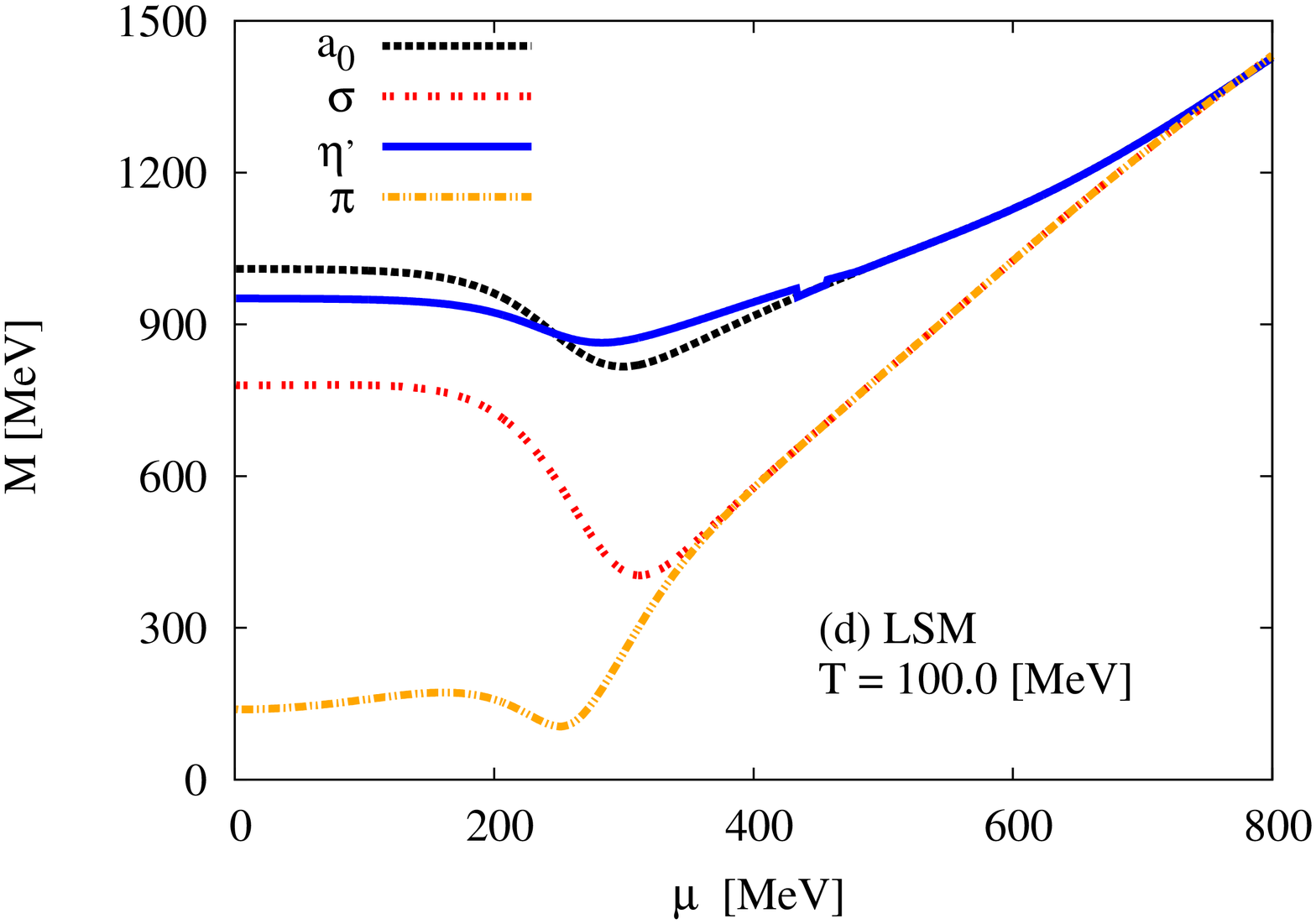}\\
\includegraphics[width=3.5cm,angle=-90]{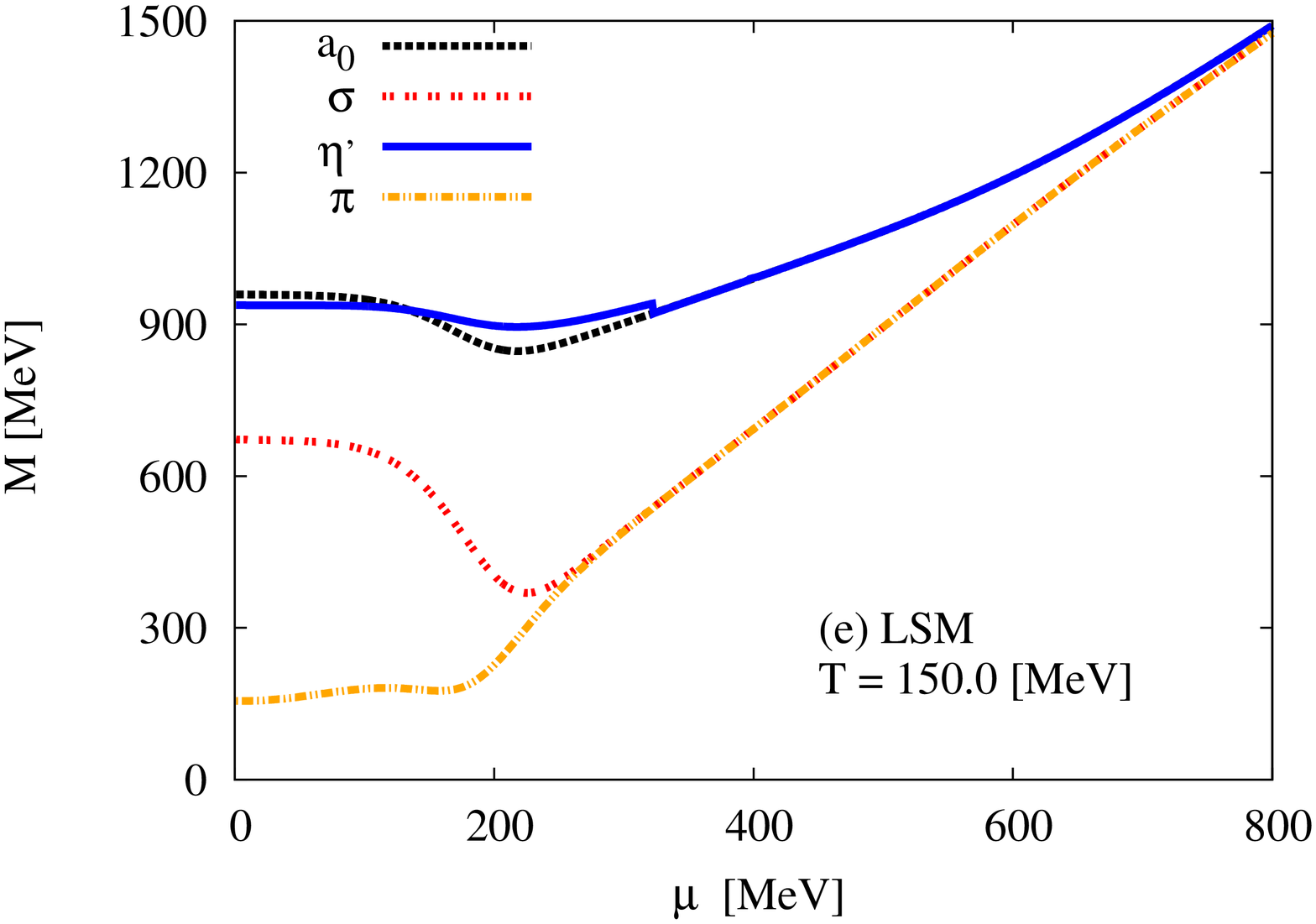}
\includegraphics[width=3.5cm,angle=-90]{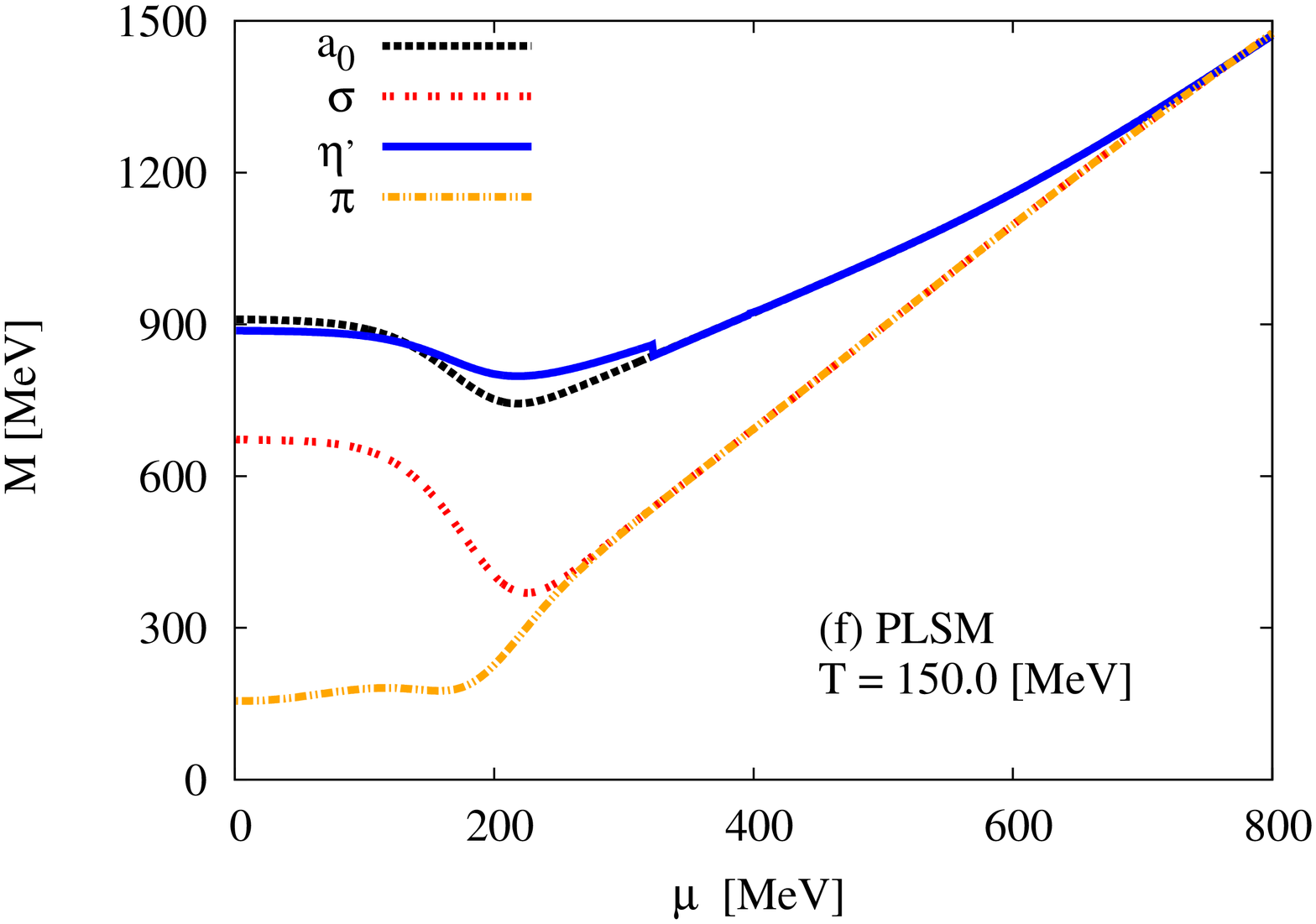}\\
\includegraphics[width=3.5cm,angle=-90]{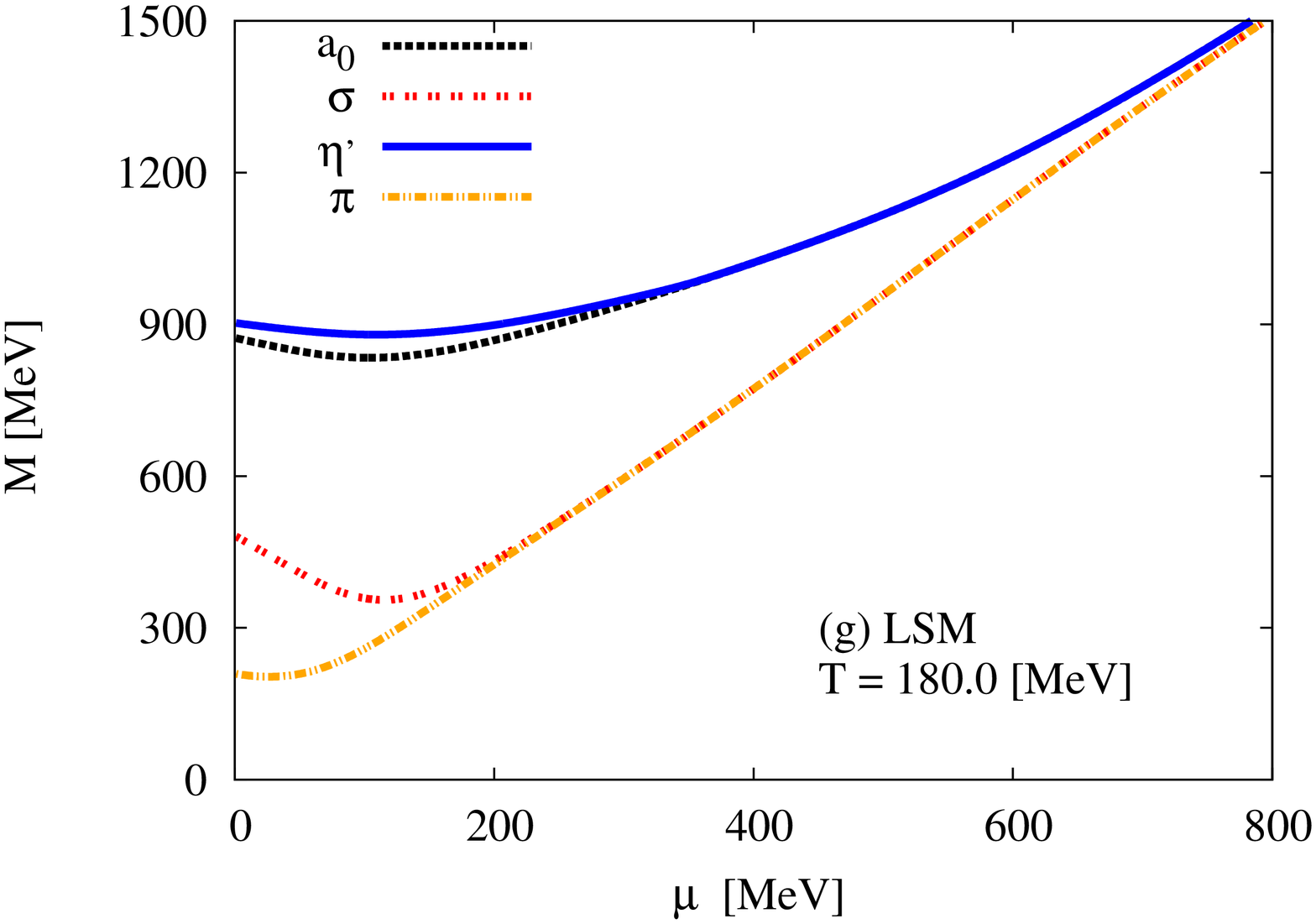}
\includegraphics[width=3.5cm,angle=-90]{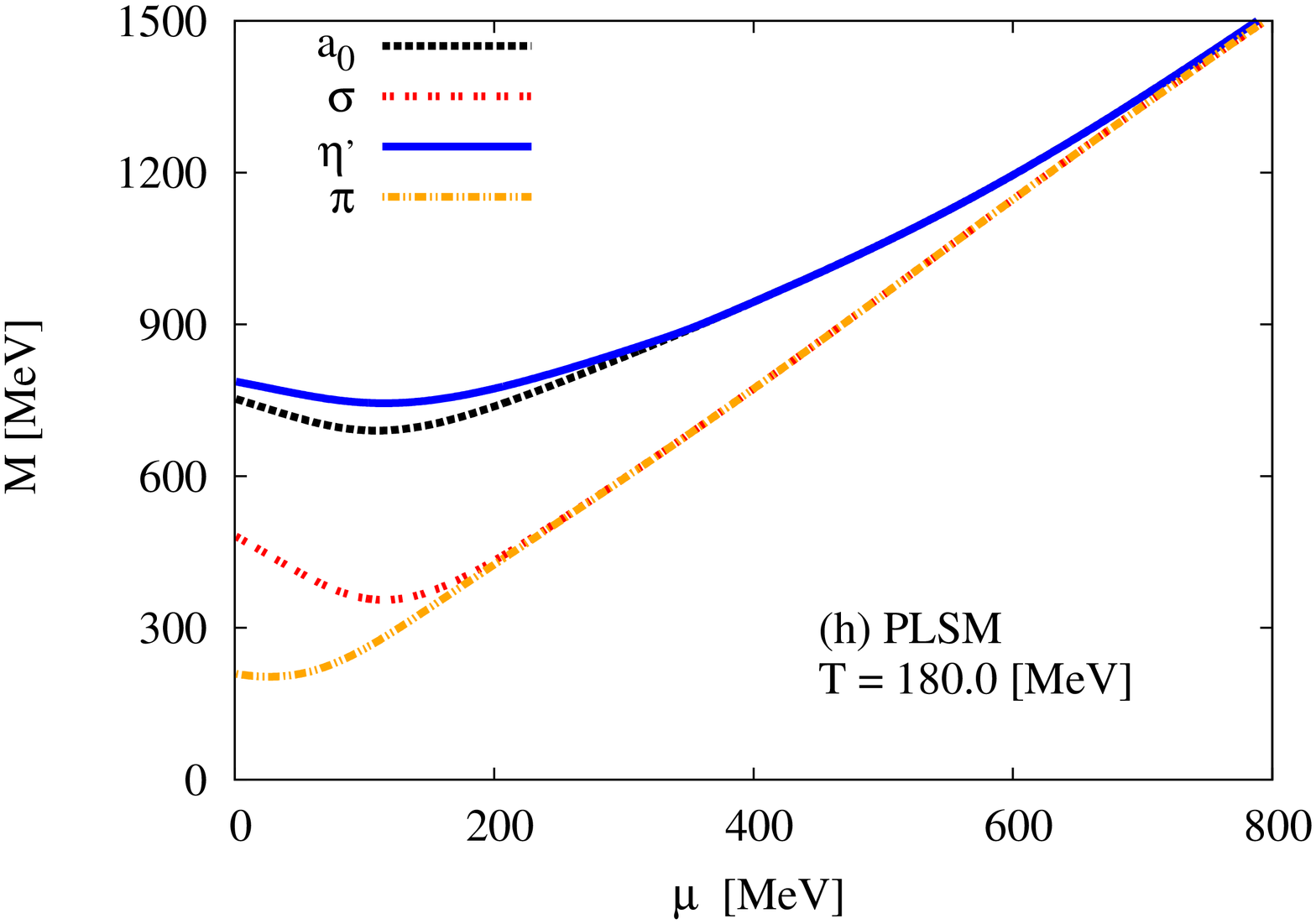}\\
\includegraphics[width=3.5cm,angle=-90]{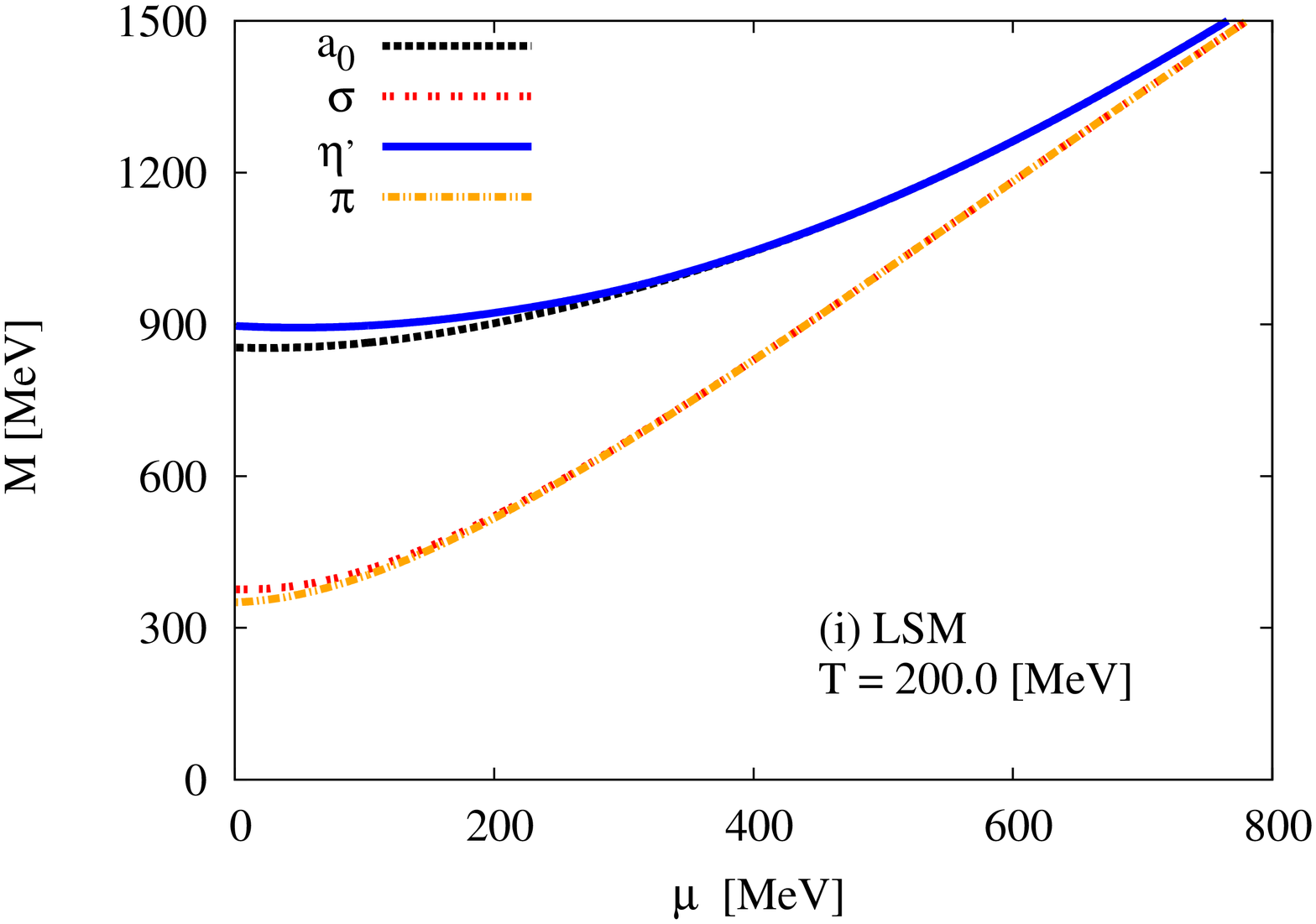}
\includegraphics[width=3.5cm,angle=-90]{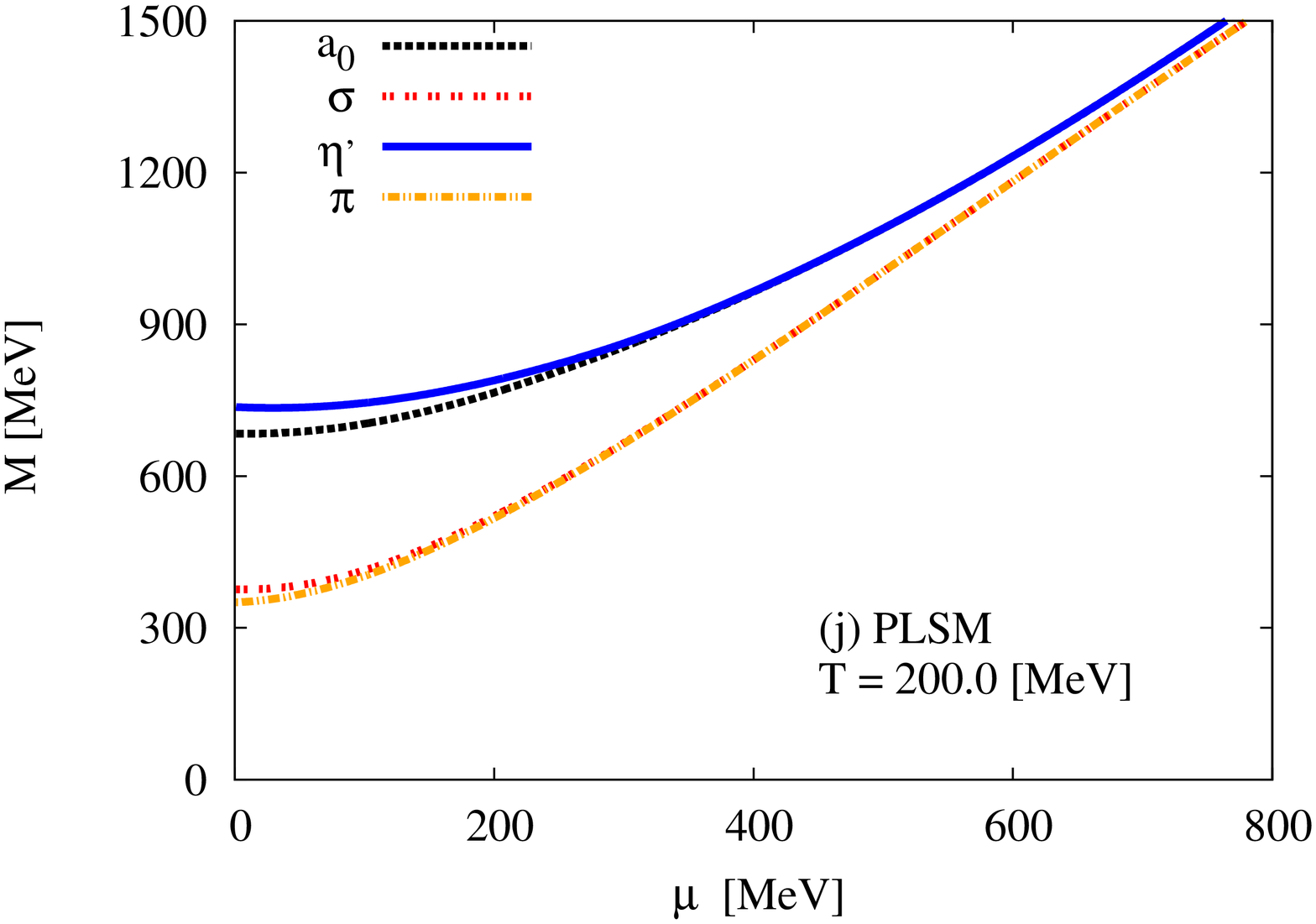}
\caption{(Color online) Left-hand panel (\lsm$\,$) and right-hand panel (P\lsm$\,$) presents scalars $a_0$ (dashed curve) and $\sigma$ (dotted curve) and pseudoscalars $\eta^{,}$ (solid curve)  and $\pi$  (dashed-dotted curve) in hadronic dense medium at fixed temperatures $T=10$, $100$, $150$, $180$ and $200\,$MeV.
\label{SPMU1}
}}
\end{figure}

\begin{figure}[hbt]
\centering{
\includegraphics[width=3.5cm,angle=-90]{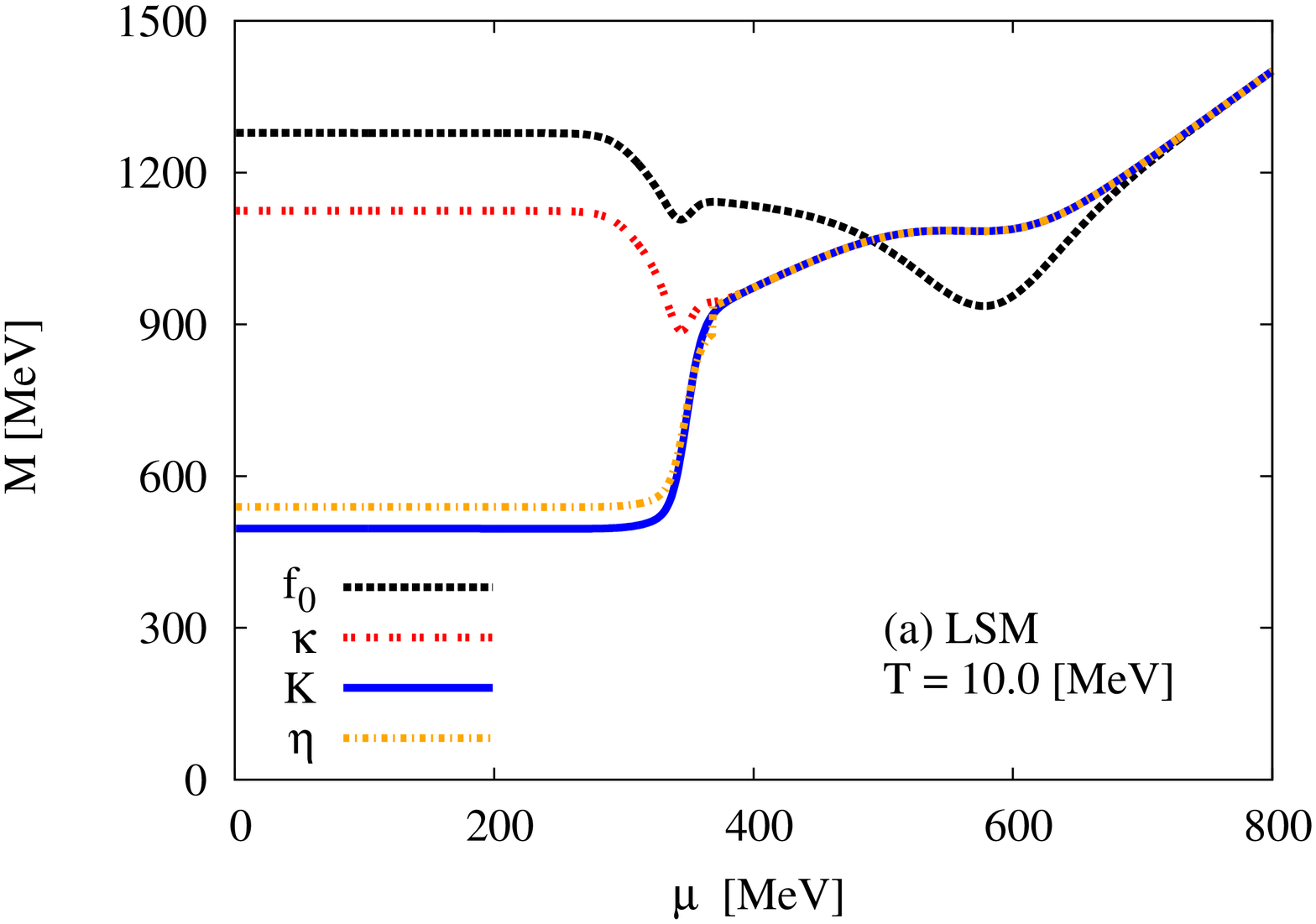}
\includegraphics[width=3.5cm,angle=-90]{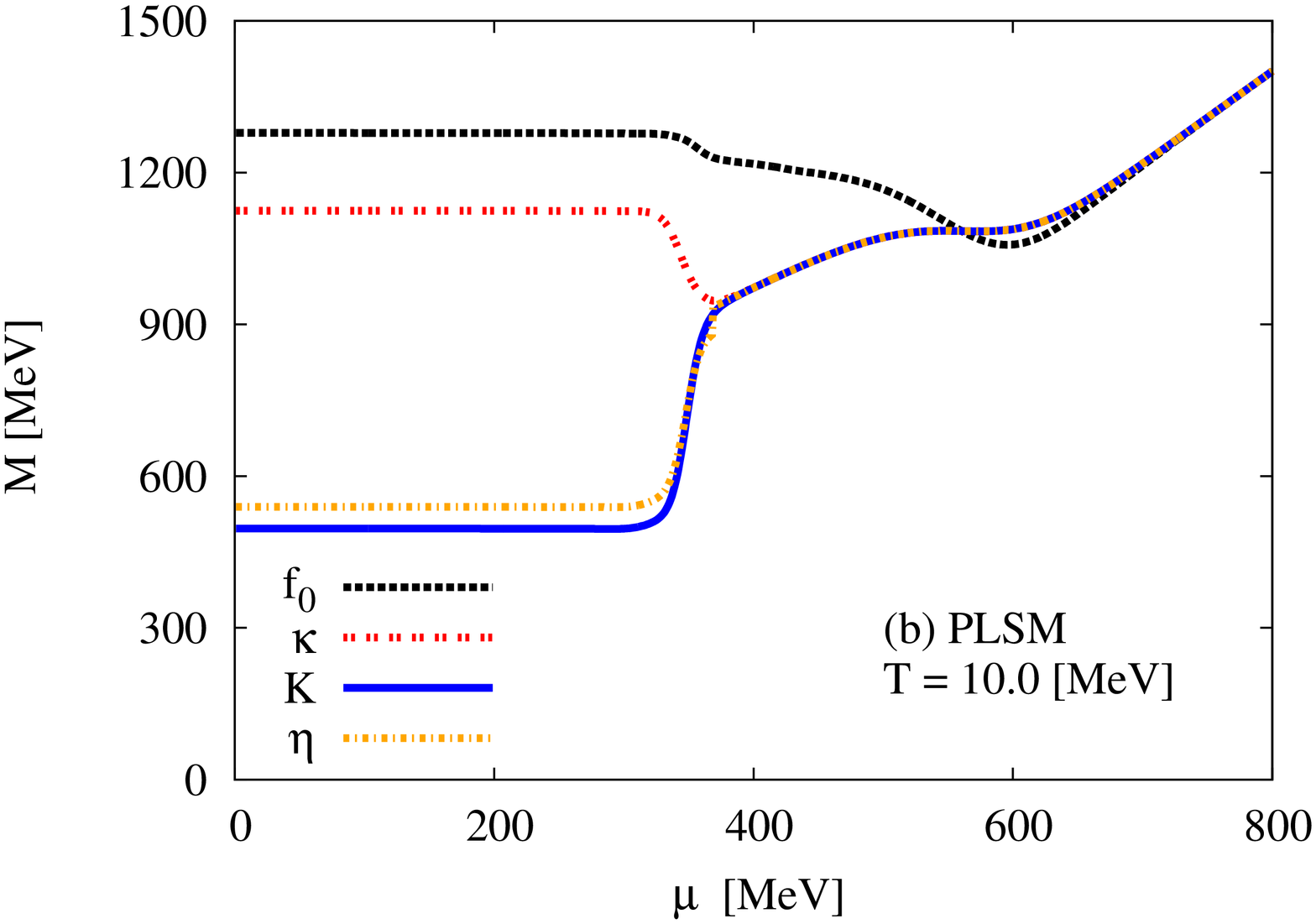} \\
\includegraphics[width=3.5cm,angle=-90]{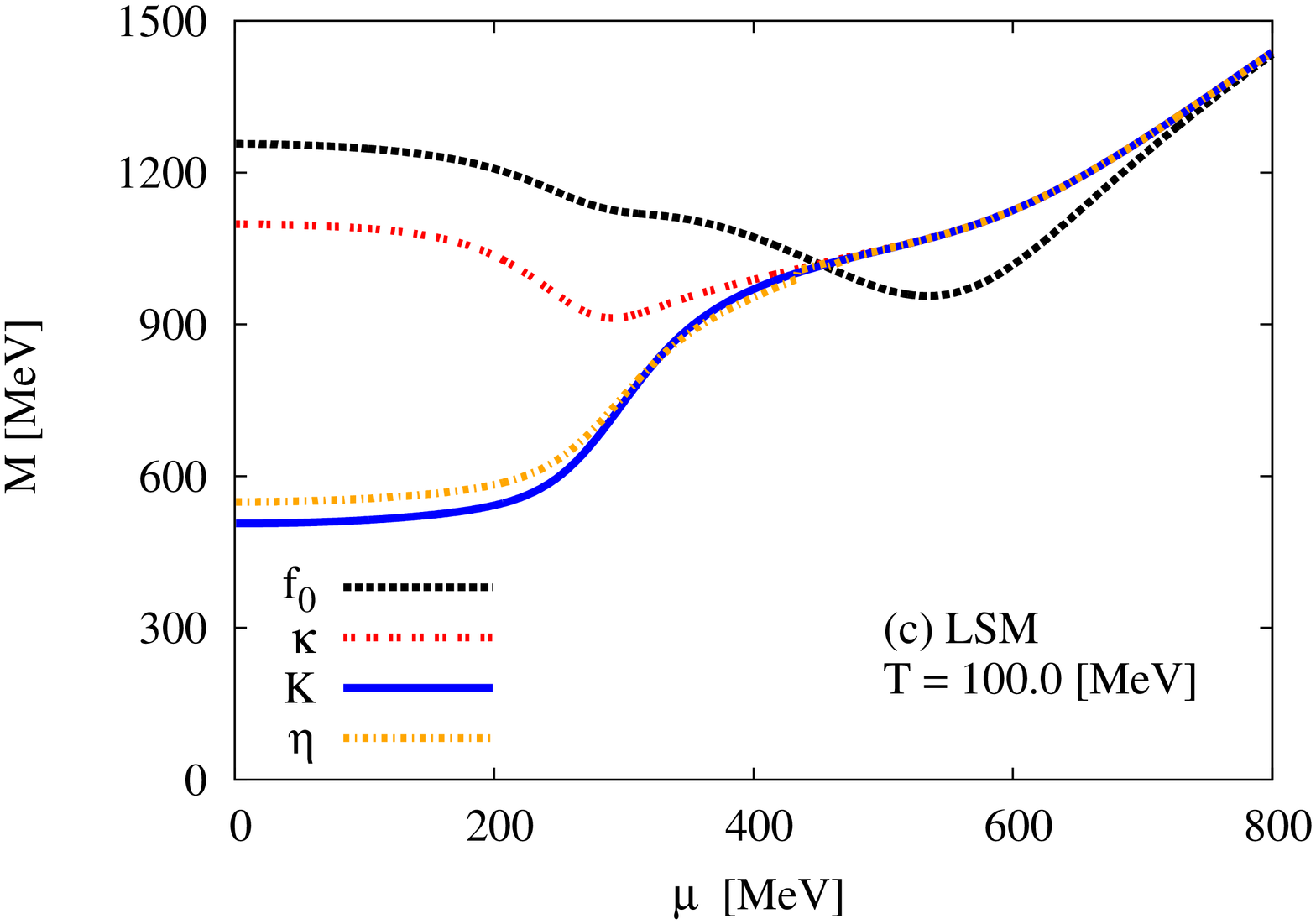}
\includegraphics[width=3.5cm,angle=-90]{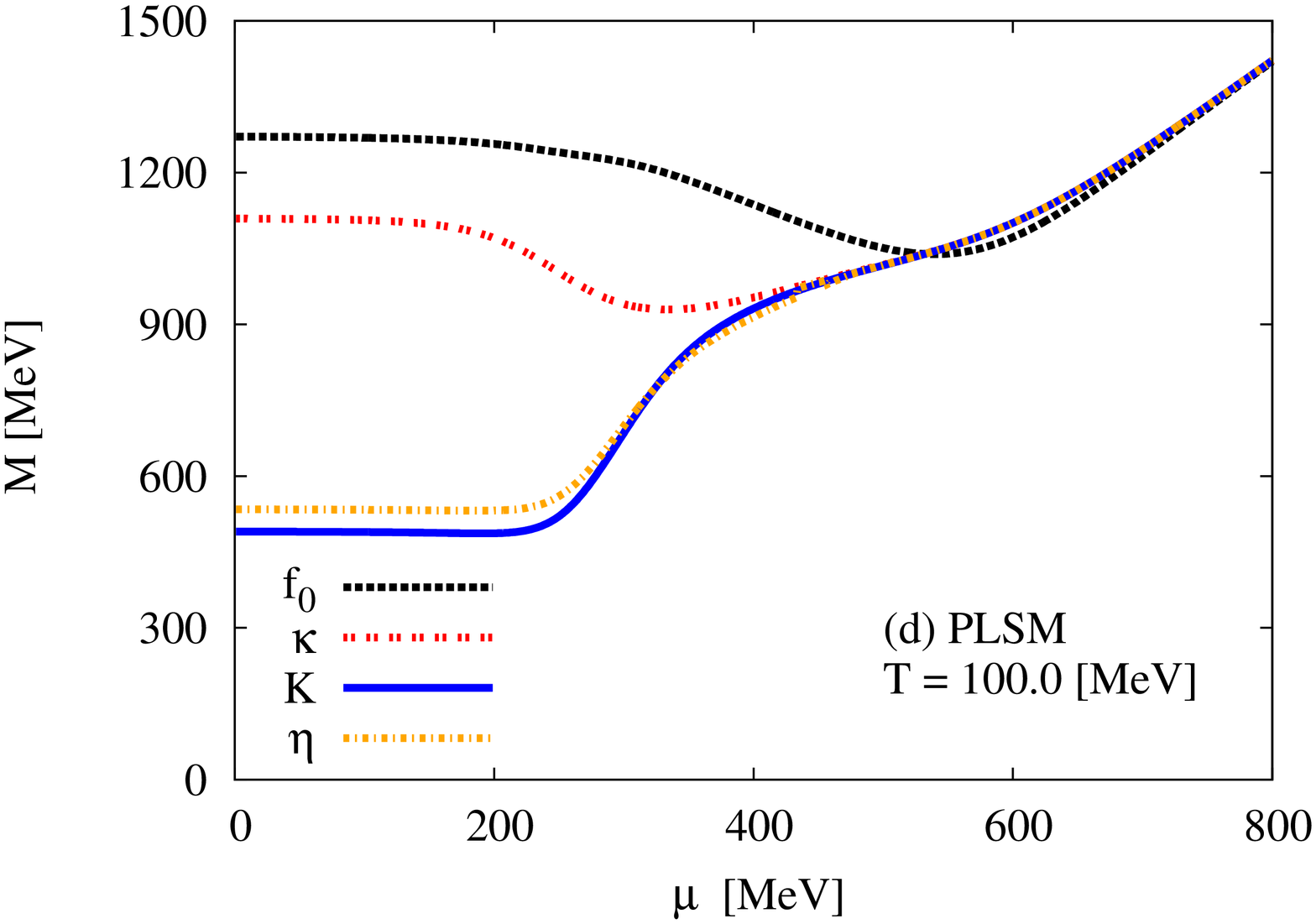}\\
\includegraphics[width=3.5cm,angle=-90]{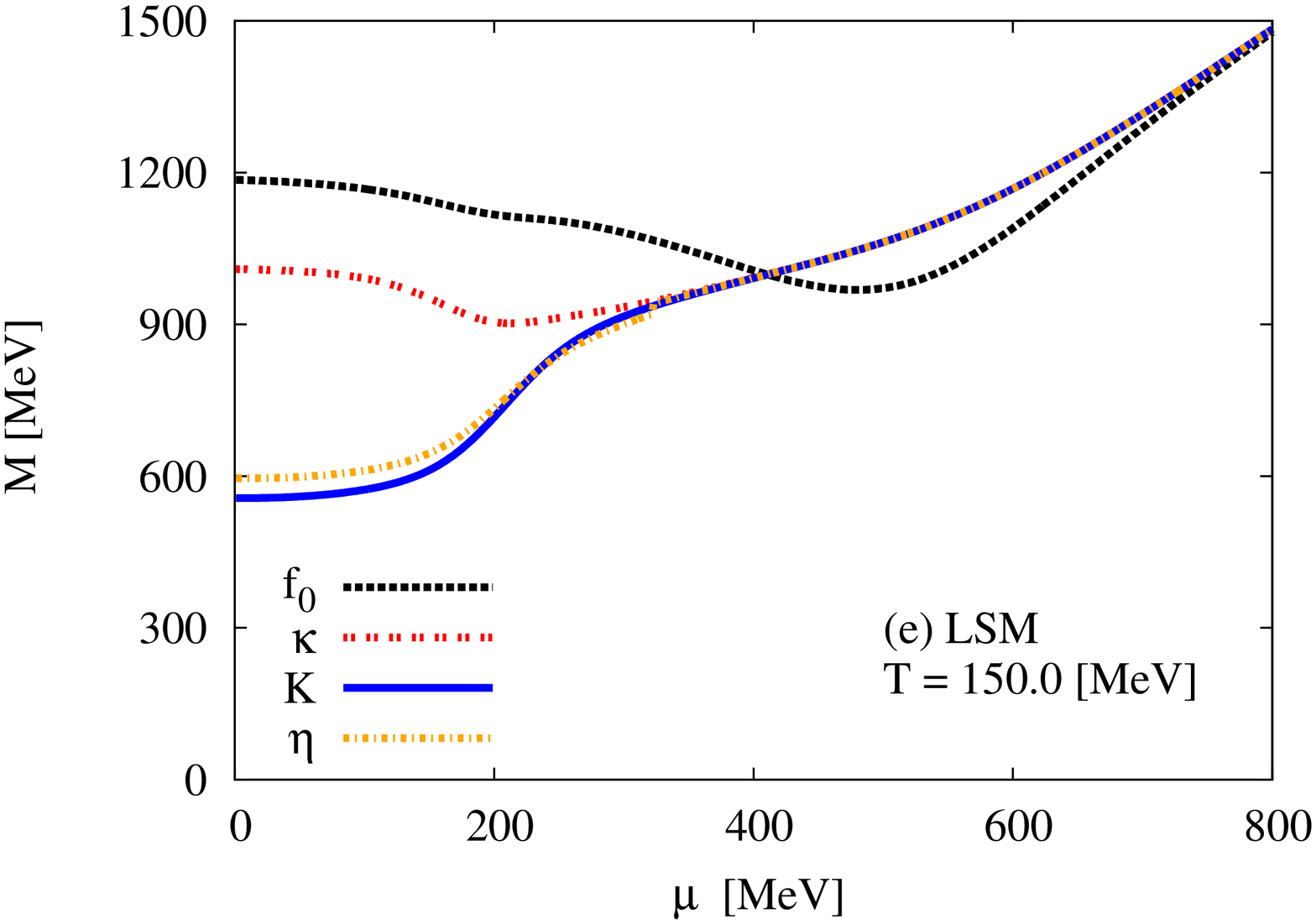}
\includegraphics[width=3.5cm,angle=-90]{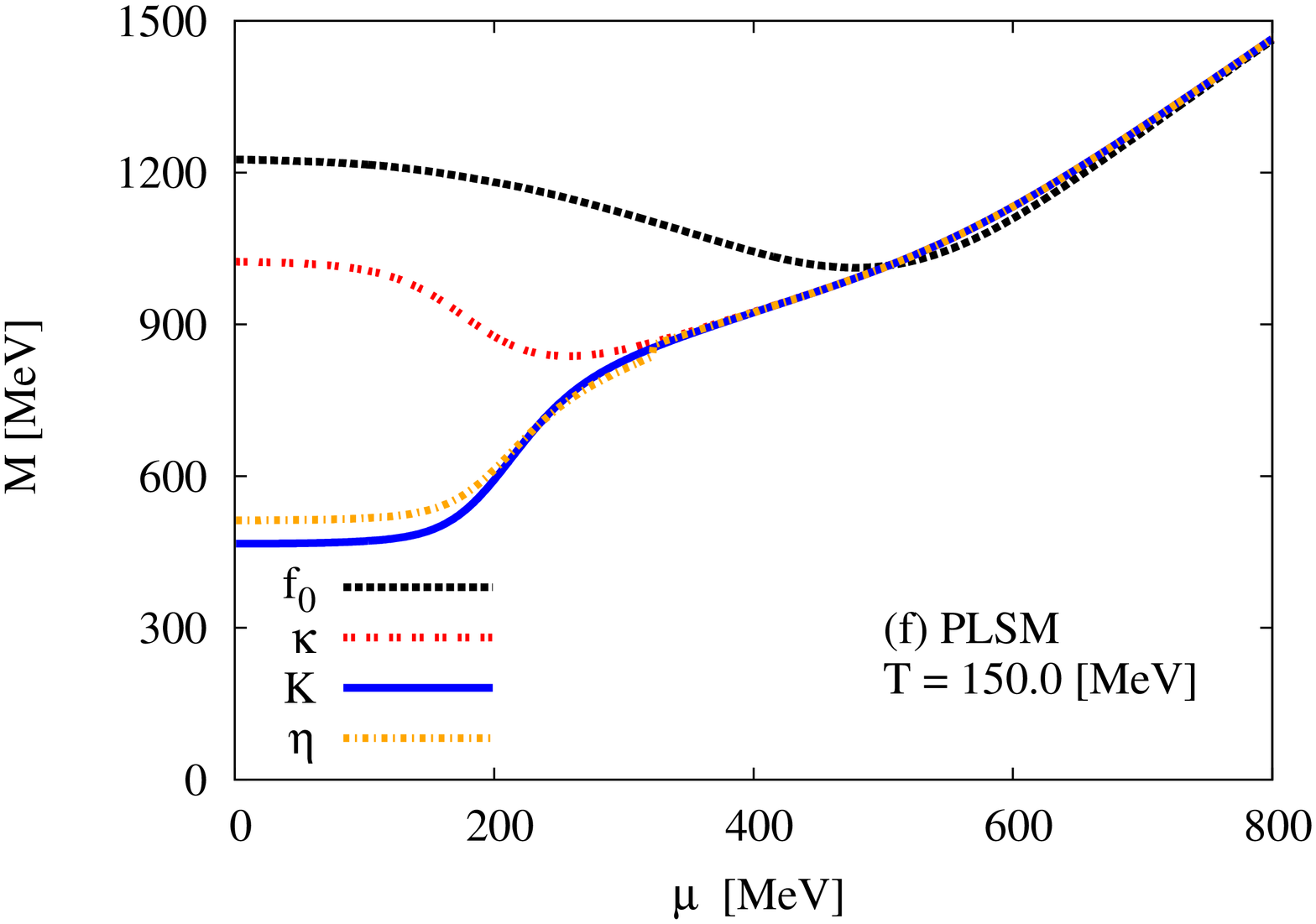}\\
\includegraphics[width=3.5cm,angle=-90]{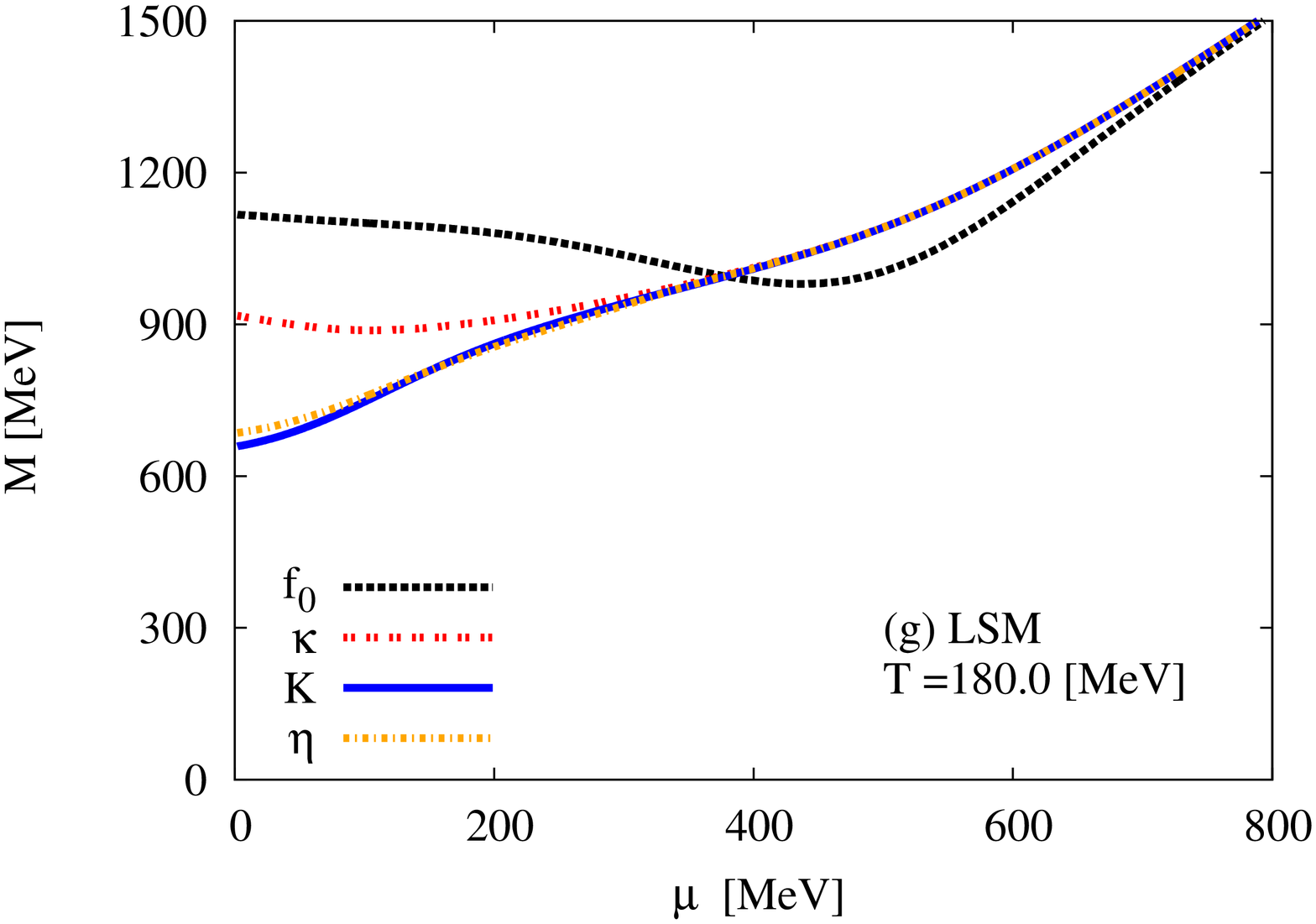}
\includegraphics[width=3.5cm,angle=-90]{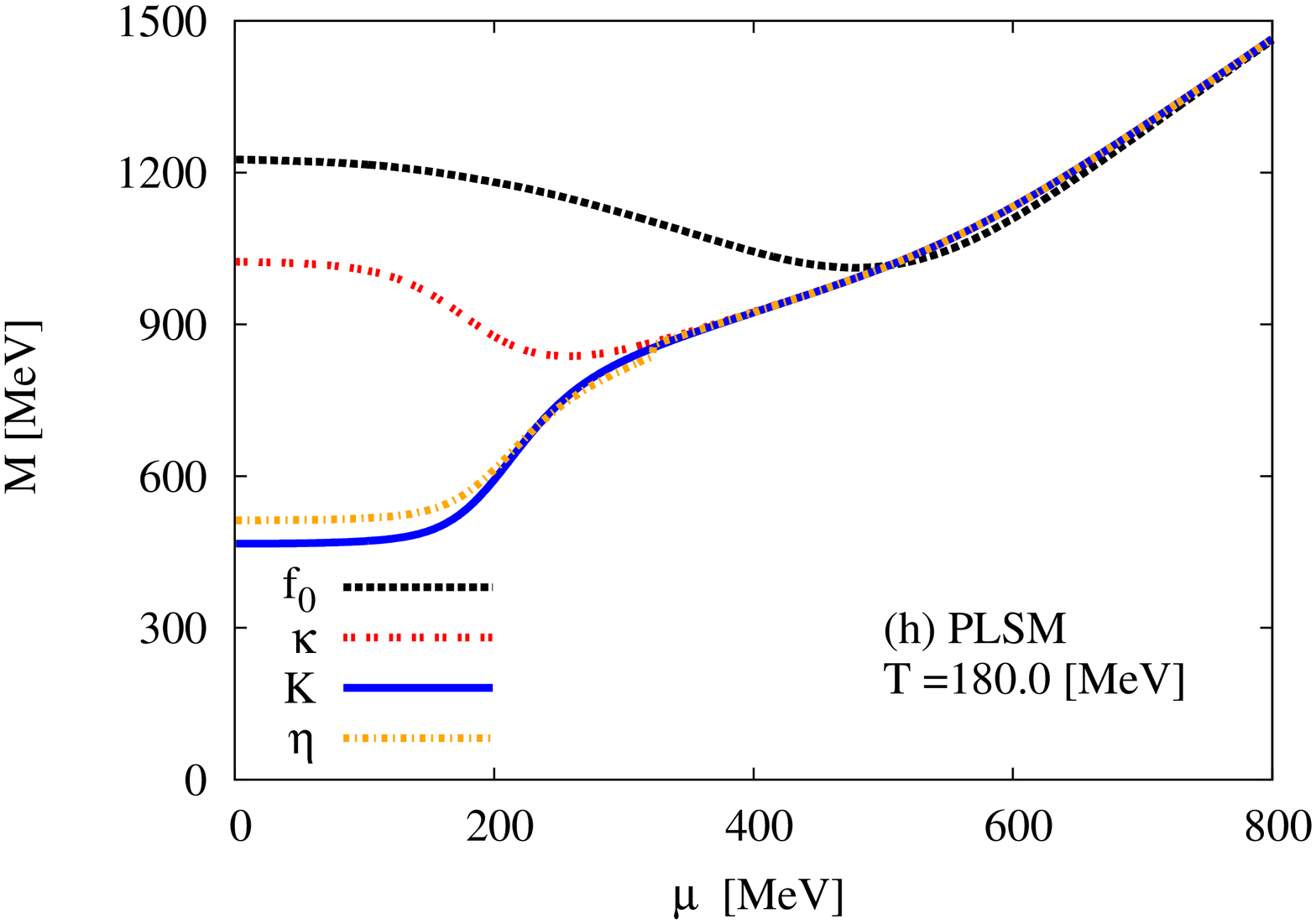} \\
\includegraphics[width=3.5cm,angle=-90]{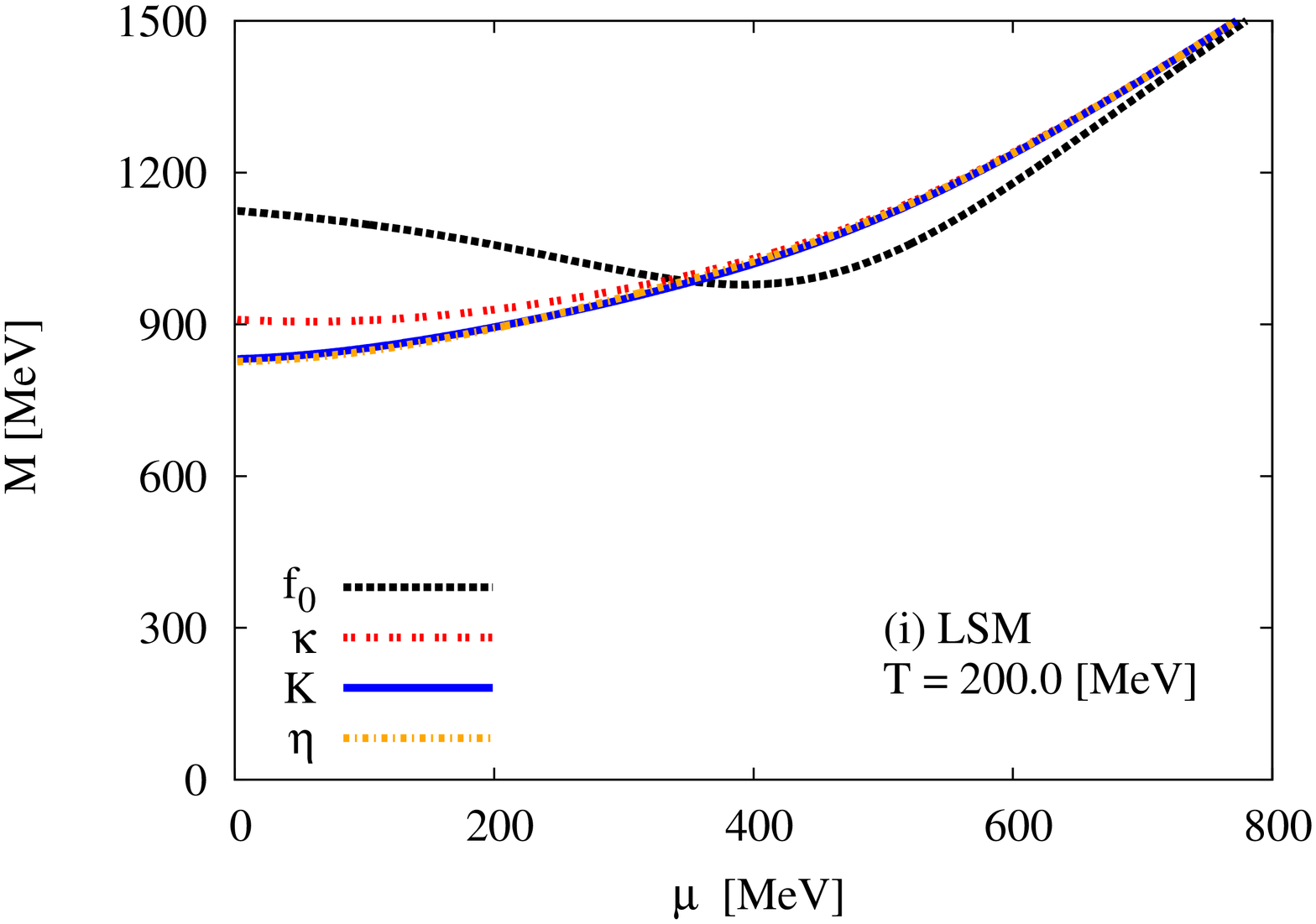}
\includegraphics[width=3.5cm,angle=-90]{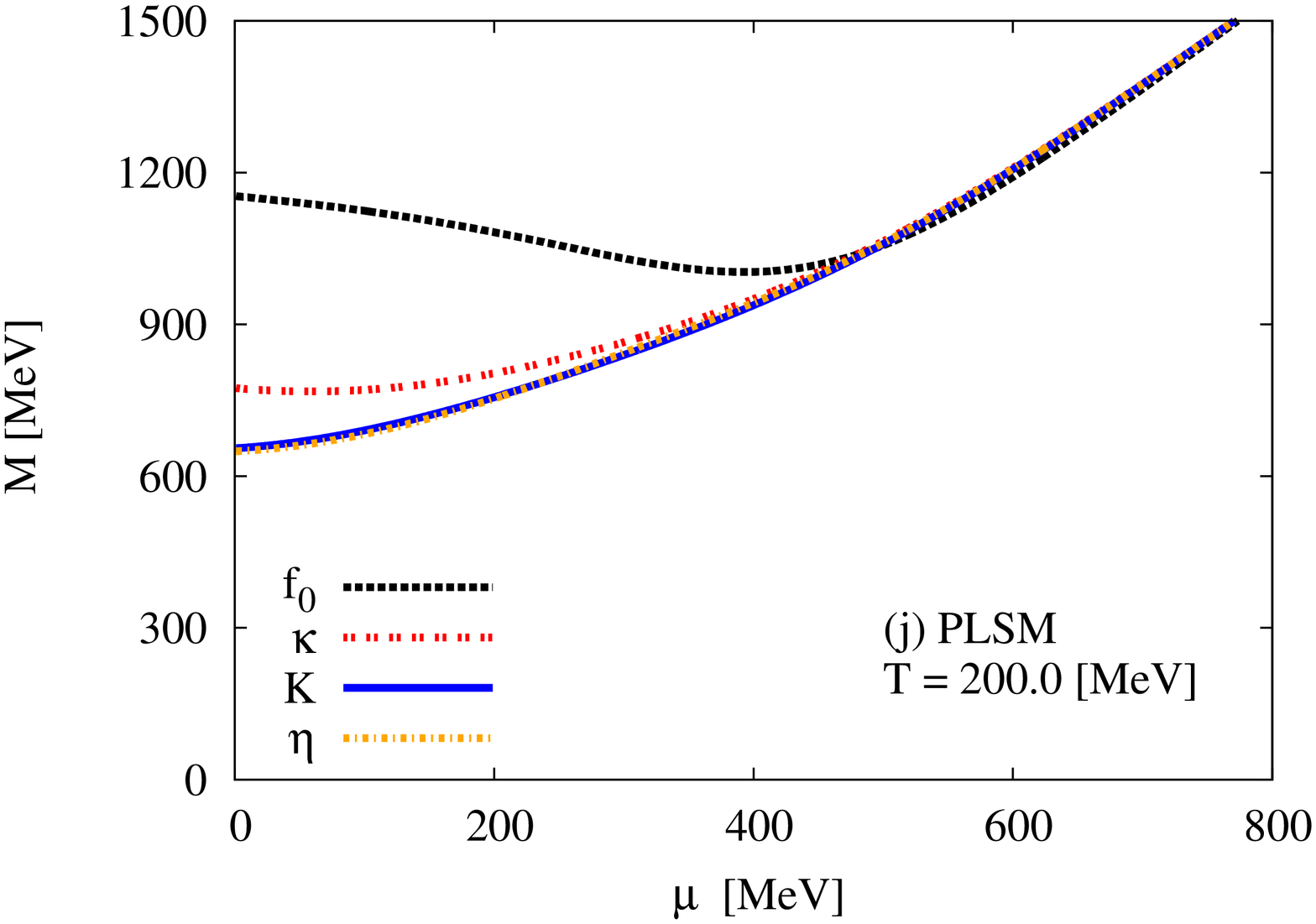}
\caption{(Color online) Left-hand panel (\lsm$\,$) and right-hand panel (P\lsm$\,$) show scalars $f_0$ (horizontal dashed curve) and $\kappa$ (vertical dashed curve)  and pseudoscalars $\eta$ (dotted curve)  and $K$  (solid curve)  in hadronic dense medium at fixed temperatures $T=10$, $100$, $150$, $180$ and $200\,$MeV.
\label{SPMU2}
}}
\end{figure}

\begin{figure}[htb]
\centering{
\includegraphics[width=3.5cm,angle=-90]{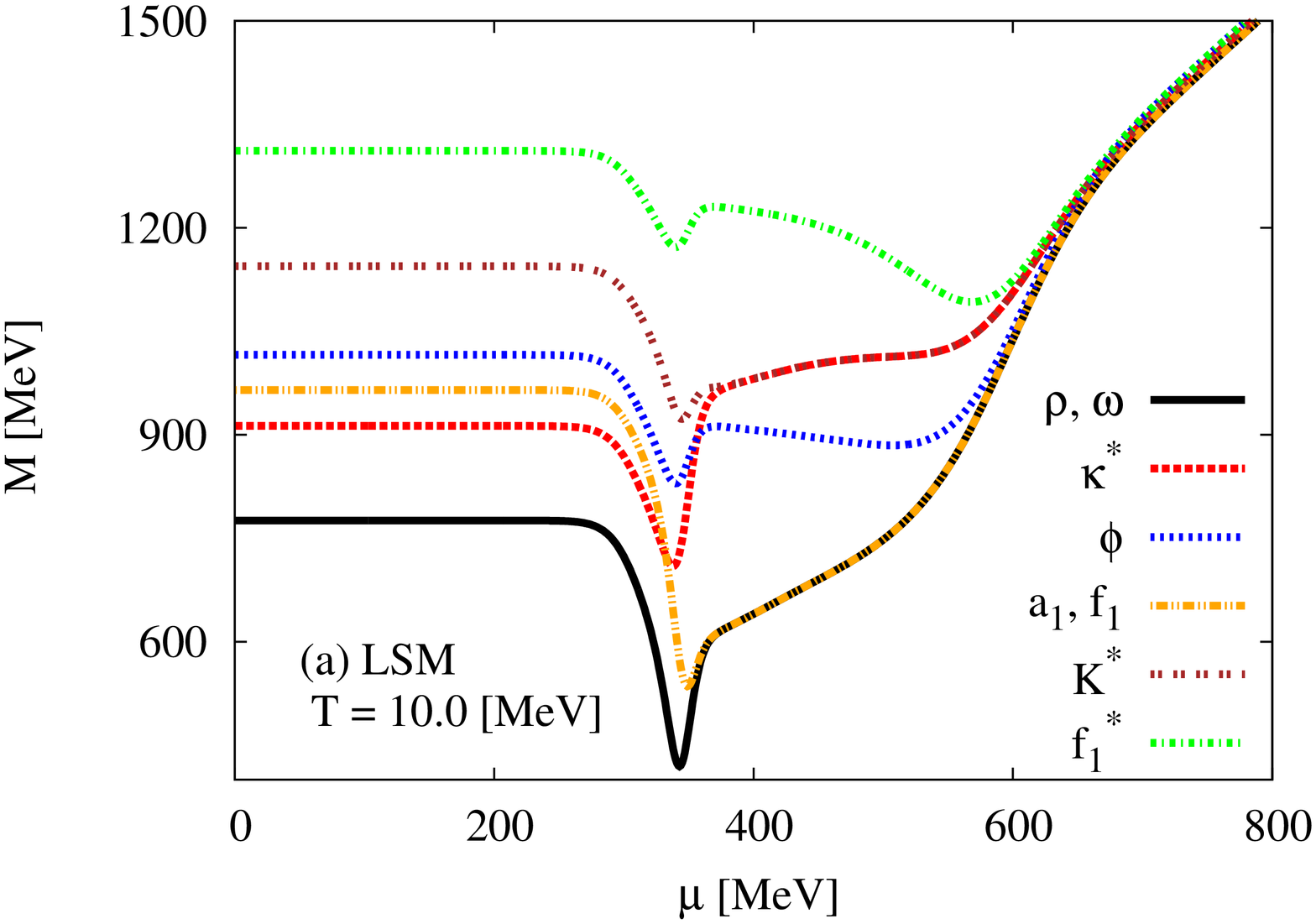}
\includegraphics[width=3.5cm,angle=-90]{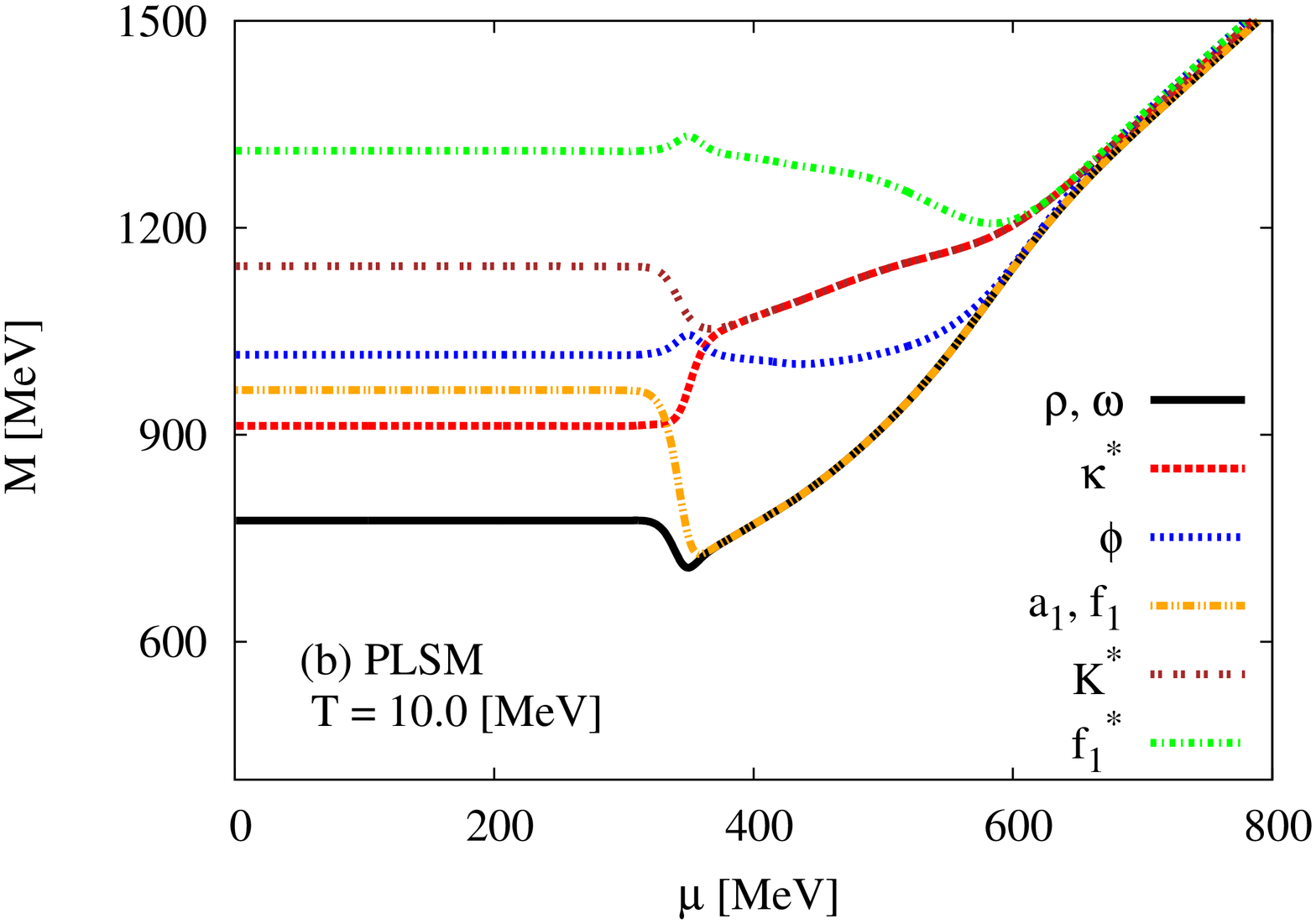}\\
\includegraphics[width=3.5cm,angle=-90]{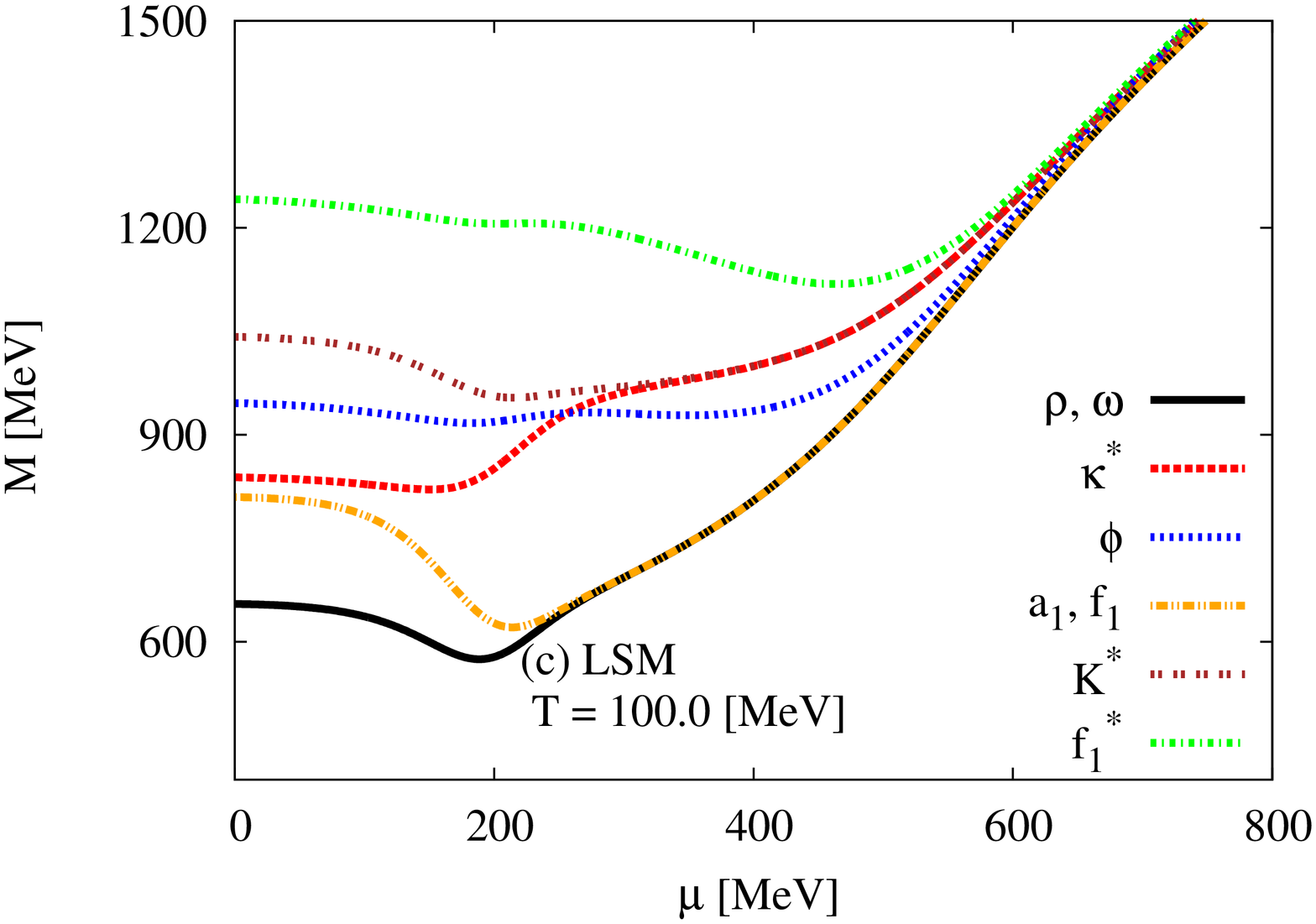}
\includegraphics[width=3.5cm,angle=-90]{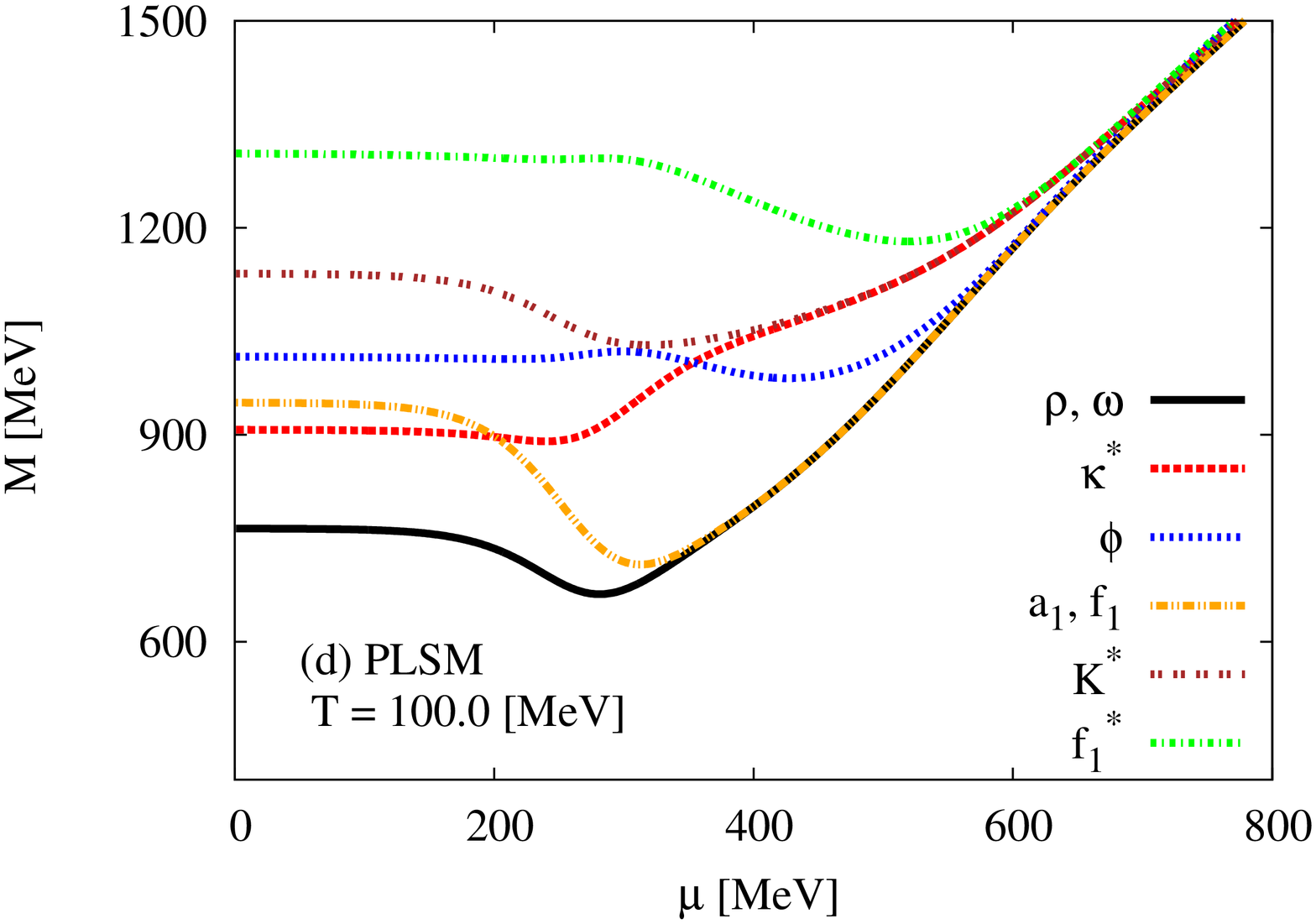}\\
\includegraphics[width=3.5cm,angle=-90]{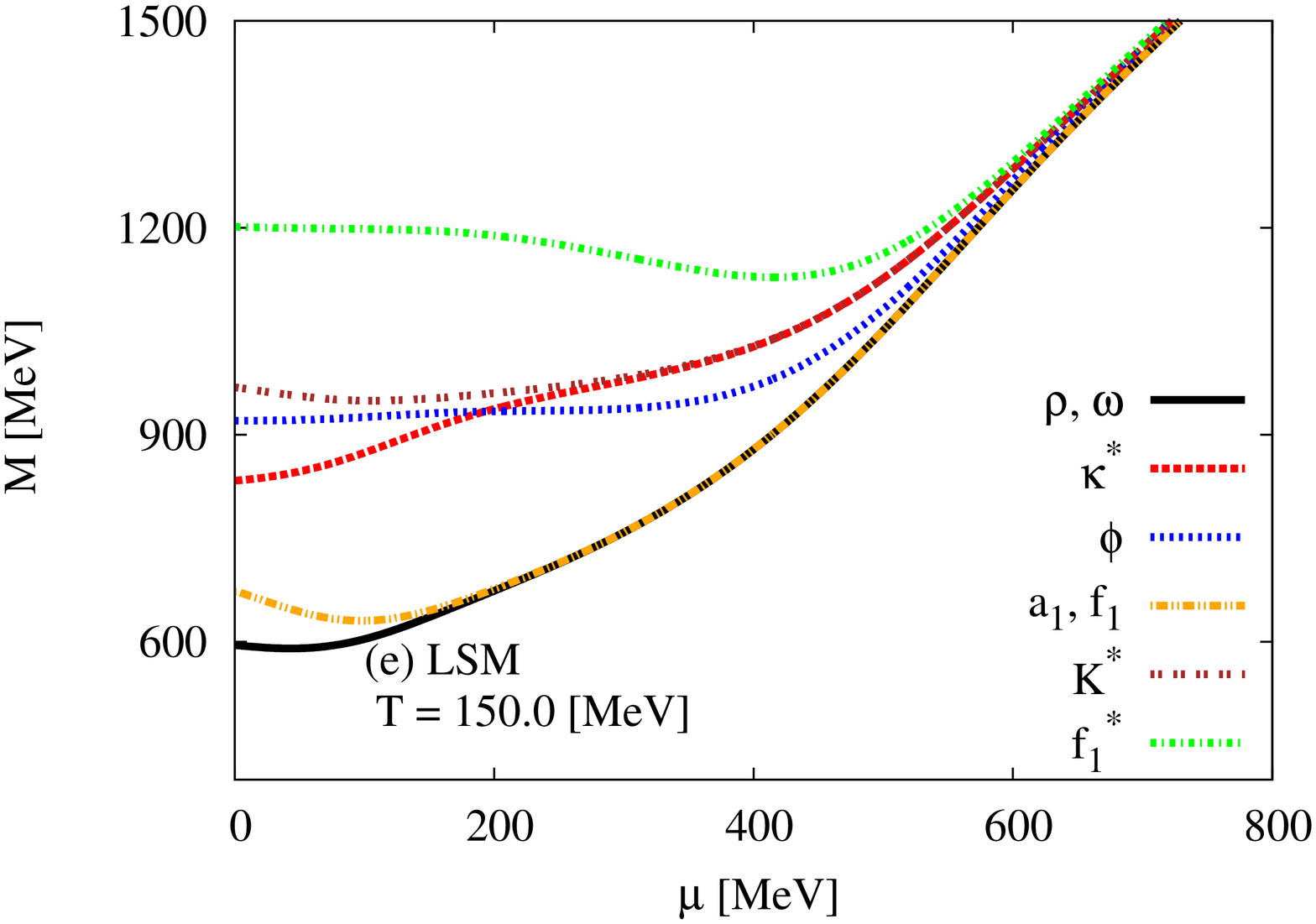}
\includegraphics[width=3.5cm,angle=-90]{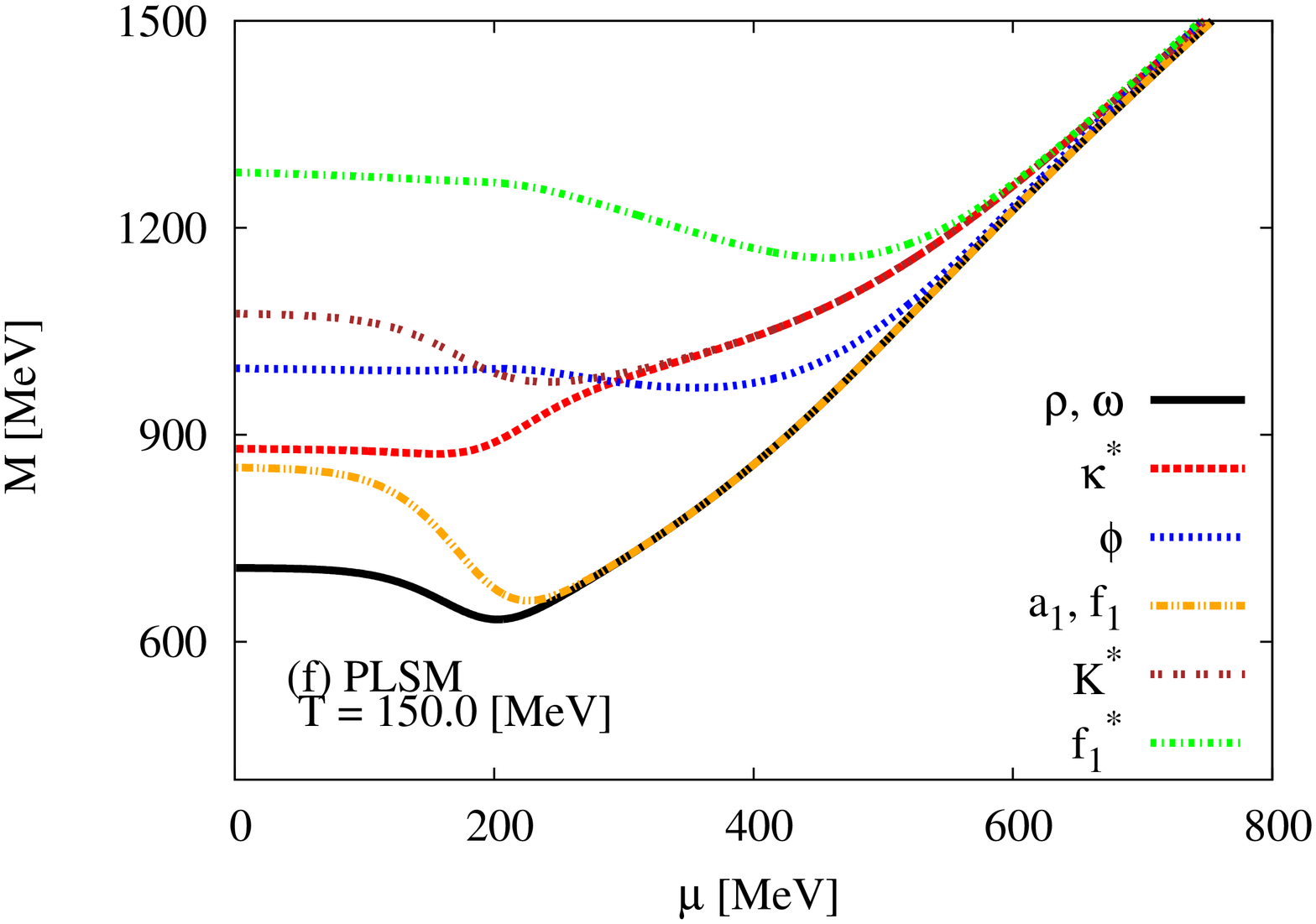}\\
\includegraphics[width=3.5cm,angle=-90]{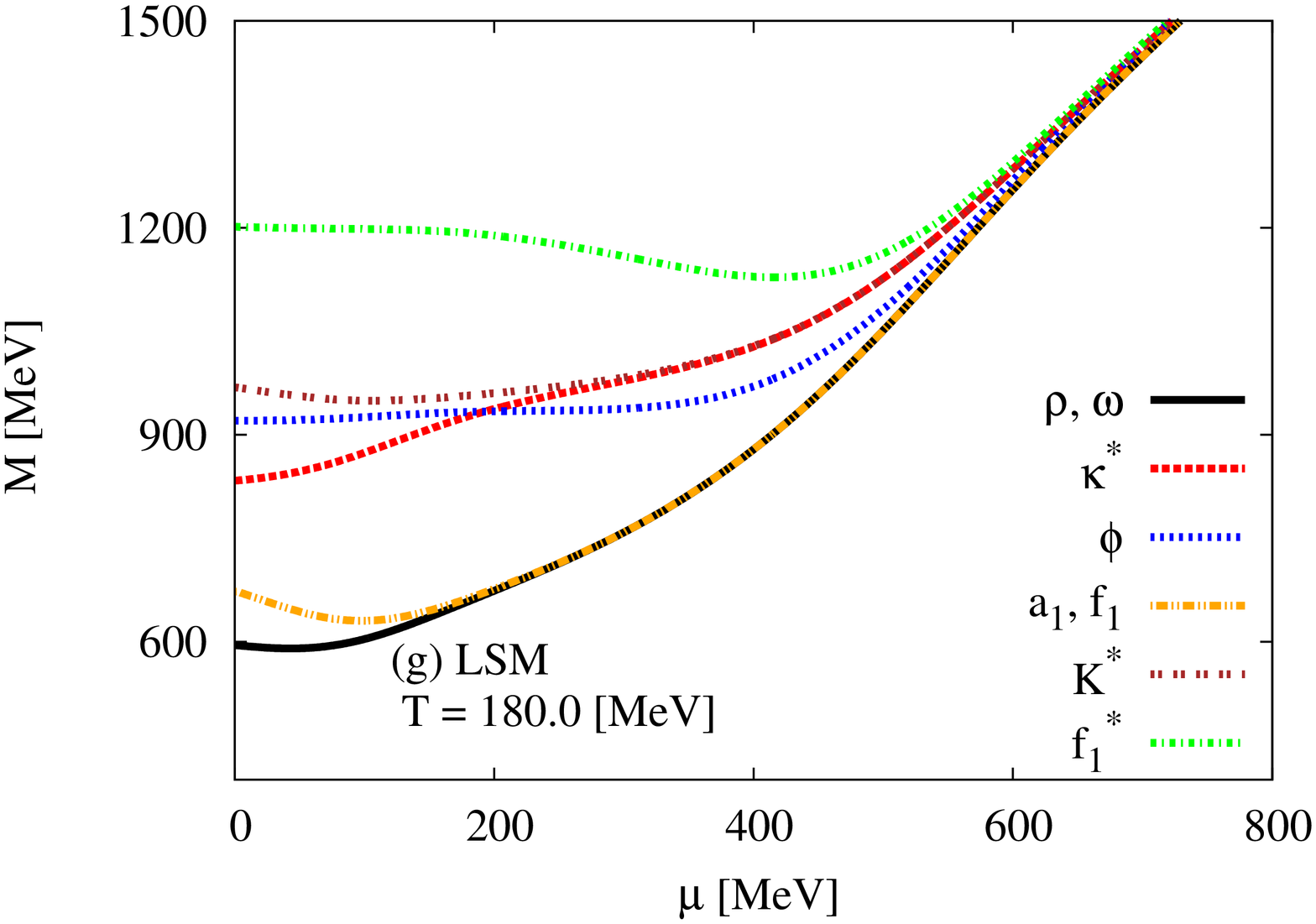}
\includegraphics[width=3.5cm,angle=-90]{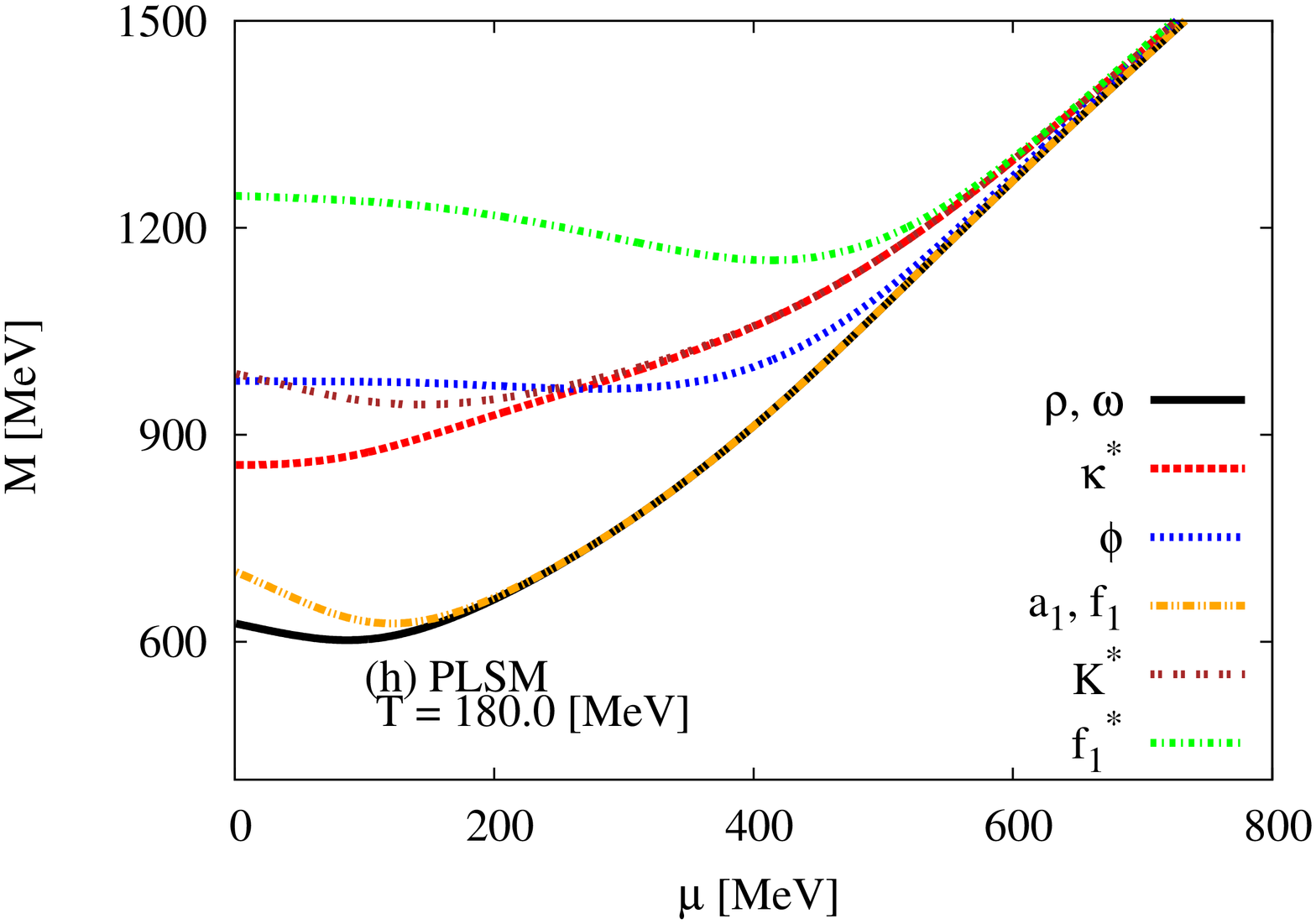}\\
\includegraphics[width=3.5cm,angle=-90]{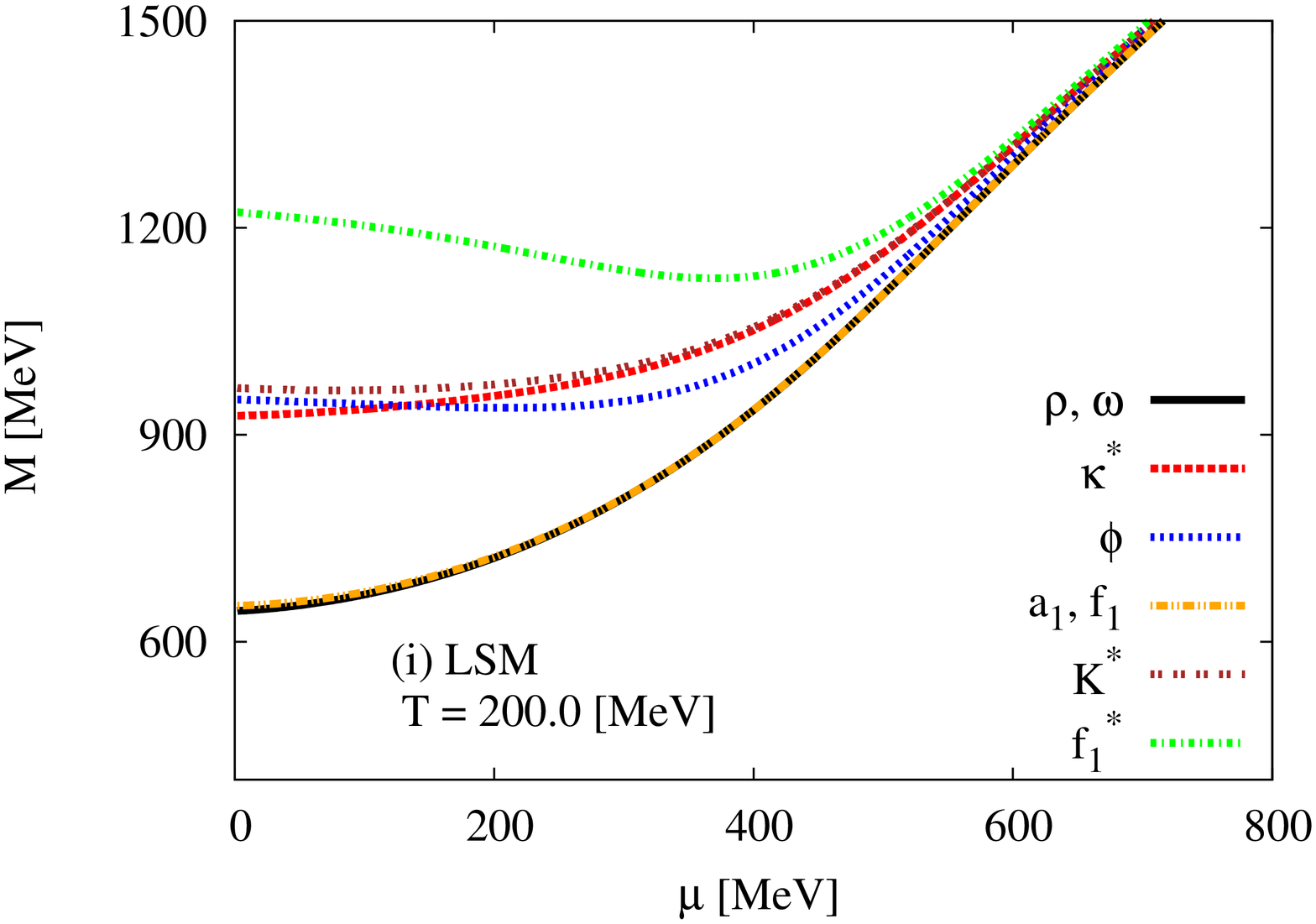}
\includegraphics[width=3.5cm,angle=-90]{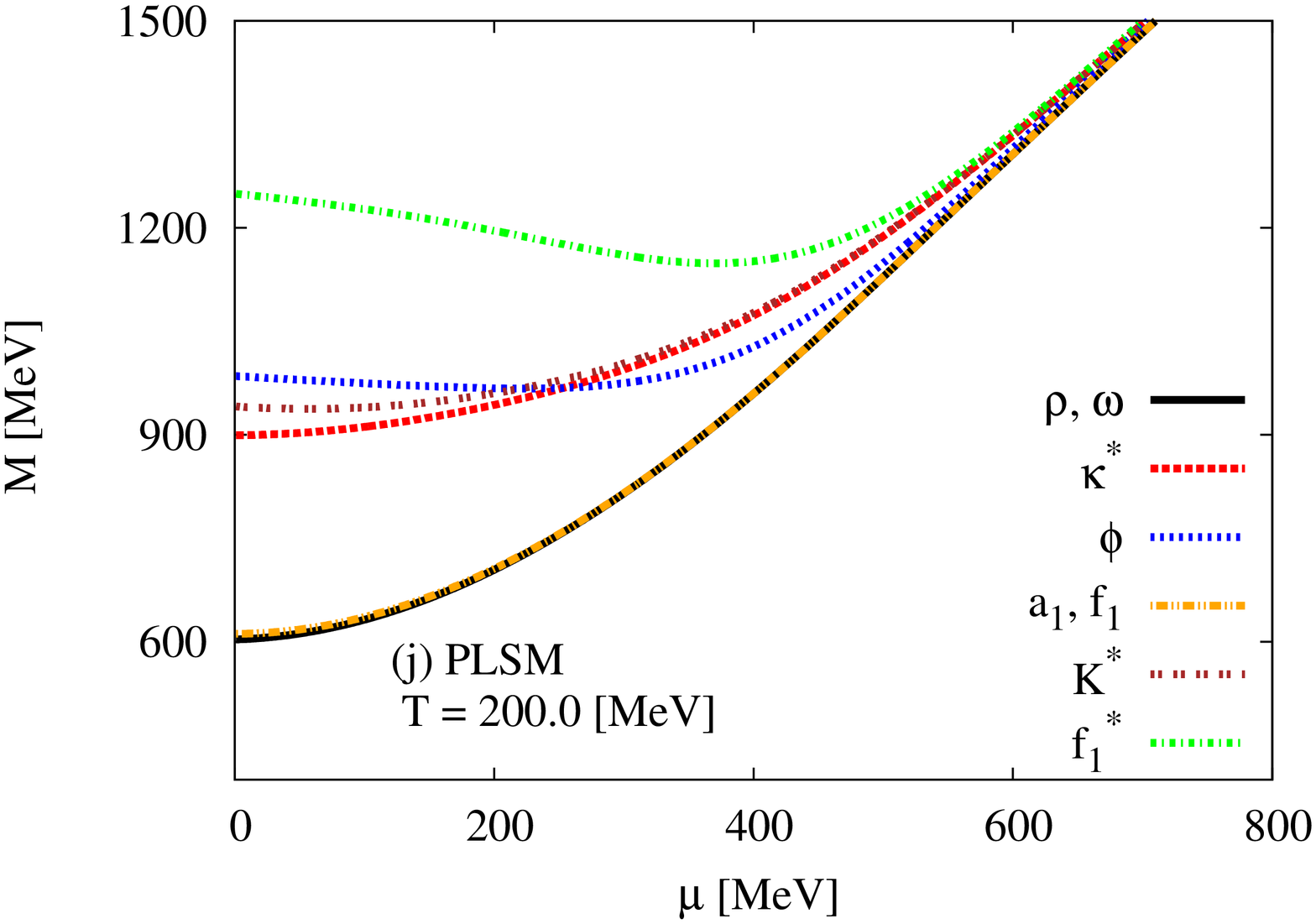}
\caption{(Color online) Left-hand panel (\lsm$\,$) and right-hand panel (P\lsm$\,$) shows vector mesons $\rho$, $\omega$ (solid  curve), $\kappa^{*}$ (long-dotted curve) and $\phi$ (dotted curve)  and axial-vector mesons $a_1=f_1$ (dashed-dotted curve), $K^*$  (dotted curve) and $f_{1}^*$ (short dashed-dotted) in hadronic dense medium at different temperatures $T=10$, $100$, $150$, $180$ and $200\,$MeV.
\label{AVMU1}
}}
\end{figure}

The meson masses are shown for the case with U(1)$_A$ anomaly as a function of baryon-chemical potential at different temperatures in the \lsm$\,$ (left-hand panel) and the P\lsm$\,$ (right-hand panel), where the scalar and pseudoscalar are presented in Figs. \ref{SPMU1} and \ref{SPMU2} while the vector and axial-vector mesons are depicted in Fig. \ref{AVMU1}. 

In left-hand panel of Fig. \ref{SPMU1}, we notice that all masses keep their vacuum values almost unchanged until the baryon-chemical potential reaches the Fermi surface for the light quarks \cite{Mathias Thesis:2009} at $\mu\sim 350\,$MeV. The mass of $\sigma$ meson drops below the mass of $\pi$ meson at the value where the first-order transition should be positioned \cite{Mathias  Thesis:2009}.  This means that the masses of pseudoscalar mesons stay nearly constant until the  phase transition takes place, Fig. \ref{SPMU1}, while the scalar mesons show a stronger melting behavior above the Fermi surface for the light quarks \cite{Mathias Thesis:2009}. The right-hand panel presents the effects of the Ployakov-loop correction introduced to the quark dynamics. This  causes a sharp transition in the mass degeneration. The increase of the melting behavior above $T_c$ derives the masses to be compacted with each other.

In left-hand panel of Fig. \ref{SPMU2}, we find again that all masses stay at their vacuum values until the baryon-chemical potential reaches the Fermi surface for the light quarks at $\mu\sim 350\,$MeV. The meson  masses drops at the first-order transition and $\kappa$ meson drops below to the masses of $K$ and $\eta$ mesons. Only in the curve for $f_0$ meson, the Fermi surface for the strange quarks is clearly visible. The mass of $f_0$ decreases below $\kappa$. Masses of $K$ and $\eta$ decreases only after the light quark phase transition (this the second phase-transition) and degenerates with other meson masses at very high chemical potential $\mu =700\,$MeV. This value decreases as the melting point of the system increases. The first slight drop of $f_0$ meson takes place at $\mu\sim 350\,$MeV, due to the induced drop in the strange condensate. The right-hand panel shows the Ployakov-loop correction introduces quark dynamics. Apparently, this enhances the mass degeneration through the deconfinement phase-transition to appear sharper and faster than in the \lsm.

Fig. \ref{AVMU1} shows the \lsm$\,$ (right-hand panel) and P\lsm$\,$  (right-hand panel) results of vector and axial-vector mesons as function of the baryon-chemical potentials at different fixed temperatures. This gives a systematic study for the variation of heating effect on the hadronic dense medium. In the left-hand panel we find that the axial-vector, $a_1$ and $f_1$ keeps their vacuum values till  $\mu=350\,$MeV. Then, they drop below to the vector mesons $\rho$ and $\omega$. This is accompanied by a strong phase-transition (first-order) and a degeneration in the masses. The axial-vector meson $K^*$ keeps its vacuum value till the same value of baryon-chemical potential. Then, it drops below to vector meson $\rho$ and $\omega$. In this case, this is accompanied by a rapid phase transition (first order). The strange meson states  $f_{1}^*$ and $\phi$ degenerate only at very high chemical potential, $\mu\sim700\,$MeV. These $\mu$-values decrease with increase $T$. Increasing $T$ reduces the baryon-chemical potential, at which the mass degeneration gets compatible with the previous cases and easily gaps the Fermi surface for the light quarks. These would mean that the masses of vector mesons stay nearly constant until the phase transition takes place, while the masses of the axial-vector mesons show a stronger melting above the Fermi surface for the light quarks.

The right-hand panel Fig. \ref{AVMU1} shows the in-medium effect of the baryon-chemical potential (density) on the vector and axial-vector mesons in the presence of Ployakov-loop correction and symmetry breaking. We find that the deconfinement phase-transition has considerable effects on the chiral phase-transition in meson masses, where the restoration of the chiral symmetry breaking becomes sharper and faster than in the \lsm. For example, very close to the critical temperature, $T=180 \,MeV$, the axial-vector mesons $a_1$ and $f_1$ keep their vacuum values till $\mu \sim 180\,$MeV. Then, the two masses become smaller than that of the vector mesons $\rho$ and $\omega$. The axial-vector meson, $K^*$, keeps its vacuum value till $\mu\sim 300\,$MeV. Then, it mass drops below the ones of the vector mesons $\rho$ and $\omega$. At a characteristic value of the baryon-chemical potential, the masses of all mesons degenerate with each other.

\begin{table}[htb]
\begin{center}
\begin{tabular}{ c c c c c }
\hline 
  & Scalar mesons & Pseudoscalar  mesons & Vector  mesons & Axial-vector mesons  \\ 
\hline  \hline
meson 
&
\begin{tabular}{c c c c}
$a_{0}$~& ~$\kappa $~&~$\sigma$~&~$ f_{0}$~\\
\end{tabular}
 & 
\begin{tabular}{c c c c}
~$\pi$~&~ $K $~ &~ $\eta$ ~& ~$\eta ^{'}$~\\
\end{tabular} 
 &
 \begin{tabular}{c c c c}
~$\rho$~& ~$K _{0}^{*}$~ & ~$\omega$~ &~ $ \phi$~\\
\end{tabular}
 & 
\begin{tabular}{c c c c}
~$a_{1}\;$ & $K_1\;$ & $f_1\;$ &  $ f_1^*\;$\\ 
\end{tabular} \\ 
\hline 
$T_{Dissolving}^{Meson}$ [MeV]
&
\begin{tabular}{c c c c}
$200$ & $250$ & $320$ & $320$ \\
\end{tabular}
 & 
\begin{tabular}{c c c c}
$320$ &$230$ & $235$ &  $300$ \\
\end{tabular} 
 &
\begin{tabular}{c c c c}
$195$ &$300$ & $195$& $ 300$\\
\end{tabular}
 & 
\begin{tabular}{c c c c}
$205$ & $250$ & $205$ & $350$\\ 
\end{tabular} \\ \hline 
\end{tabular}
\caption{The approximative dissolving temperature corresponding to the different meson states. \label{dissolve}}
\end{center}
\end{table}

\subsection{Exclusion of Anomalous Terms}
\label{anamoly}

The axial anomalous term U(1)$_A$ is considered by an effective t' Hooft determinant in the Lagrangian, which breaks U(1)$_A$ symmetry \cite{E. Witten:1979,C. Vafa:2007}. This term appears in the anomaly Lagrangian, Eq. (\ref{eq:Lagrangian}), and in the pure mesonic potential, Eq. (\ref{Upotio}), through the parameter $c$. Eliminating this term likely affects the chiral phase-transition and plays an essential role on the phenomenology of scalar and pseudoscalar masses at finite temperature and density. The vector and axialvector masses are not affected by the anomalous term, Eqs. (\ref{m_rho})-(\ref{AV}). It is conjectured that the axial anomaly-breaking term is constant (not depending on temperature and chemical potential) \cite{Schaefer:2009}. In this section, we introduce the influence of the axial anomaly on the meson masses. 

In the case that the anomalous terms depend on the temperature, a fast effective restoration of the axial symmetry takes place \cite{Schaefer:2009}.  It was found that the anomalous term decreases with increasing temperature \cite{Schaefer:2009}.  At very high temperatures, both chiral condensates $\sigma_x$ and $\sigma_y$  degenerate \cite{Schaefer:2009}.

In the case that the chiral condensates depend on the baryon-chemical potential, we find that the upper Fermi surface of the light quarks coincides with the light quark mass, $\mu \approx m_l =300~$MeV, where the chiral condensates are in the broken phase (below the phase transition) and the strange condensate has no influence on the axial anomaly \cite{Schaefer:2009}. The phase transition is mainly estimated by the non-strange condensate $\sigma_x$, while the leap in the strange condensate $\sigma_y$ can be neglected.  Below Fermi surface (above the phase transition), the strange condensate should be taken into account $\mu \approx m_ g=433~$MeV.

\subsubsection{Temperature Dependence}
\label{TDepenanomaly}

\begin{figure}[hbt]
\centering{
\includegraphics[width=5cm,angle=-90]{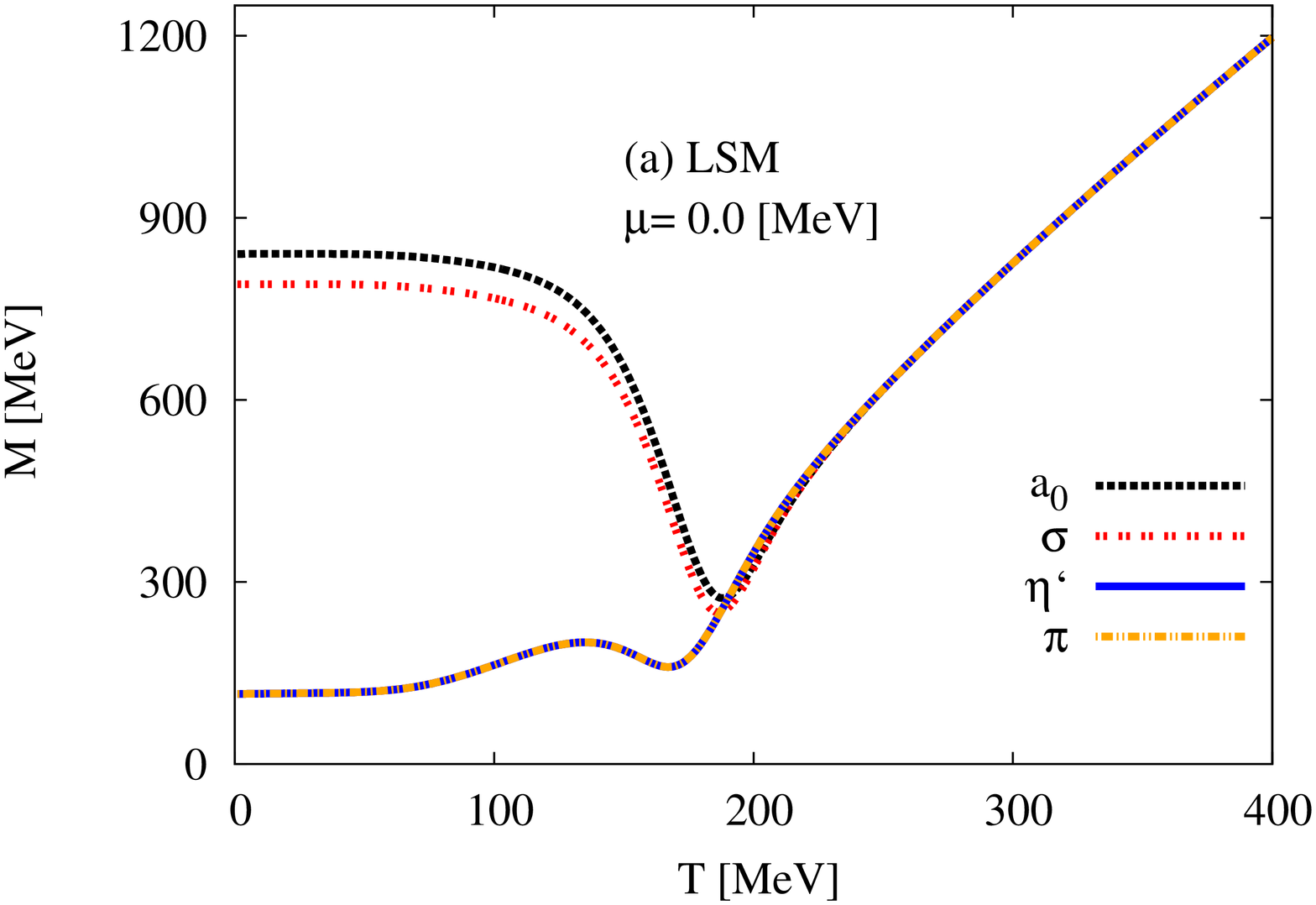}
\includegraphics[width=5cm,angle=-90]{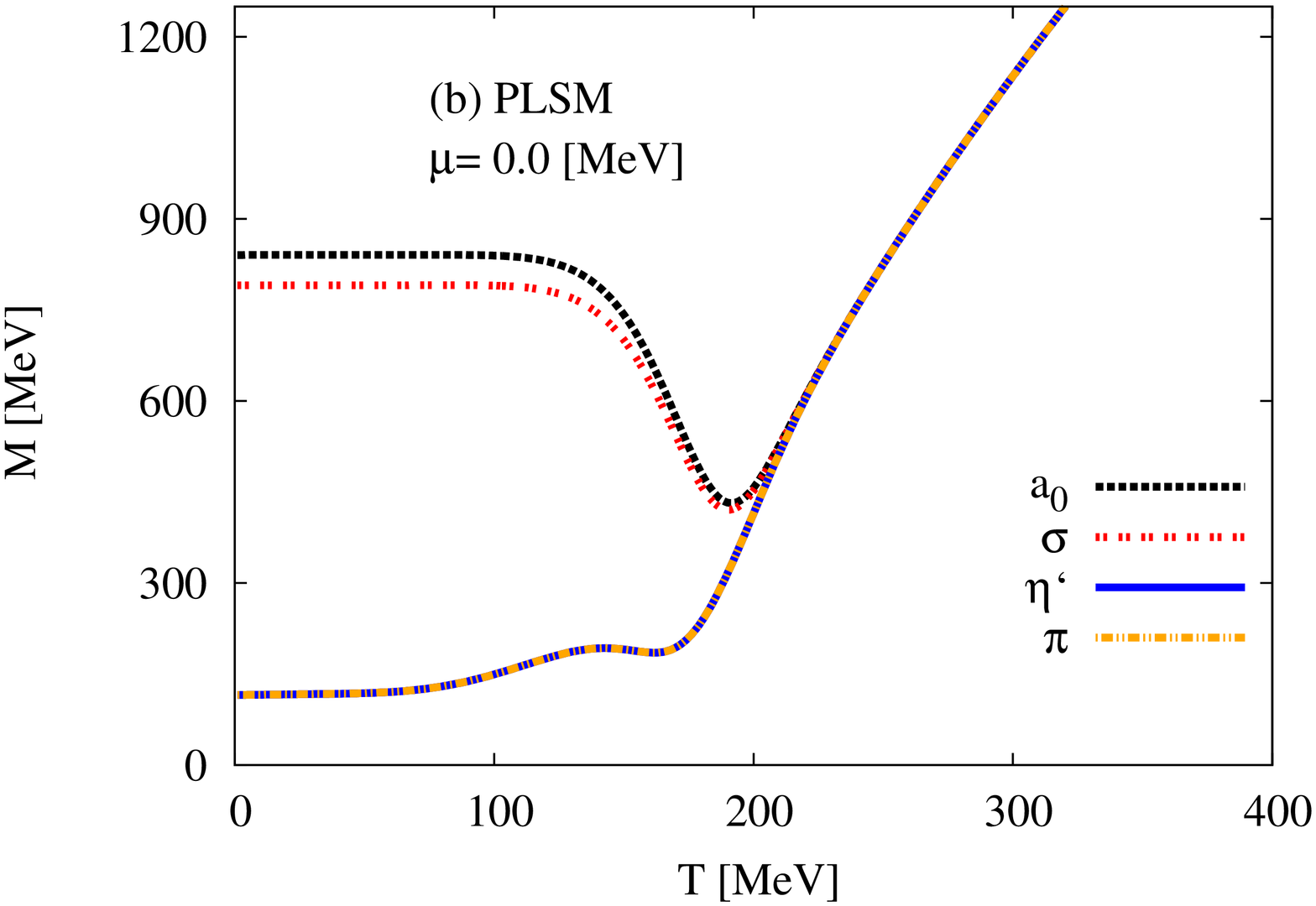}
\caption{\footnotesize (Color online) Left- (\lsm$\,$) and right-hand panel (P\lsm$\,$) show scalar $a_0$ (dashed curve) and $\sigma$ (dotted curve)  and pseudoscalar states $\eta^{,}$ (solid curve)  and $\pi$  (dashed-dotted curve) as function of temperature at vanishing baryon-chemical potentials $\mu=0.0~$MeV.
\label{SMU1wanomaly}
}}
\end{figure}

The thermal evolution of the meson states in case of negligible influence of  the axial anomaly term U(1)$_A$ at vanishing baryon-chemical potential $\mu = 0.0~$MeV, \lsm$\;$ (left-hand panel) and P\lsm$\;$ (right-hand panel), in Fig. \ref{SMU1wanomaly}, show that the critical temperature $T_c$ remains unchanged, the mass gap between the chiral partners vanishes in the restored phase and all meson states begin to degenerate at the chiral restoration temperature $T_{c}^q$ of light quarks. This value of $T_c^q$ does not change when introducing the anomaly term. The introduction of color and gluon dynamics in form of Polyakpov-loop corrections to $a_0$ and $\sigma$. Both drop to $\eta^`$ and $\pi$.

\begin{figure}[hbt]
\centering{
\includegraphics[width=5cm,angle=-90]{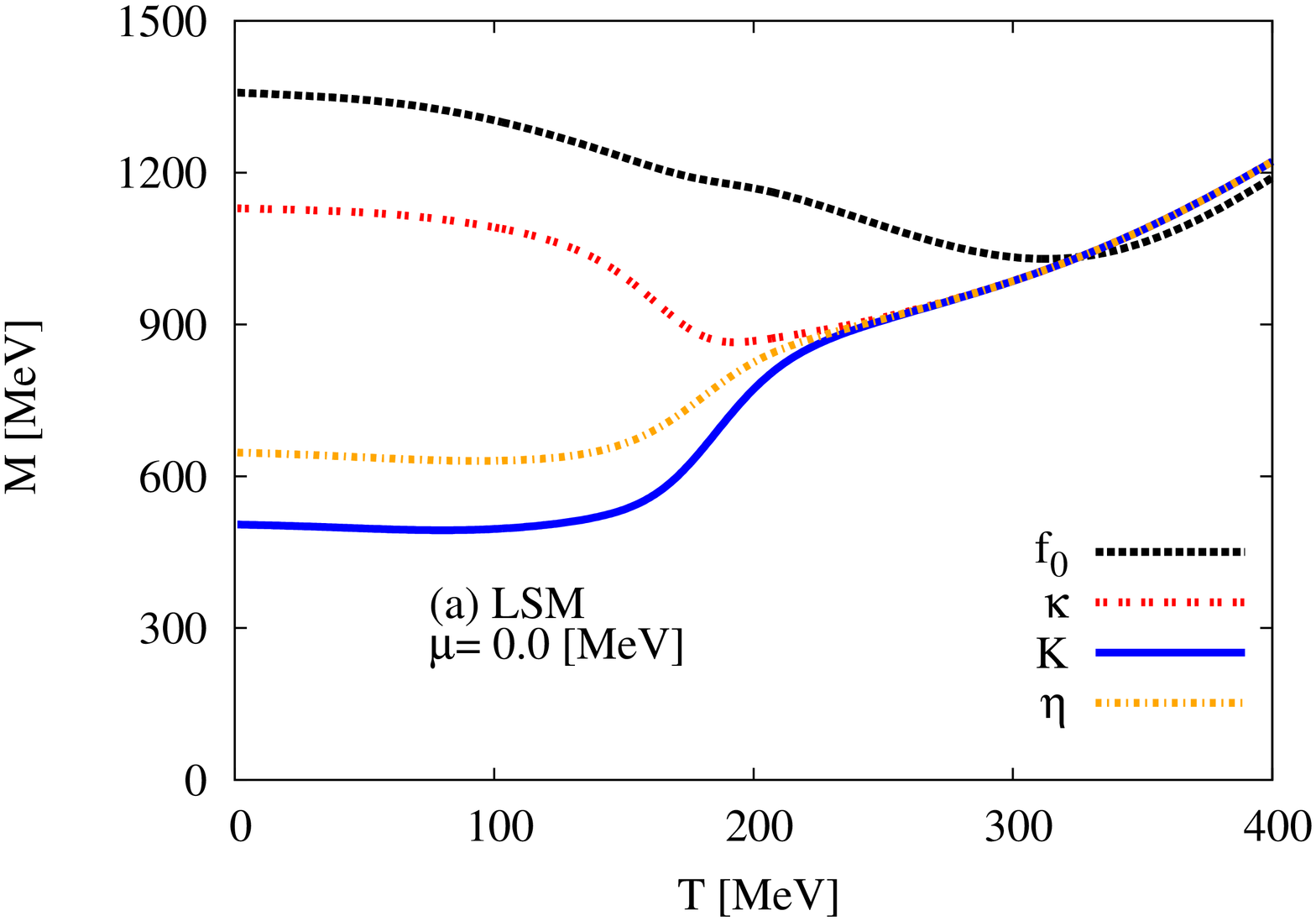}
\includegraphics[width=5cm,angle=-90]{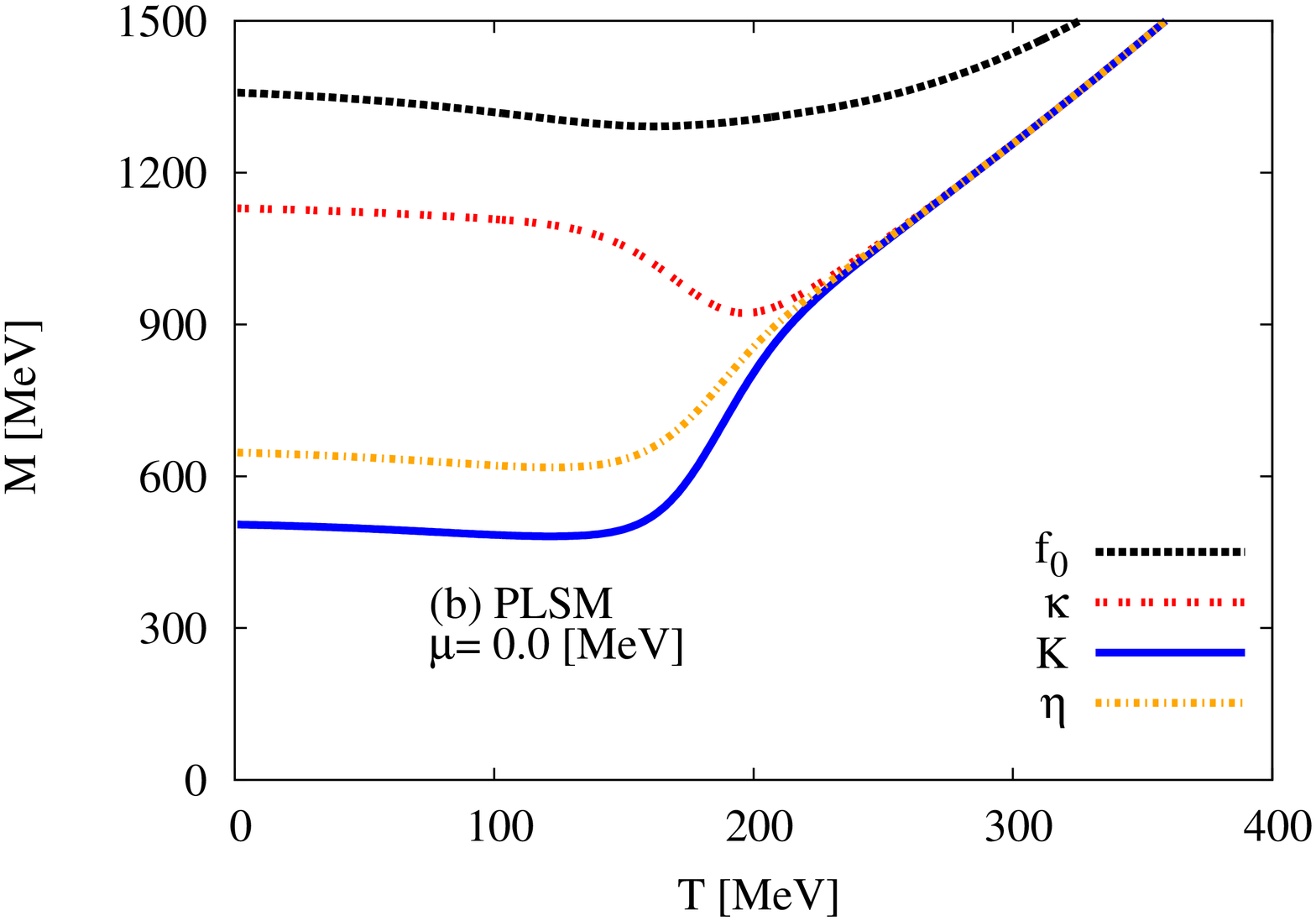}
\caption{\footnotesize (Color online) Left- (\lsm$\,$) and right-hand panel (P\lsm$\,$) show scalars $f_0$ (horizontal dashed curve) and $\kappa$ (vertical dashed curve)  and pseudoscalars $\eta$ (dotted curve)  and $K$  (solid curve)  as function of temperature at vanishing baryon-chemical potentials $\mu=0.0~$MeV.
\label{SMU2wanomly}
}}
\end{figure}

Figure \ref{SMU2wanomly} shows that the chiral restoration remains uncompleted till the temperature exceeds the critical one corresponding to the chiral restoration for light quarks. In presence of an axial anomaly term, it is obvious that the four meson states degenerate at same approximative temperature. The chiral restoration for strange quarks is not fully completed because $\eta$ degenerates with  $\kappa$ and $K$ at values larger than that of the chiral restoration of the light quarks, $T_{c}^q$. These values are not changed in both cases, i.e. with/without anomaly. But they are increased when introducing color and gluon interaction.  

\subsubsection{Density Dependence}

\begin{figure}[hbt]
\centering{
\includegraphics[width=5cm,angle=-90]{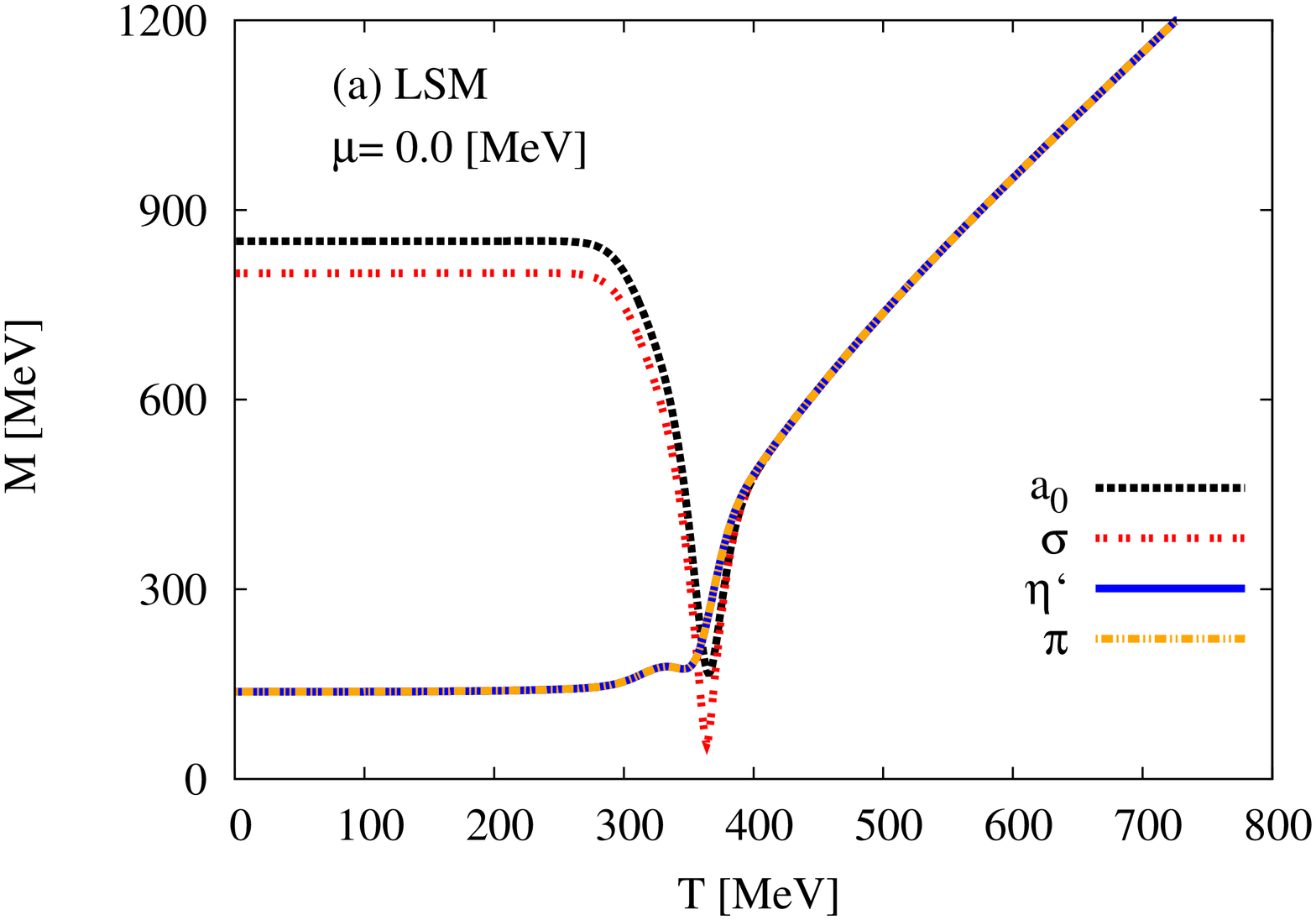}
\includegraphics[width=5cm,angle=-90]{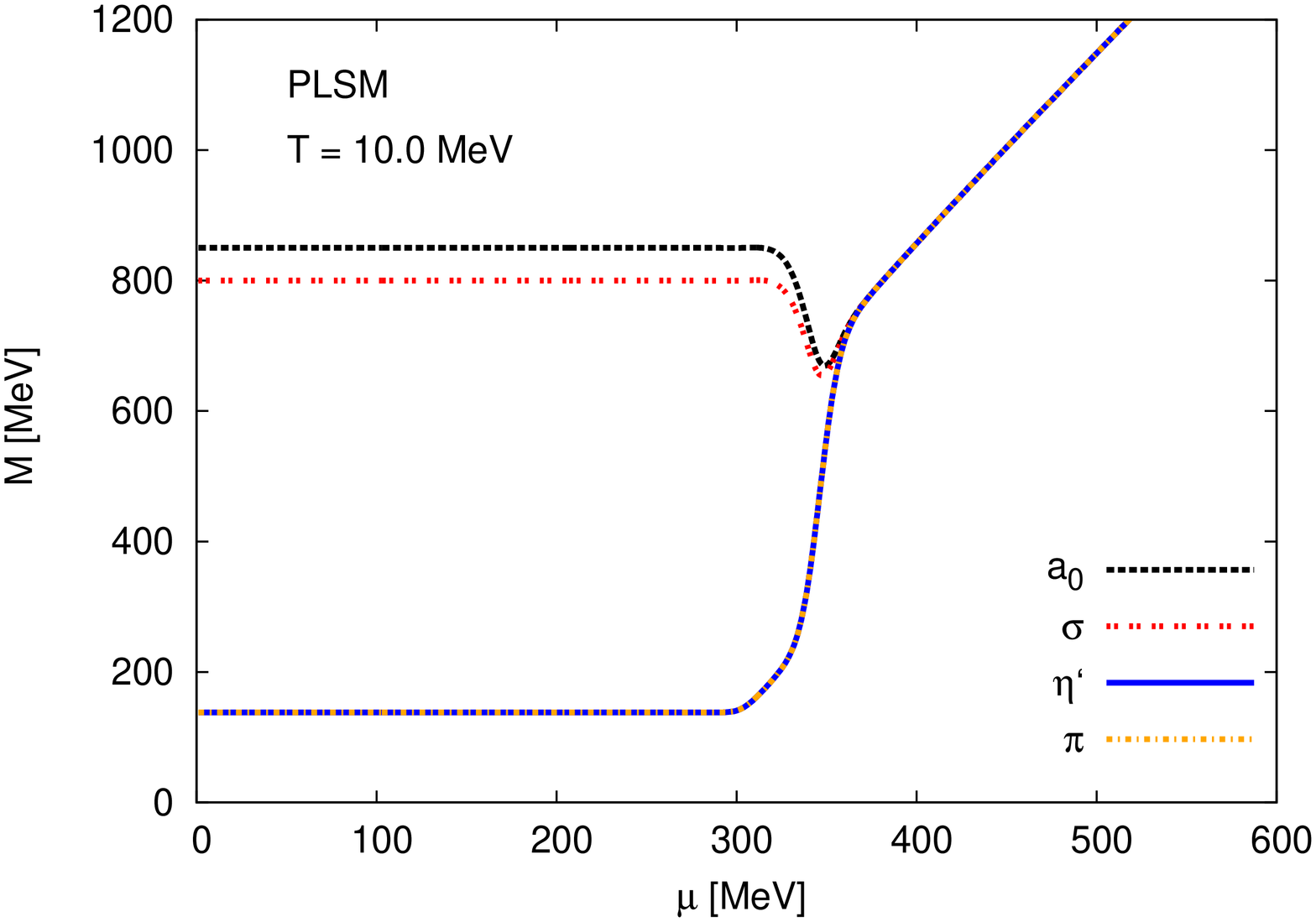}
\caption{\footnotesize (Color online) Left- (\lsm$\,$) and right-hand panel (P\lsm$\,$) present scalars $a_0$ (dashed curve) and $\sigma$ (dotted curve) and pseudoscalars $\eta^{,}$ (solid curve)  and $\pi$  (dashed--dotted curve) in dense medium at fixed temperature $T=10~$MeV.
\label{SMU1Canom}
}}
\end{figure}

The density evolution of mesonic states in case of negligible influence of the axial anomaly term U(1)$_A$ at finite temperate is evaluated at $T=10~$MeV in \lsm$\;$ (left-hand panel) and P\lsm$\,$ (right-hand panel) in Fig. \ref{SMU1Canom}. The critical temperature does not change from the case, in which the axial anomaly term is included. But the introduction of the color dynamics in absence of the axial anomaly term appears in the left-hand panel of Fig. \ref{SMU1Canom}. The limit of the Fermi surface is unchanged in both cases, i.e. with and without anomaly. In Fig. \ref{SMU1Canom}, the drops of $a_0$ and $\sigma$ states to $\eta^`$ and $\pi$ states are slowly. This is sharp and localized in a small region around the critical $\mu_c$. The phase transition is first order. This means that the scalar mesons show a stronger melting behavior, while the introduction of the color dynamics of quarks bears out the pseudoscalar states to have a large melting point as shown in the left-hand panel of Fig. \ref{SMU1Canom}. The degenerate states between all four meson masses are assumed to take place at second-order phase-transition. 

\begin{figure}[hbt]
\centering{
\includegraphics[width=5cm,angle=-90]{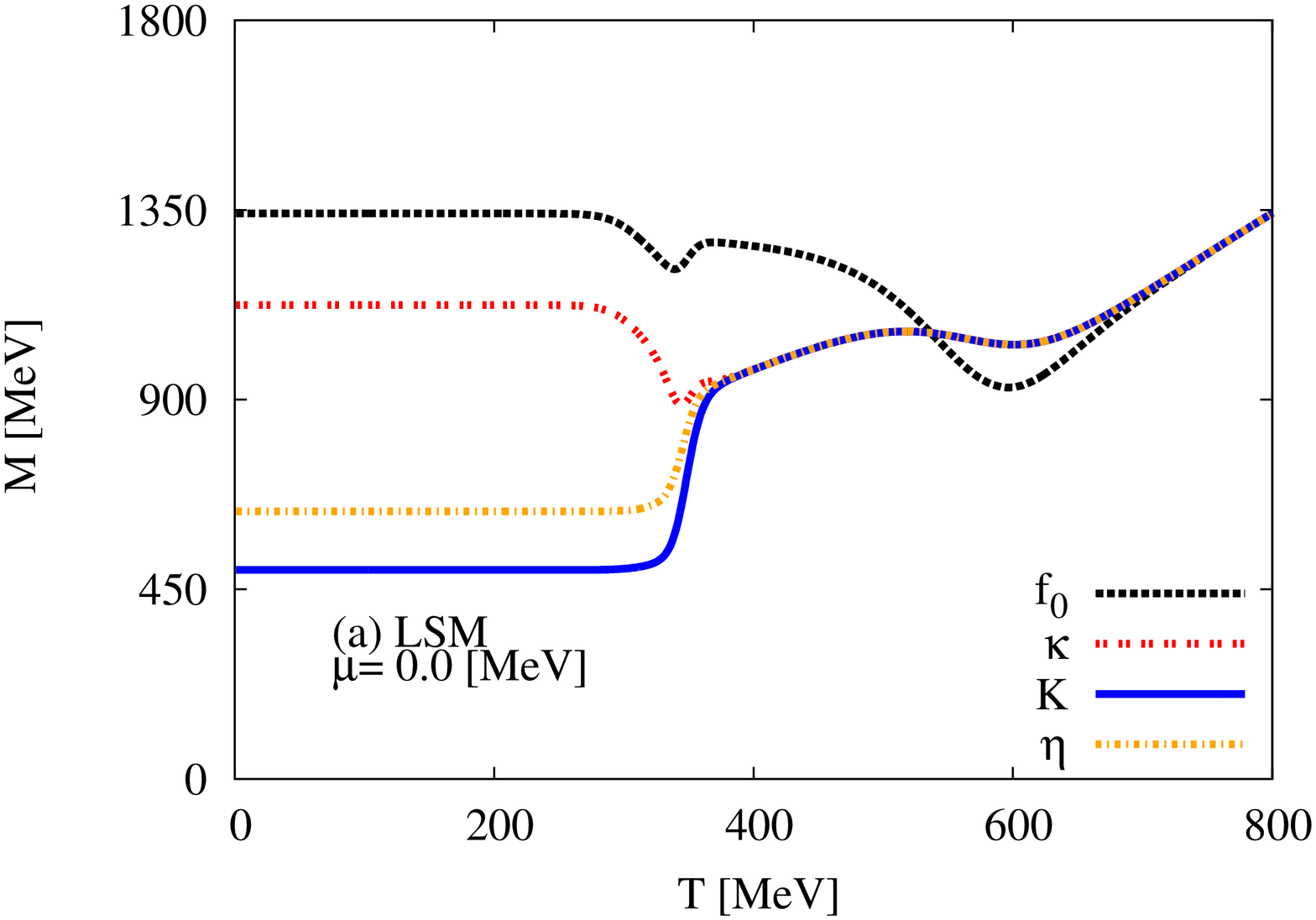}
\includegraphics[width=5cm,angle=-90]{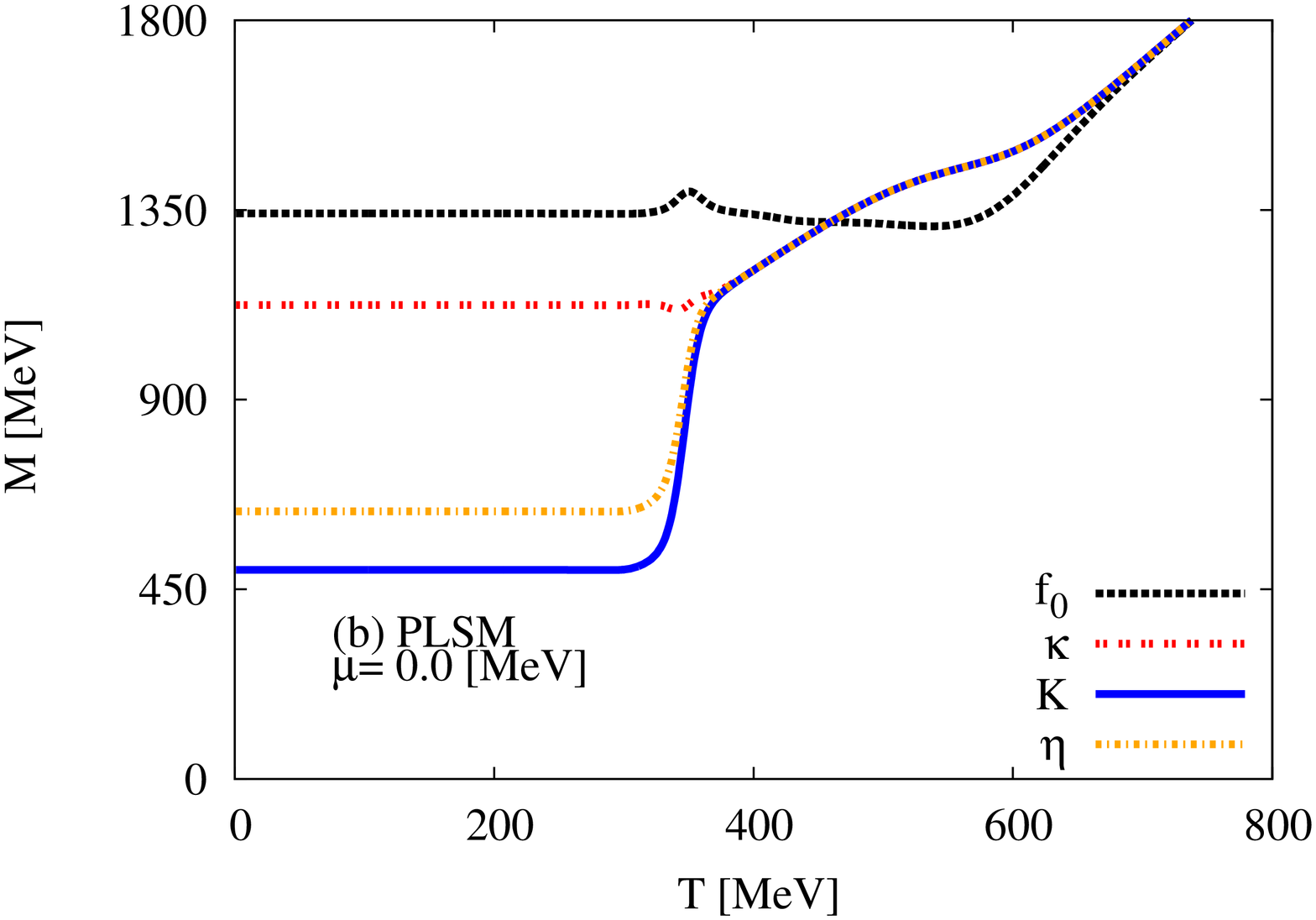}
\caption{\footnotesize (Color online) Left- (\lsm$\,$) and right-hand panel (P\lsm$\,$) show scalars $f_0$ (horizontal dashed curve) and $\kappa$ (vertical dashed curve)  and pseudoscalars $\eta$ (dotted curve)  and $K$  (solid curve)  in dense medium at fixed temperature $T=10~$MeV.
\label{SMU22Canom}
}}
\end{figure}

In Fig. \ref{SMU22Canom}, $\kappa$ state drops to $K$ and $\eta$ states in a first-order phase-transition, but the chiral phase-restoration will not be completed till $f_0$ degenerates at a higher-order phase-transition. All properties obtained in case of including an anomaly case are also observed in the case without anomalous terms.

\subsection{Numerical Parameters of the Model}

Table \ref{tab:sum1} summarizes the numerical values of the various parameters of the present work. These have be deduced from the thermal and density evolution of the scalar and pseudoscalar meson masses \cite{Schaefer:2009}. Here, it is distinguished between the case where the anomalous terms, $c$, are finite and vanishing. 

Table \ref{tab:sum2} summarizes the numerical values of the various parameters of the model used in this work. They have been deduced from the thermal and density evolution of the vector and axial-vector meson masses  \cite{Schaefer:2009}.

\begin{table}[htb]
\begin{center}
\begin{tabular}{ c  c  c  c  c  c  c  c }
\hline 
 & $c\,$[MeV] & $h_x\,$[MeV$^3$] & $h_y\,$[MeV$^3$] & $m^2\,$[MeV$^2$] & $\lambda_1$ & $\lambda_2$ & $g$\\ 
\hline \hline 
With anomaly & $4807.84$ & $(120.73)^3$ & $(336.41)^3$ & $-(306.26)^2$ & $13.49$& $46.48$&$6.5$\\ 
Without anomaly & $0$ & $(120.73)^3$ & $(336.41)^3$& $-(503.55)^2$ & $-4.55$ &$82.47$ &$6.5$\\ 
\hline
\end{tabular}
\caption{Scalar/pseudoscalar: the numerical values of the parameters used in the calculations \cite{Schaefer:2009}. \label{tab:sum1}}
\end{center}
\end{table} 

\begin{table}[htb]
\begin{center}
\begin{tabular}{ c  c  c  c  c  c  c  c }
\hline 
 & $h_1$ & $h_2$ & $h_3$ & $m_1^2 [MeV^2]$ & $\delta_x [MeV^2]$ &  $\delta_y [MeV^2]$ & $g_1$\\ \hline\hline
Vector/axial-vector  & $0$ & $9.87$ & $4.8667 $ & $(0.4135)^2$ & $0$ & $(0.1511)^2$ & $6.5$ \\ 
\hline 
\end{tabular}
\caption{Vector/axial-vector: the numerical values of the parameters used in the calculations  \cite{Rischke:2012}. \label{tab:sum2}}
\end{center}
\end{table}

\section{Normalization to Lowest Matsubara Frequency}
\label{QGP}

\begin{figure}[htb]
\centering{
\includegraphics[width=5.cm,angle=-90]{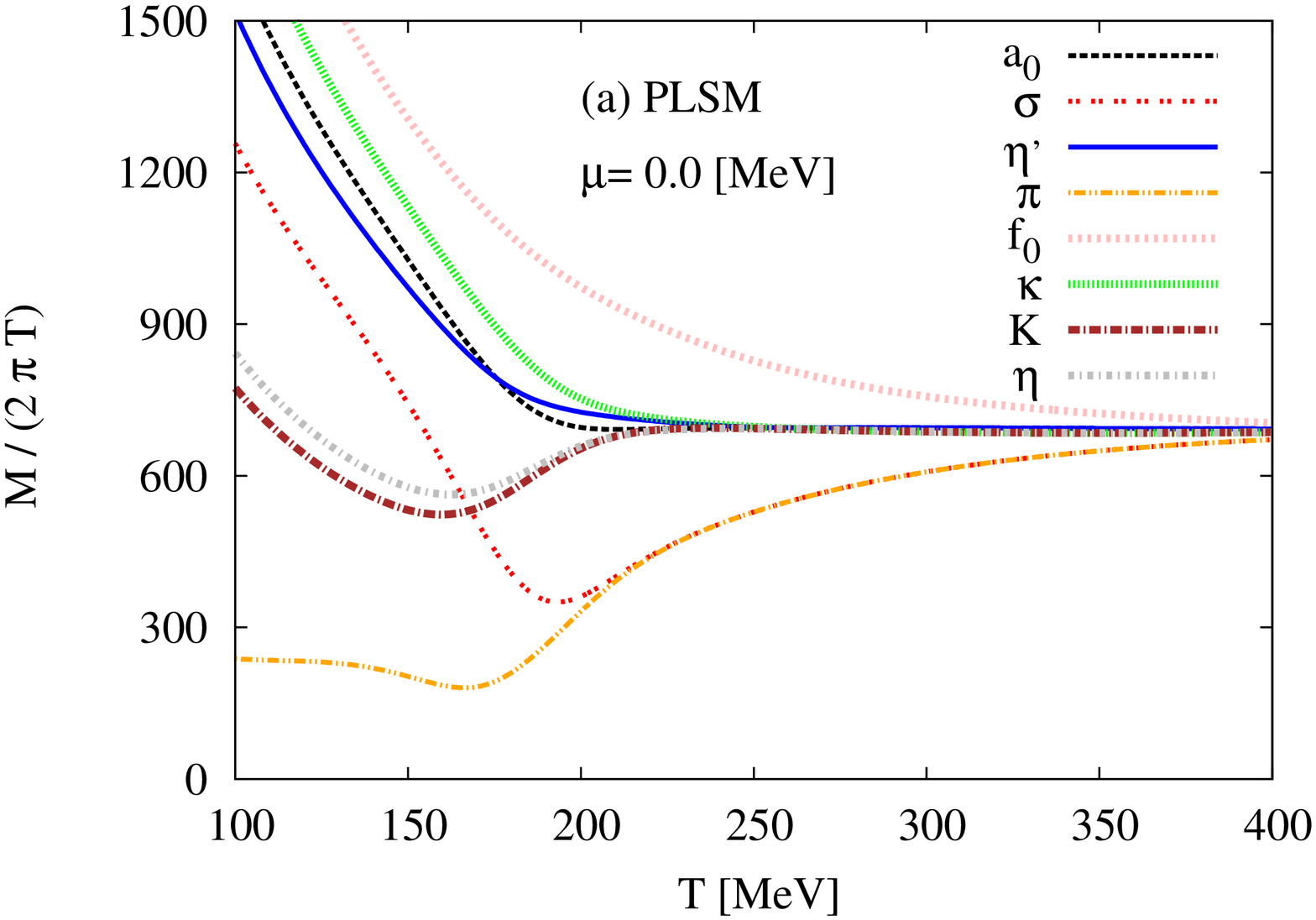}
\includegraphics[width=5.cm,angle=-90]{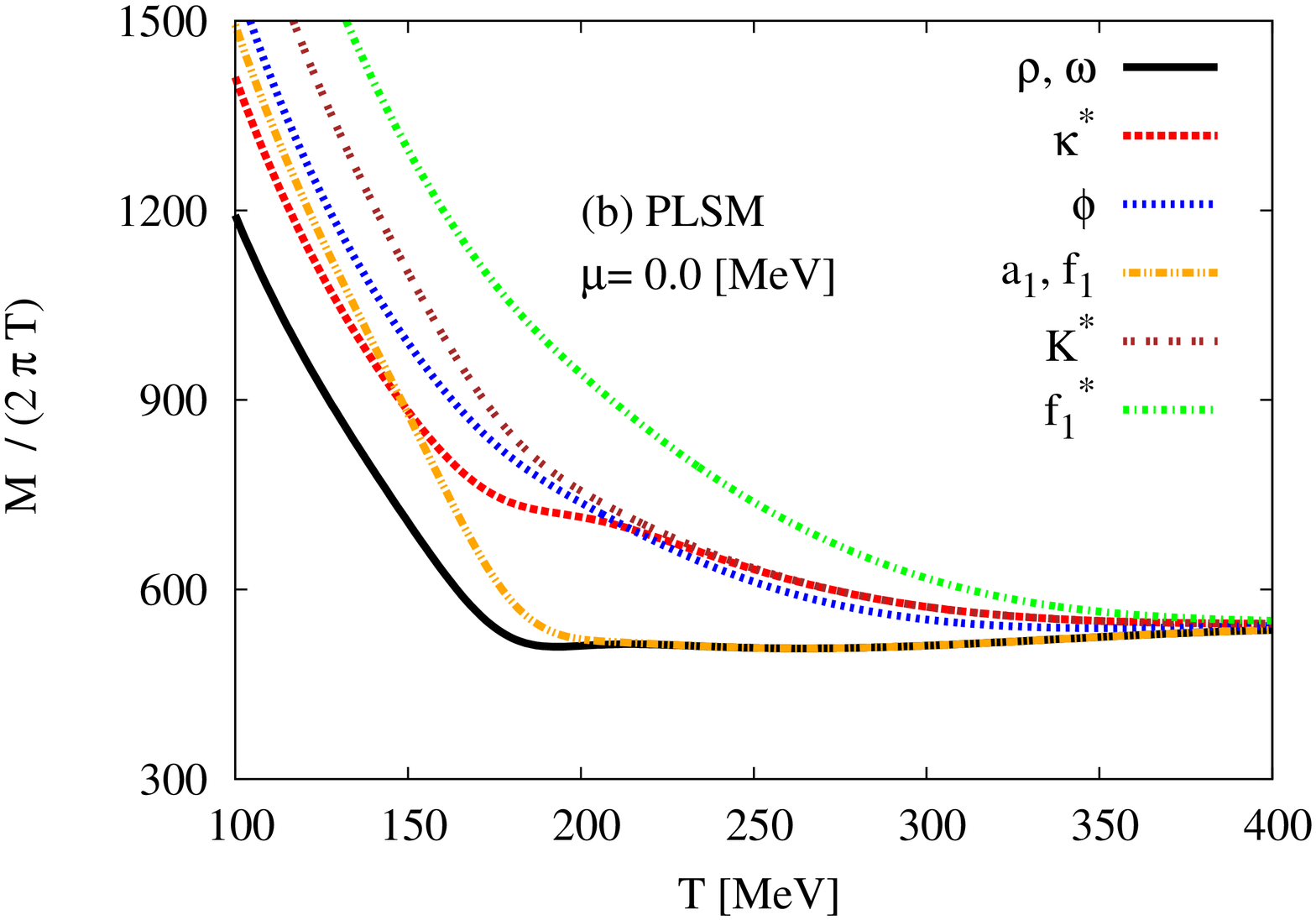}
\caption{(Color online) The left-hand panel shows scalar/pseudoscalar meson sectors in thermal medium at vanishing baryon-chemical potential, while the right-hand panel refers to the vector/axial-vector meson sectors. 
\label{freqMu0}
} }
\end{figure}

In finite temperature field theory, the Matsubara frequencies are a summation over the discrete imaginary frequency, $S_{\eta}=T \sum_{i \omega_n} g(i \omega_n)$, where $g(i \omega_n)$ is a rational function, $\omega_n=2 n \pi T$ for bosons and $\omega_n=(2 n+1)\, \pi\, T$ for fermions and $n=0,\pm 1,\pm 2, \cdots$ is an integer (plays the role of a quantum number). By using Matsubara weighting function $h_{\eta}(z)$, which has simple poles exactly located at $z=i \omega_n$, then
\bea
S_{\eta} &=& \frac{T}{2 \pi i} \oint g(z) \, h_{\eta}(z) \, dz,
\eea
where $\eta=\pm$ stands for the statistic sign for bosons and fermions, respectively. $h_{\eta}(z)$ can be chosen depending on which half plane the convergence is to be controlled, 
\bea
h_{\eta}(z) &=& \le\{\begin{array}{l} \eta \frac{1+n_{\eta}(z)}{T}, \\ \\ \eta \frac{n_{\eta}(z)}{T}, \end{array} \ri.
\eea
where $n_{\eta}(z)=\le(1+\eta e^{z/T} \ri)^{-1}$ is the single-particle distribution function.

The mesonic masses are conjectured to have contributions from the Matsubara frequencies \cite{lmf1}. Furthermore, at high temperatures ($\geq T_c$), the behavior of the thermodynamics quantities, including the quark susceptibilities, besides the masses is affected by the interplay between the lowest Matsubara frequency and the Polyakov-loop correction \cite{lmf2}. We apply normalization for the different mesonic sectors  with respect to the lowest Matsubara frequency \cite{Tawfik:2006B} in order to characterize the dissolving temperature of the mesonic bound states. It is found that the different mesonic states have different dissolving temperatures. This would mean that the different mesonic states have different $T_c$'s, at which the bound mesons begin to dissolve into quarks. Therefore, the masses of different meson states should not be different at $T>T_c$. To a large extend, their thermal and dense dependence should be removed, so that the remaining effects are defined by the free energy \cite{lmf1}, i.e. the masses of {\it free} bosons are defined by $m_l$.

That the masses of almost all mesonic states become independent on $T$, i.e. constructing a kind of a universal line, would be seen as a signature for meson dissociation into quarks. It is a deconfinement phase-transition, where the quarks behave almost freely. In other words, the characteristic temperature should not be universal, as well. So far, we conclude that the universal $T_c$ characterizing the QCD phase boundary is indeed an approximative average (over various bound states).    
 
\subsection{Critical Temperatures and Critical Chemical Potentials}
 
In left-hand panel of Fig. \ref{freqMu0}, it is obvious that each scalar/pseudoscalar meson normalized to the lowest Matsubara frequency begins to dissolve into its quark constituents, individually. At very high temperatures, we expect a universal line independent on temperatures, where many bound particles dissolve, entirely. For example, $\kappa$, $K$, $a_0$, $\eta$, $\eta^,$, $f_0$, $\sigma$ and $\pi$ dissolve, slowly. The right-hand panel shows the same behavior but corresponding to vector/axial-vector mesons, where $\rho$, $\omega$, $a_1$, $f_1$ dissolve, rapidly, while $f_{1}^{*}$ is the last bound state, which seems to survive the typical $T_c$. In Tab. \ref{dissolve}, different meson states are listed corresponding to their dissolving temperatures.

\begin{figure}[htb]
\centering{
\includegraphics[width=5.0cm,angle=-90]{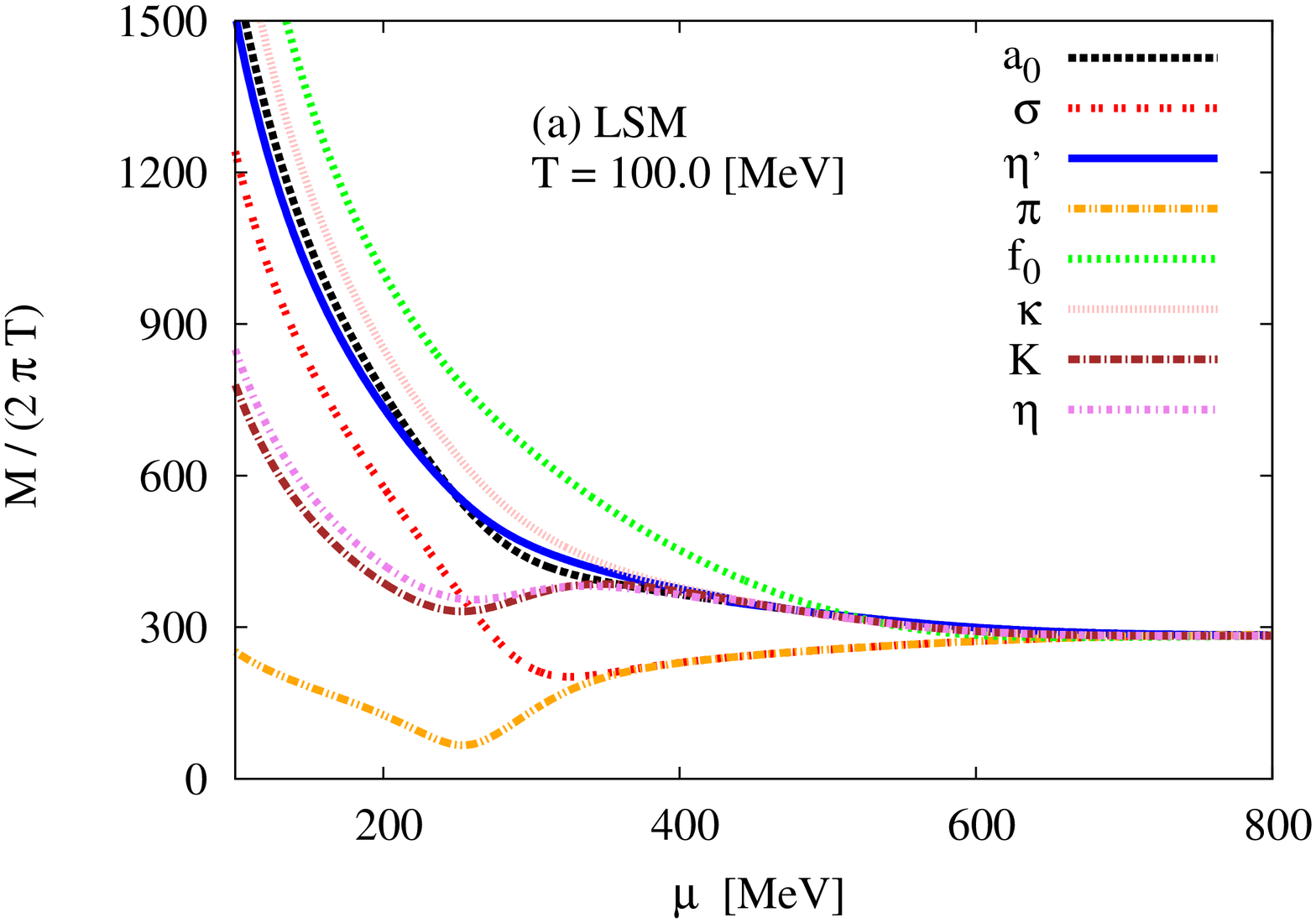}
\includegraphics[width=5.0cm,angle=-90]{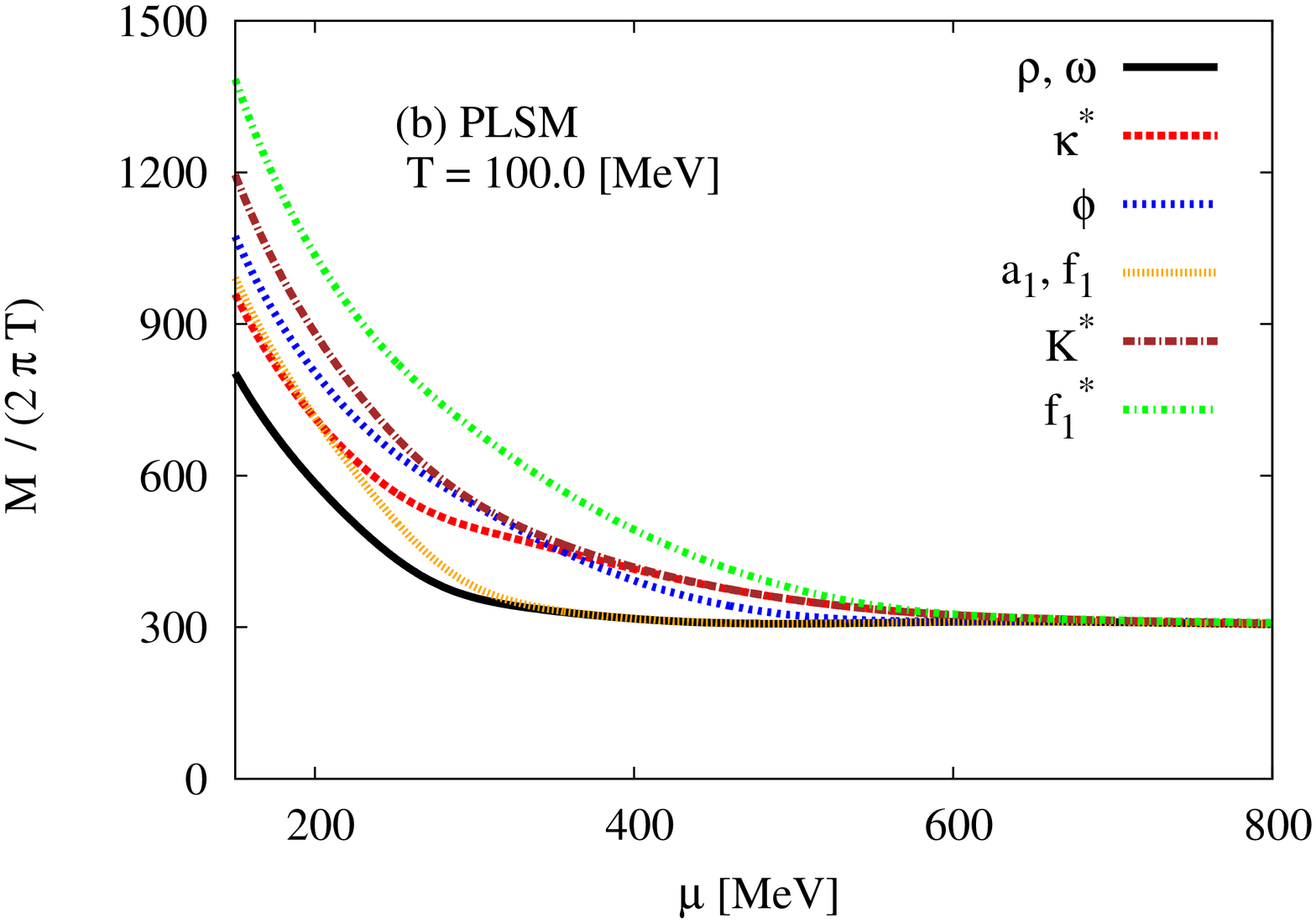}
\includegraphics[width=5.0cm,angle=-90]{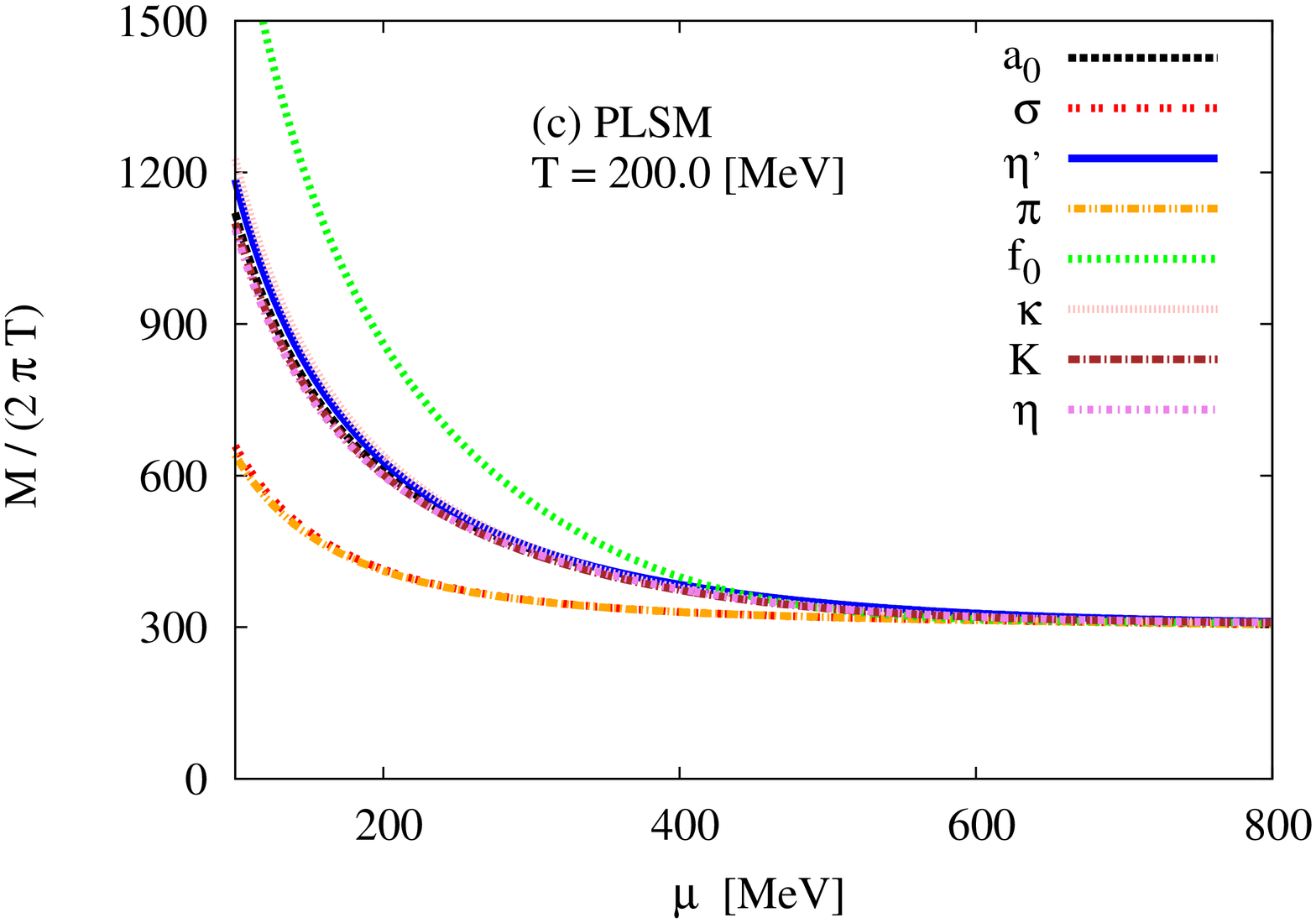}
\includegraphics[width=5.0cm,angle=-90]{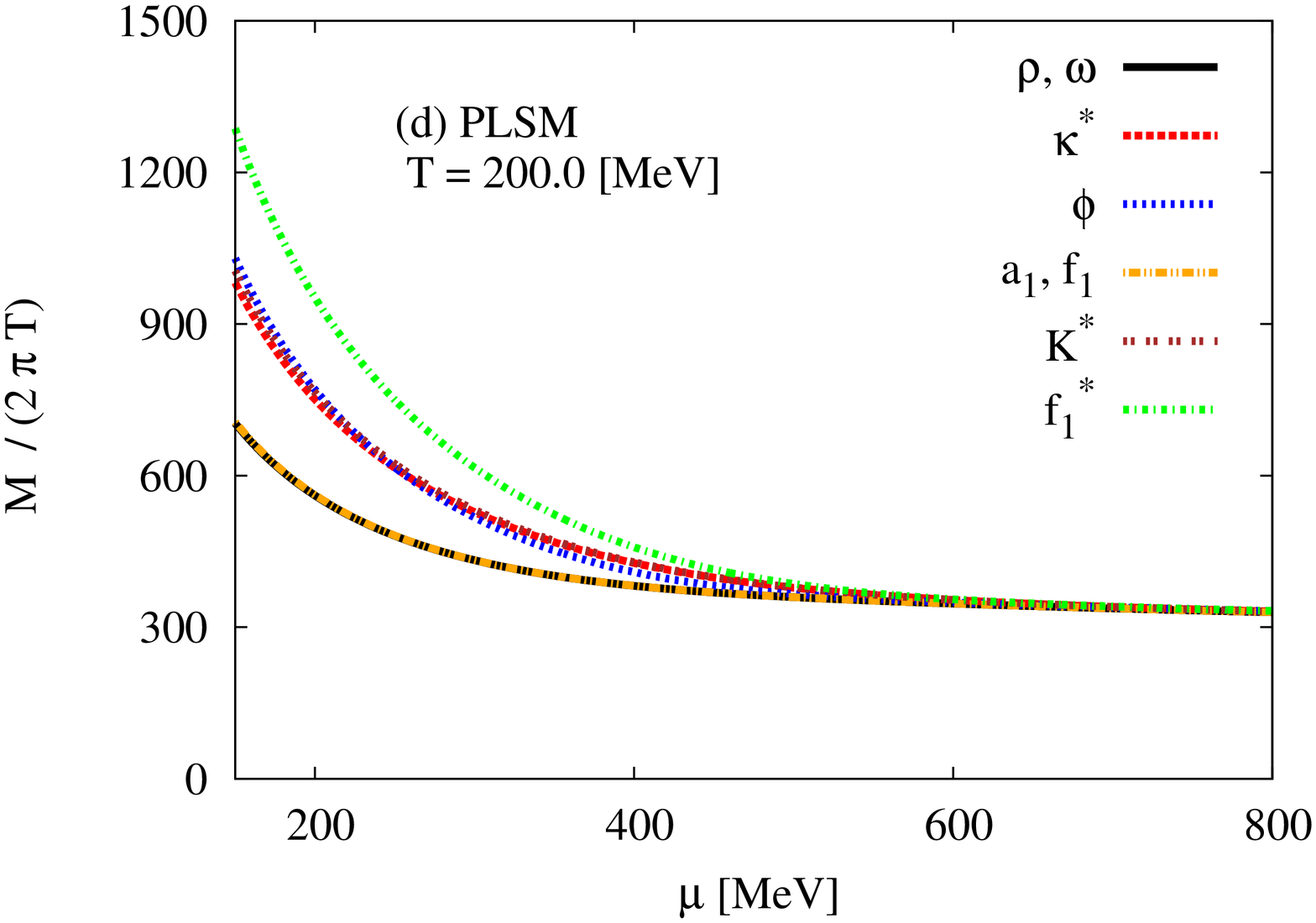}
\caption{(Color online) Left-hand panel shows scalar/pseudoscalar mesons at $T<100\,$MeV and at $T>200\,$MeV. The right-hand panel presents to vector/axial-vector mesons.
\label{freqT}
} }
\end{figure}

In Fig. \ref{freqT}, the top panels show the in-medium effects of the baryon-chemical potential (density) on the masses of mesonic states normalized to the lowest Matsubara frequency at a fixed temperature lower than the typical $T_c$. It is obvious that increasing $\mu$ also brings the masses very close to a universal value, i.e. free energy. The bottom panels show the same but at a fixed temperature higher than he typical $T_c$. Here, increasing $\mu$ seems to bring the masses very close to a universal value in faster and easier way. Finally, it is apparent that the temperature (an essential quantity in the lowest Matsubara frequency) should be corrected/weighted in order for the matrix model to reproduce the mean field results, correctly \cite{lmf2}.

\section{Meson masses in large-$N_c$ limit}
\label{enlargeNc}

When replacing the QCD gauge symmetry SU(3) by SU($N_c$), where $N_c\gg 3$ is the number of colors, we obtain a simpler QCD-theory. In other words, such a large-$N_c$ limit offers an effective approach to study the QCD \cite{largenc}. The relevant quantities can be given in $N_c^{-n}$-series, so that large-$N_c$ dominant can be separated from suppressed terms. In doing this and in order to guarantee consistent large-$N_c$ approach, the QCD coupling $g_{QCD}$ must be scaled \cite{Heinz:2012}; $g_{QCD}\, N_c \rightarrow \text{finite}$, if $N_c \rightarrow \infty$. Accordingly, it was concluded in Ref. \cite{Heinz:2012} that the meson masses scale with $N_c$ while the interaction scales with $N_c^{-(k-2)/2}$. The decay amplitudes are suppressed as $1/\sqrt{N_c}$ \cite{Heinz:2012}. In this limit, the meson masses will be stable and non-interacting. At finite $T$, a non-interacting gas of mesons is realized for $N_c\gg 3$.

In defining the quarkyonic phase \cite{Giacosa:20111} which is conjectured to separate the hadronic from the partonic phases in the $T$-$\mu$ phase diagram, the large-$N_c$ approach has been implemented \cite{Pisarski:2007}.  Accordingly, the limits for the chiral models should be corrected for low-energy hadrons (having densities close to that of the nuclear matter) \cite{Giacosa:20111}. At very low temperatures, this should agree with the Walecka limit \cite{Walecka}. The properties of nuclear matter and chiral phase-transition have been investigated in the large-$N_c$ limit \cite{largenc, Giacosa:20111}. There is only one case in which
nuclear matter does not disappear by increasing $N_c$. This is the naive quarkonium assigned to the lightest scalar resonance \cite{Giacosa:20110}. The low-energy hadrons (light scalar states below $1~$GeV) do not formulate quarkonium states, predominantly. On the other hand, the resulting nucleon-nucleon attraction in the scalar channels is not strong enough to bind nuclei  \cite{largenc,Giacosa:20111}.

\begin{figure}[htb]
\centering{
\includegraphics[width=6cm,angle=-90]{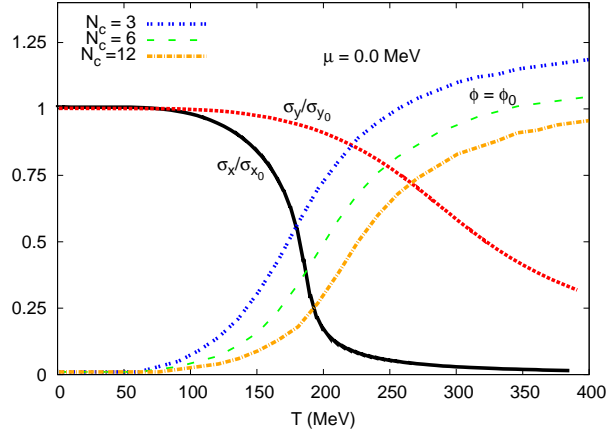}
\label{cond_enlarge_Nc}
\caption{\footnotesize (Color online) The normalized chiral condensates $\sigma_{x}$ and $\sigma_{y}$ (solid and dotted curves, respectively)  and the expectation values of the Polyakov-loop fields, $\phi$ and $\phi ^{*}$, dotted curve ($N_c=3$), dashed curve ($N_c=6$), dotted dash ($N_c=12$) respectively, are given as function of the temperature at vanishing baryon-chemical potential.
}}
\end{figure}

\begin{table}[htb]
\begin{center}
\begin{tabular}{ c c c c }
\hline  
 $N_c$ & $3$ & $6$ & $12$ \\ 
\hline \hline 
$T_c^l$ [MeV] & $181$ & $189$  & $195$ \\ 
\hline 
$T_c^s$ [MeV] & $ 225$ & $ 245$ & $ 270$ \\ 
\hline  
\end{tabular} 
\caption{Dependence of the critical temperatures for light- $T_c^l$ and strange-quark $T_c^s$ on $N_c$.}
\label{TC_NC}
\end{center}
\end{table}

In order to study the behavior of the meson masses with varying $N_c$, we start with the PLSM normalized chiral-condensates, $\sigma_x$ and $\sigma_y$, and the Polyakov-loop fields, $\phi$ and $\phi ^{*}$, at finite temperatures and vanishing baryon-chemical potential, Fig. \ref{cond_enlarge_Nc}. We find that  $\phi$ and $\phi ^{*}$ are good indicators for the deconfinement phase-transition. Both order parameters possess information about the confining glue-sector to the effective chiral-model, the LSM. From the quarks-antiquarks potential, Eq. (\ref{phais2}) and (\ref{PloykovPLSM}), it is obvious that the Polyakov-loop expectation values vary with $N_c$. We expect that the deconfinement phase-transition moves to higher critical-temperatures with increasing $N_c$ and $T_c \rightarrow \infty$ when $N_c \rightarrow \infty$.  Tab. \ref{TC_NC} summarizes $T_c$ for light and strange quarks at different $N_c$.

\begin{figure}[htb]
\centering{
\includegraphics[width=8cm,angle=-90]{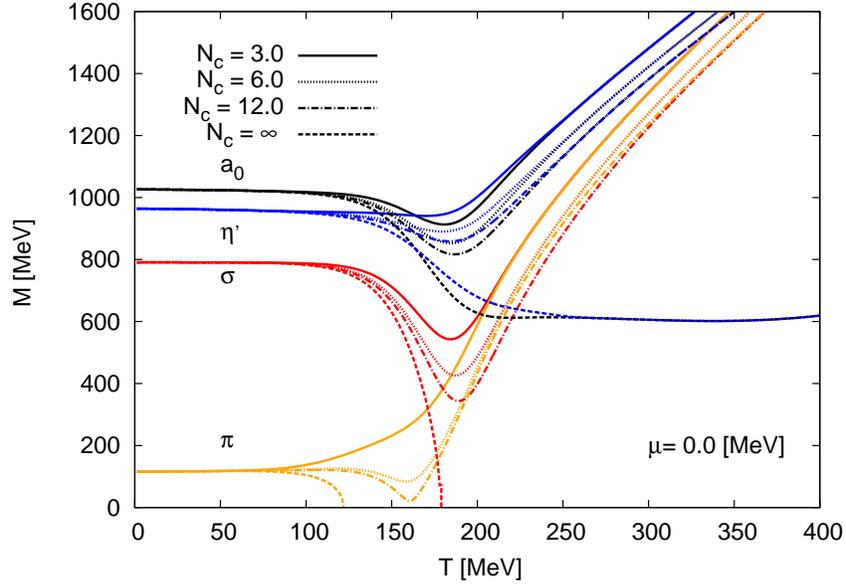}
\caption{\footnotesize (Color online) The scalar  meson masses are given as function of  $T$ at $\mu=0$ and $N_c= 3$ (solid curves), $N_c= 6$ (dotted curves), $N_c= 12$ (dash-dotted curves ) and  $N_c\rightarrow \infty$ (dashed curves).
\label{Mass_enlarge_Nc1}}}
\end{figure}

\begin{figure}[htb]
\centering{
\includegraphics[width=8cm,angle=-90]{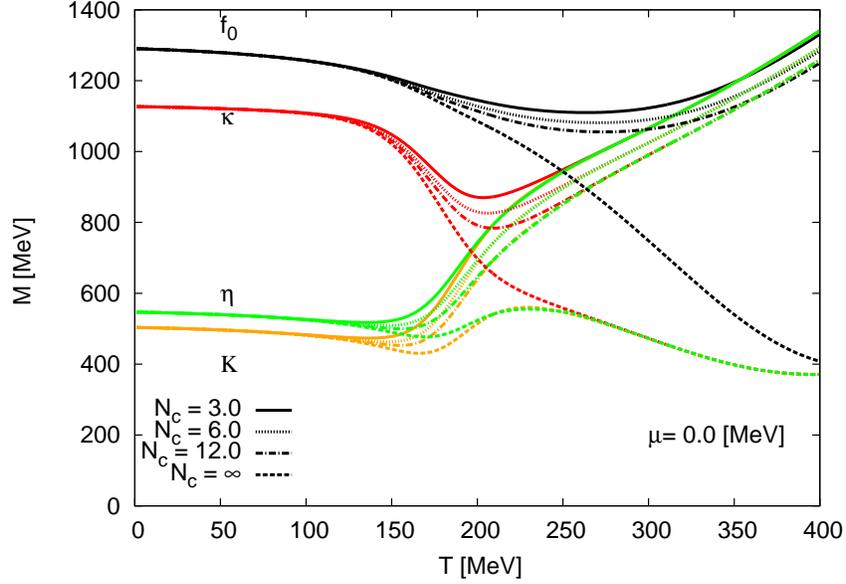}
\caption{\footnotesize (Color online) The same as in Fig. \ref{Mass_enlarge_Nc1} but for pseudoscalar meson masses.
\label{Mass_enlarge_Nc2}}}
\end{figure}

\begin{figure}[htb]
\centering{
\includegraphics[width=8cm,angle=-90]{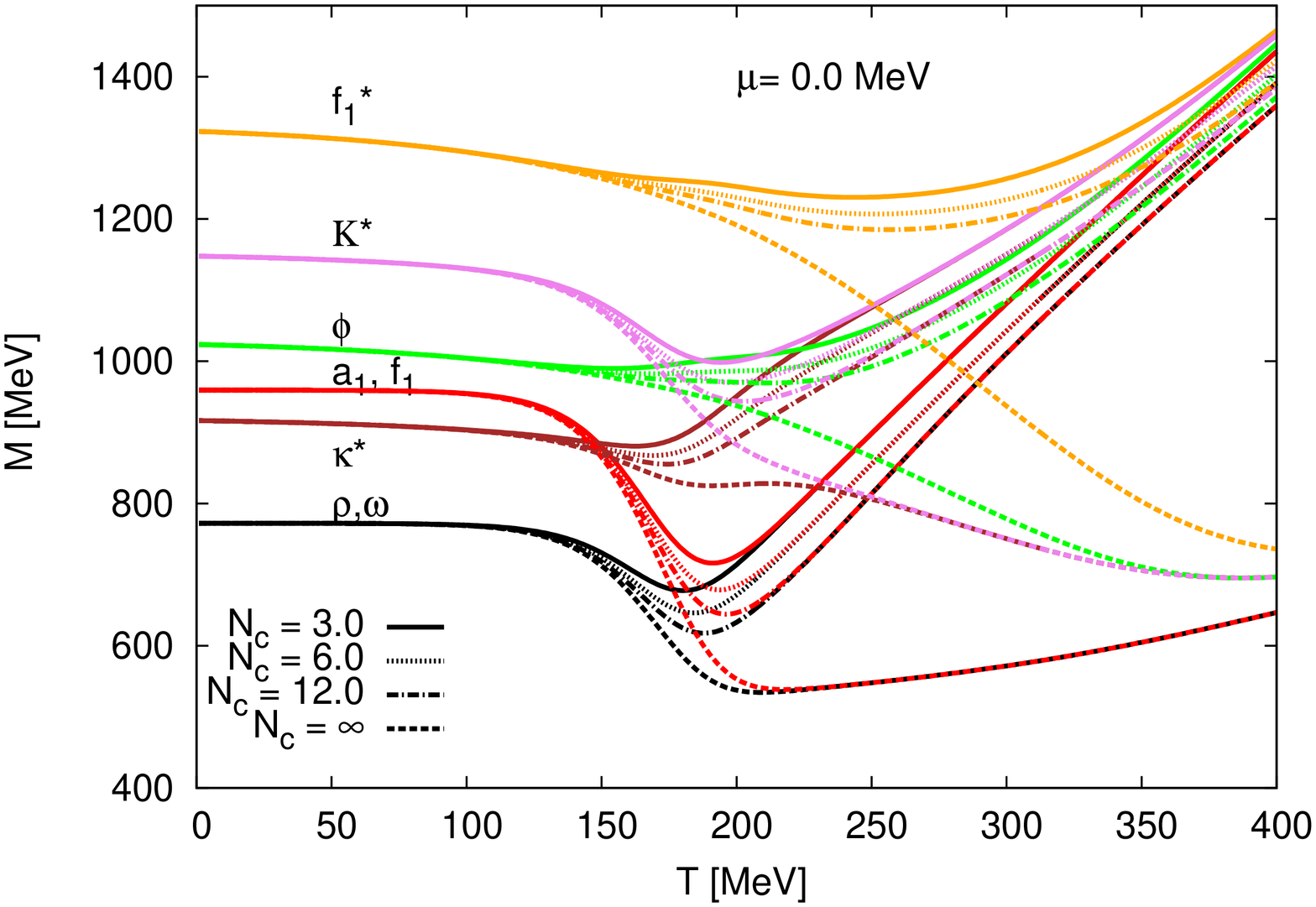}
\caption{\footnotesize (Color online) The same as in Fig. \ref{Mass_enlarge_Nc1} but for axial and axialvector meson masses.
\label{Mass_enlarge_Nc3}}}
\end{figure}

Fig. \ref{Mass_enlarge_Nc1} shows the scalar meson sectors at different $N_c$ as function of  $T$ at $\mu=0$ and $N_c= 3$ (solid curves), $N_c= 6$ (dotted curves), $N_c= 12$ (dash-dotted curves ) and $N_c\rightarrow \infty$ (dashed curves). The masses of all mesons are not influenced when varying $N_c$. It seems that the mesons are stable and non-interacting, especially at densities close to that of the nuclear matter. At very low temperatures, the results seem to agree with a Walecka-like model \cite{Walecka}. The meson channels can be divided into three regions; one at low $T$, one around $T_c$ and one at very high $T$.
\begin{itemize}
\item  The first region is established,  where the strong force between quarks should be dominant and the mass degeneration appears despite of the variation of  $N_c$. This can be interpreted as the effect of the vacuum contributions on the chiral symmetry-restoration. 
\item The second region takes place due to fluctuations in the variation of colors $N_c$ relating to the deconfinement phase-transition at $T_c$. 
\item In the last region, the bosonic thermal-contributions are dominant and the mass gap between mesons seems to disappear. The mesonic states degenerate at large  $N_c$. 
\end{itemize}
In the large-$N_c$ limit, the meson masses are stable and noninteracting at low $T$. They keep the mass gap between the different meson channels. At high $T$, this gap disappears and the masses become $T$-independent.  Except $\pi$ and $\sigma$, the other scalar meson masses are $T$-independent at large $N_c$ and high $T$. For the pseudoscalar meson masses, Fig. \ref{Mass_enlarge_Nc2}, the large $N_c$ limit unifies the $T$-dependence of all states in a universal bundle. The same is also observed for axial and axialvector meson masses in the large-$N_c$ limit, Fig. \ref{Mass_enlarge_Nc3}.

\section{Conclusions}
\label{sec:disc}

There are various approaches implementing theoretical descriptions of the hadron masses in thermal and hadronic dense medium \cite{Rischke:2007,Lenaghan:2000ey,Schaefer:2009,V. Tiwari:2009,V. Tiwari:2013}.  The NJL (or PNJL) studies the thermal spectrum of eight mesons; four scalars and four pseudoscalars at vanishing and finite baryon-chemical potential \cite{NJL:2013,P. Costa:PNJL}. Previous works using \lsm$\,$ (or PQM) focused on the study of (pseudo)-scalar mesons at finite temperature but vanishing density (baryon-chemical potential) \cite{Rischke:2007,Lenaghan:2000ey,Schaefer:2009,V. Tiwari:2009,V. Tiwari:2013} and described the vacuum phenomenology of some states in scalar and vector meson nonets, besides the comparison with the experimental measurements for the decay width and the scattering length \cite{Dirk Hparameters:2010,Rischke:2010 vacuum,Rischke:2012,Wolf:2011 vacuum,Rischke:2011 vacuum,Rischke:2010 decay}. 

In the present work, a systematic study using the chiral symmetric linear $\sigma$-model is  introduced. The scalar, pseudoscalar, vector, and axial-vector fields are included. The representation of all these four categories in dependence on the temperature and on the baryon-chemical potential is taken into consideration. This allows us to define the characteristics of the chiral phase-structure for all these mesonic states, i.e. in thermal and hadronic dense medium and determine the critical temperature and density at which each mesonic state breaks into its free quarks.

At vanishing temperature, the scalar, pseudoscalar, vector and axial-vector meson nonets are confronted to the experimental measurements reported by PDG \cite{PDG:2012}. Also, we compare the results with the lattice QCD calculations \cite{HotQCD,PACS-CS} for pseudoscalar and vector mesons. The scalar and pseudoscalar spectrum calculated from PNJL \cite{NJL:2013,P. Costa:PNJL} is compared with the present work, as well. We first want to highlight that the uncertainties are deduced from the fitting for the parameters used in calculating the equation of states and some other thermodynamic quantities. The fitting requires experimental inputs for axial/axialvector and scale/pseudoscalar states. Thus, we conclude that the results are very precise for some light hadron resonances. The effects of the chiral condensate and the deconfinement phase-transition would play an important role in charactering the chiral phase-structure of many hadrons and therefore, explain the differences seen in the heavy states. The PNJL model is limited to study (pseudo)-scalar meson states. Only pseudoscalar and vector meson masses are available in the lattice QCD calculations (HotQCD Collaboration) \cite{HotQCD} and (PACS-CS Collaboration) \cite{PACS-CS}.  Relative to these two approaches, it can be concluded that the present work reproduces well the mesonic spectrum.

In order to investigate the influence of the Polyakov-loop potential on the chiral symmetry-restoration, the present results are compared with P\lsm$\,$. The P\lsm$\,$ mainly describes the chiral condensates in non-strange $\sigma_x$ and strange $\sigma_y$ condensate in additional of the deconfinement phase-transition, $\phi$ and $\phi^{*}$, in temperature- and  density- (baryon-chemical potential-) dependence. This allows the estimation of the spectrum of some mesonic states in SU(3) as a result from the chiral phase-structure of scalar/pseudoscalar and axial/axialvector states at various densities and temperatures. First, we compare the critical temperatures estimated from at the phase transition and from the order parameters. We found that the chiral phase-transition gets shifted to higher temperatures as a result of the inclusion of the Polyakov loop in \lsm$\,$. In the mesonic masses,  the thermal bosonic contributions decrease with increasing the temperature, while the fermionic contributions increases at high temperature. At low temperatures, the fermionic contributions are negligible. The early (related to low critical temperature and/or small chemical potential) melting of the strange condensate $\sigma_y $ relative to the non-strange one can be interpreted due to the mass degeneration at larger values of temperature and/or chemical potential.  In the phase, where the symmetry is explicitly broken in P\lsm$\,$, the meson masses generated  by P\lsm$\,$ have a good agreement with the experimental results.

We have illustrated that the P\lsm$\,$ can be used to check which mesonic states degenerate with (an)other one(s) and which states degenerate faster relative to the other ones, especially near the Fermi surface. The limitation that all hadrons should melt at a universal critical temperature (QCD phase boundary) can be understood as an approximation. We conclude that each bound state would have a characteristic temperature and density (baryon-chemical potential) at which it dissolves to its free quarks. We plan to extend this study by including more mesonic states and characterizing their thermal and dense evolution. Also, we want to introduce some low-lying baryonic states. Such a plan requires a basic modification of the Lagrangian. The normalization of various meson masses to the lowest Matsubara frequency removes all thermal dependence of the bound mesons and estimates the individual dissolving temperatures. It has been found that the various mesonic states have different dissolving temperatures and baryon-chemical potentials, i.e. they survive the {\it typically-averaged} QCD phase boundary, defined by the QCD critical temperatures with varying baryon-chemical potentials.

We have studied the thermal behavior of meson masses in the large-$N_c$ limit. At low temperatures, we find that the meson masses are stable and non-interacting. With increasing temperature, they keep the mass gap between the different meson channels. At high $T$, this gap disappears and the masses become $T$-independent. The scalar meson masses are $T$-independent at large $N_c$ and high $T$ (except $\pi$ and $\sigma$). For the pseudoscalar meson masses, the large $N_c$ limit unifies the $T$-dependence of all states in a universal bundle. The same is also observed for axial and axialvector meson masses in the large-$N_c$ limit.

\section*{Acknowledgements}
This research has been supported by the World Laboratory for Cosmology And Particle Physics (WLCAPP), Cairo-Egypt, http://wlcapp.net/.  The authors are very grateful to the anonymous referee for her/his constructive suggestions. 
AT would like to thank Dirk H. Rischke and Denis Parganlija for the fruitful discussions the careful reviewing of the script!



\end{document}